\documentclass[review,14pt]{elsarticle}
\usepackage{verbatim,amsfonts}
\usepackage{color,ulem}
\usepackage{graphicx}
\usepackage[utf8]{inputenc}
\usepackage{bm}
\usepackage{epigraph}
\usepackage{amsmath}

\newcommand{\dfr}[2]{\frac {\displaystyle #1}{\displaystyle #2}}

\begin{document}

\begin{frontmatter}

    \title{Gapped Momentum States}

    \author{Matteo Baggioli} \ead{matteo.baggioli@uam.es}

    \address{Instituto de Fisica Teorica UAM/CSIC,
c/ Nicolas Cabrera 13-15, Cantoblanco, 28049 Madrid, Spain}
    \author{Mikhail Vasin} \ead{professorvasin@gmail.com}
    \author{Vadim Brazhkin} \ead{brazhkin@hppi.troitsk.ru}

    \address{Institute for High Pressure Physics, RAS, 142190, Moscow, Russia}
    \author{Kostya Trachenko} \ead{k.trachenko@qmul.ac.uk}

    \address{School of Physics and Astronomy, Queen Mary University of London, Mile End Road, London, E1 4NS, UK}

    \begin{abstract}

  Important properties of a particle, wave or a statistical system depend on the form of a dispersion relation (DR). Two commonly-discussed dispersion relations are the gapless phonon-like DR and the DR with the energy or frequency gap. More recently, the third and intriguing type of DR has been emerging in different areas of physics: the DR with the gap in momentum, or $k$-space. It has been increasingly appreciated that gapped momentum states (GMS) have important implications for dynamical and thermodynamic properties of the system. Here, we review the origin of this phenomenon in a range of physical systems, starting from ordinary liquids to holographic models. We observe how GMS emerge in the Maxwell-Frenkel approach to liquid viscoelasticity, relate the $k$-gap to dissipation and observe how the gaps in DR can continuously change from the energy to momentum space and vice versa. We subsequently discuss how GMS emerge in the two-field description which is analogous to the quantum formulation of dissipation in the Keldysh-Schwinger approach. We discuss experimental evidence for GMS, including the direct evidence of gapped DR coming from strongly-coupled plasma. We also discuss GMS in electromagnetic waves and non-linear Sine-Gordon model. We then move on to discuss the recently developed quasihydrodynamic framework which relates the $k$-gap with the presence of a softly broken global symmetry and its applications. Finally, we review recent discussions of GMS in relativistic hydrodynamics and holographic models. Throughout the review, we point out essential physical ingredients required by GMS to emerge and make links between different areas of physics, with the view that new and deeper understanding will benefit from studying the GMS in seemingly disparate fields and from clarifying the origin of potentially similar underlying physical ideas and equations.

    \end{abstract}


\end{frontmatter}

\tableofcontents
\listoffigures

\section{Introduction}
\label{section:introduction}
A dispersion relation describes several fundamental properties of a particle, quasiparticle or a wave, by providing a relationship between their energy (frequency) and momentum (wavenumber). The dispersion relation (DR) also provides important insights into the medium where particles and quasiparticles propagate. We rely on a particular form of DR to calculate many observables such as phase and group velocity as well as statistical properties such as density of states, system energy and its derivatives. This includes diverse systems such as Fermi and Bose gases,
electromagnetic radiation and condensed matter phases including solids and superfluids. In all these systems, most important properties such as energy depend on the form of a dispersion relation \cite{landau}.

There are two commonly discussed forms of dispersion relations. The first one is the gapless DR describing a wave such as a photon or phonon:

\begin{equation}
E=p\,c
\label{nogap}
\end{equation}

\noindent where $c$ is the propagation velocity. A similarly gapless DR, $E=\frac{p^2}{2m}$, describes a non-relativistic particle.

The second one is the DR which has the energy or mass gap on y-axis. A common example is a relativistic DR for a massive particle

\begin{equation}
E=\sqrt{p^2+m^2}
\label{mgap}
\end{equation}
\noindent where $m$ is particle mass and $c=1$.

The energy gap implies that a particle or a system have no states between zero and the gap. In several systems this circumstance is notably a major factor determining, for example, electrical
conductivity of semiconductors and superconductors. In the latter, emergence of the gap at the critical temperature
marks the superconducting transition and governs other properties such as electronic heat capacity and interaction with electromagnetic field. Understanding the origin and nature of energy gaps is
currently the topic of wide discussion including, for example, the gap in high-temperature superconductors
and strongly-interacting field theory. One of these areas include the emergence of the mass gap in the non-linear Yang-Mills theory.

Gapless DR and DR with the energy gap are shown in Fig. 1.

\begin{figure}
\begin{center}
{\scalebox{0.4}{\includegraphics{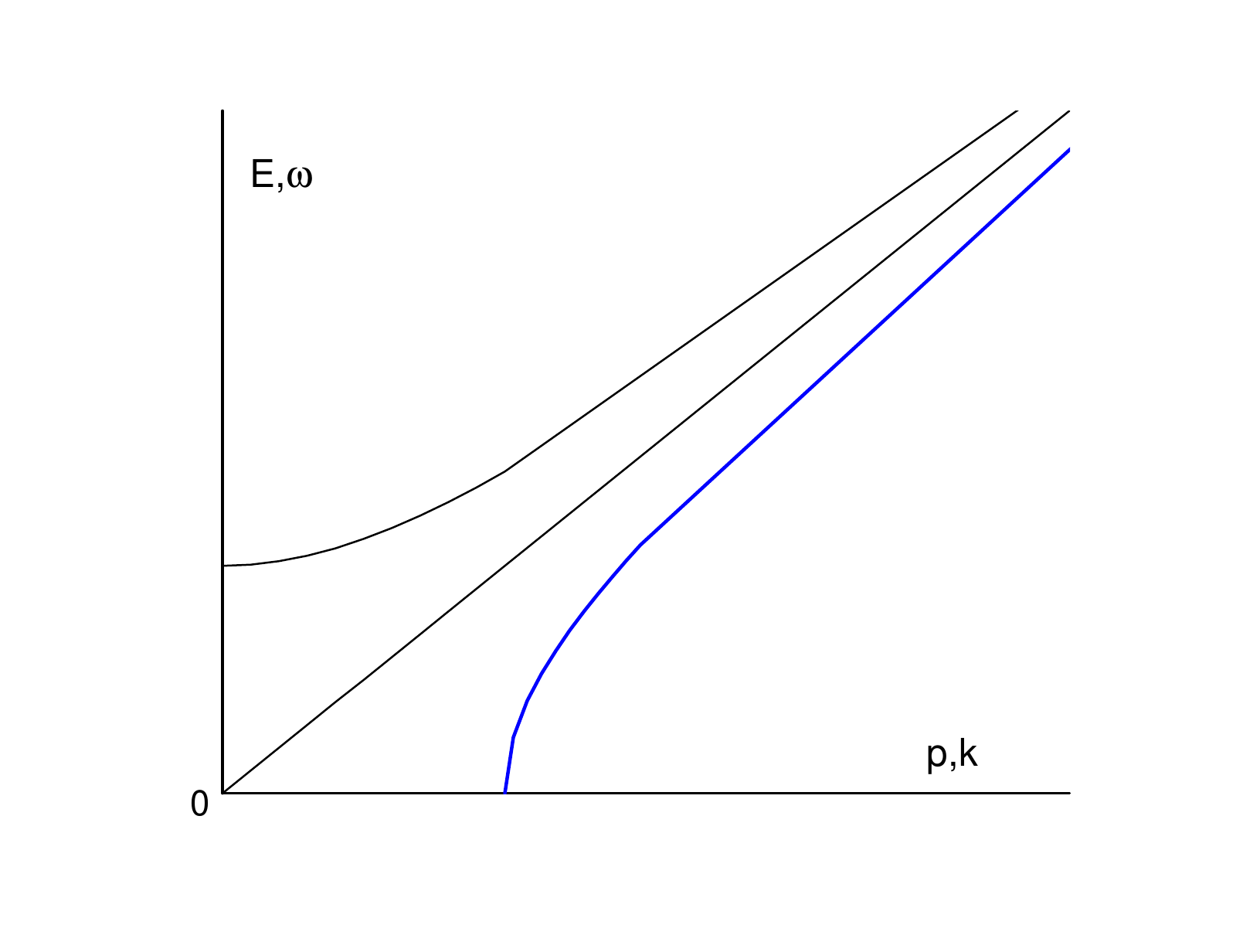}}}
\end{center}
\caption{Two commonly discussed dispersion relations showing the dependencies of energy $E$ or frequency $\omega$ on momentum $p$ or wavevector $k$: top and middle curves show dispersion relations
for a massive particle with the energy (frequency) gap and for a massless particle (photon) or an acoustic phonon. The bottom curve shows the dispersion relation (\ref{pgap}) with the gap in momentum
space. This curve is symmetric to the top curve relative to the gapless $E=p$ line. Schematic illustration.}
\label{disp}
\end{figure}

It is interesting to ask whether a third type of dispersion relation exists which is symmetric relative to the $E=p$ line and which has the form

\begin{equation}
E=\sqrt{p^2-p_g^2}
\label{pgap}
\end{equation}

For the energy to be real in (\ref{pgap}), $p>p_g$ should hold. Hence, $p_g$ is identified with the gap in momentum space. This implies that no momentum states exist between 0 and $p_g$ (similarly to no energy states existing between 0 and $m$ for the DR with the energy gap).

The DR with the gap in momentum space is illustrated in Fig. 1. Formally, (\ref{pgap}) follows from swapping $E$ and $p$ in (\ref{mgap}) (and setting $p_g=m$), implying the symmetry of curves (\ref{mgap}) and (\ref{pgap}) relative to the gapless $E=p$ line. Alternatively, (\ref{pgap}) follows from setting $m$ in (\ref{mgap}) to an imaginary value $m=ip_g$, as is done in the discussion of a tachyonic DR \cite{tachyons}.

Dispersion relations with gaps in momentum space in Fig. 1 appear unusual in mainstream physics discussions and are intriguing. We will see that they appear in a surprising variety of areas, from ordinary liquids to holographic models. In some of these areas, gapped momentum states (GMS) are often viewed as a curiosity, but their physical origin is not discussed from the fundamental point of view.

It has become increasingly apparent that GMS are important from several additional perspectives. First, they are interesting in themselves. For example, the dispersion relation with the mass (energy) gap (\ref{mgap}) gives zero group velocity $\frac{d\omega}{dk}$ in the limit $k\rightarrow 0$ and a diverging phase velocity $\frac{\omega}{k}$ in the same limit. As we will show below, this is interestingly {\it reversed} for the DR with the $k$-gap. Second, it has been realised that one can continuously transit from the energy-gapped to momentum-gapped DR in Fig. 1, via the gapless line, by changing a single physical parameter such as temperature or dissipation. This interestingly suggests that all three curves in Fig. 1 can be parts of one physical system, extending the dispersion relation diagram into the hitherto unexplored gap in $k$-space. Finally and perhaps more importantly, it has become apparent that GMS have important implications for fundamental dynamical and thermodynamic properties of the system. Surveying how the $k$-gap is related to those properties in different areas of physics is one of the aims of this review.

On general grounds, gapped momentum states are often related to
dissipation and open systems, an interesting and challenging problem of
fundamental importance. Indeed, basic assumptions and results of
statistical physics are related to introducing and frequently exploiting
the concept of a closed or quasi-closed system or subsystem. This
includes, for example, central ideas of a statistical or thermal
equilibrium and ensuing definitions of entropy and temperature and their
changes, statistical independence of subsystems and resulting additivity
of the logarithm of the statistical distribution function and ensuing
distributions such as Gibbs distribution \cite{landau}.

A closed system is an approximation simplifying theoretical description.
If small systems or systems with small relaxation time, the concept of a
closed system may not apply. More generally, theoretical description of
{\it open} systems and dissipation is an interesting and challenging
problem, viewed as a core problem in modern physics \cite{rotter}. In
quantum-mechanical systems, this problem is related to the foundations
of quantum theory itself (see, e.g., Refs.
\cite{bender,mohsen}), although we will see that dissipation
and associated GMS manifests itself differently in different systems.
Understanding dissipation has seen renewed recent interest in areas
related to non-equilibrium and irreversible physics, decoherence effects
and complex systems as well as in the area of relativistic hydrodynamics
where dissipative terms in the action have been proposed and their
effects explored.

Starting from early work (see, e.g., \cite{feynman,leggett}), a common approach to treat dissipation is to introduce a central dissipative system of interest, its environment
modelled as, for example, a bath of harmonic oscillators and an interaction between the system and its environment enabling energy exchange (see, e.g., \cite{kamenev,weiss} for review). In this
picture, dissipative effects can be discussed by solving models using approximations such as linearity of the system and its couplings.

A distinct effect of dissipation is related to a situation where the energy of the system is not changed overall, but the propagation range of a collective mode (e.g. phonon) acquires a finite range. No dissipation takes place when a plane wave propagates in a crystal where the wave is an eigenstate \cite{ropp}. However, a plane wave dissipates in systems with structural and dynamical disorder such as liquids. As we will see, the dispersion relation becomes gapped in momentum space as a result.

The purpose of this review is to discuss how the DR with the gap in $k$-space, or gapped momentum states (GMS), emerge in different physical situations, explore their common physical origin and discuss some important implications for dynamical and thermodynamic properties of the system. Understanding GMS is fairly straightforward in condensed matter systems such as liquids, and so we start with liquids and supercritical fluids. We subsequently find that a Lagrangian formulation of liquid dynamics with GMS involves a two-field description and proceed to showing how this description also emerges in the Keldysh-Schwinger technique developed to describe dissipative processes. We then proceed to considering the emergence of GMS in other areas of physics: strongly-coupled plasma, electromagnetic waves, the non-linear Sine-Gordon model, quasihydrodynamic approach and, finally, holographic models. In addition to theory, we review modeling and experimental evidence for GMS in liquids and strongly-coupled plasma. We conclude with a tentative list of questions and challenges related to further understanding of GMS, including in theory and experiments.

As we survey different areas, we seek to uncover common physics connecting seemingly disparate physical effects and phenomena, including the interplay between propagation and dissipation effects. Finding and appropriately identifying dissipative terms such as system relaxation time and studying its temperature dependence gives new insights into generic physical behavior in very different systems such as liquids, plasma, fields and holographic models. Some of the results reviewed include our own, which we put into context of previous and current work.

Throughout this review, we will be referring to gapped momentum states and $k$-gap (gap in $k$-space) interchangeably and depending on the effect and system we consider.

\section{Gapped momentum states in liquids and supercritical fluids}\label{liquid-section}

\subsection{Liquids: problems of theoretical description}

Particle dynamics in liquids involves solid-like oscillatory motion at quasi-equilibrium positions and diffusive jumps into neighbouring locations \cite{frenkel}. These jumps enable liquid flow and endow liquids with viscosity. Describing this dynamics necessitates consideration of a non-linear interaction allowing for both oscillation and activated jumps over potential barrier of the inter-particle potential. This implies that describing liquid dynamics from first principles involves a large number of coupled non-linear oscillators. This problem is not currently tractable due to its exponential complexity \cite{ropp}.

This problem does not originate in solids and gases. The smallness of atomic displacements in solids and weakness of interactions in gases simplify their theoretical description. Liquids do not have those simplifying features (small parameter): they combine large displacements with strong interactions. For this reason, liquids are believed to be not amenable to theoretical understanding at the same level as gases and solids \cite{landau}.

Common theoretical description of liquids involves a continuum approximation and, because liquids flow, the hydrodynamic approximation is used \cite{hydro}. For example, the Navier-Stokes equation describes liquid flow and features viscosity as an important flow property. At the same time, liquid properties such as density, bulk modulus and heat capacity are close to those of solids \cite{ropp}. An important solid-like property is the liquid ability to support high-frequency solid-like shear waves. Predicted by Frenkel \cite{frenkel}, this property has been seen in experiments \cite{grim,scarponi,burkel,rec-review,hoso,mon-na1,mon-na2,mokshin,sn1,sn2} and modeling results, although with a long time lag.

Frenkel's idea was that liquid particles oscillate as in solids for some time and then diffusively move to neighbouring quasi-equilibrium positions. He introduced $\tau$ as the average time between diffusive jumps and predicted that liquids behave like solids and hence support propagating shear modes at time shorter than $\tau$, or frequency above the Frenkel frequency $\omega_{\rm F}$:

\begin{equation}
\omega>\omega_{\rm F}=\frac{1}{\tau}
\label{omegaF}
\end{equation}

Solid-like shear modes are absent in the hydrodynamic description operating when $\omega\tau<1$, whereas (\ref{omegaF}) implies the opposite regime $\omega\tau>1$. To account for the shear modes and other solid-like properties, liquid theories designate the hydrodynamic description as a starting point and subsequently generalize it to account for liquid response at large $\omega$ and wavevector $k$. Several ways of doing so have been proposed, giving rise to a large field of generalized hydrodynamics \cite{boon,hansen,march1,baluca}. This approach was used to describe non-hydrodynamic liquid properties, but faced issues related to its phenomenological character as well as assumptions and extrapolations used (see, e.g. Refs. \cite{rec-review,mon-na1}).

The traditional hydrodynamic approach to liquids is supported by our common experience that liquids flow and hence necessitate hydrodynamic flow equations as a starting point. This reflects our experience with common low-viscous liquids such as water or oil where $\tau$ is much shorter than observation time. However, flow is less prominent in liquids with large $\tau$ (e.g. liquids approaching glass transition) where properties become more solid-like and elastic \cite{dyre}. This begs the question of what should be a correct starting point of liquid description? We will return to this point in more details below. Here, we note that as far as the $k$-gap is concerned, both hydrodynamic and solid-like elastic effects enter the theory on equal footing, without a-priori designating either of them as a correct starting point. Below we will show how the interplay of solid-like propagating terms and liquid-like dissipative terms can be treated on equal footing as far as the equations are concerned, and how this treatment gives rise to GMS.

\subsection{Gapped momentum states in liquids in the Maxwell-Frenkel approach}\label{uno}

The dispersion relation for transverse modes in liquids involving a gap in $k$-space was derived in generalized hydrodynamics mentioned above. In this approach, the hydrodynamic transverse current correlation function is generalized to include large $k$ and $\omega$ \cite{boon}, as discussed in section \ref{hydrodyn} in more detail. Around the same time, the $k$-gap was first mentioned on the basis of results of molecular dynamics simulations, where the calculated peaks of transverse current correlation functions were seen at large $k$ but not at low \cite{verlet}.

However, the equation predicting the $k$-gap in liquids was written about 50 years before the generalized hydrodynamics result. The equation was derived by Frenkel \cite{frenkel}. Having written the equation, Frenkel, perhaps surprisingly, did not seek to solve it.

Frenkel's starting point was the idea of Maxwell that liquids combine viscous and solid-like elastic properties and are therefore {\it viscoelastic}. Maxwell formulated this combination as \cite{maxwell}:

\begin{equation}
\frac{ds}{dt}=\frac{P}{\eta}+\frac{1}{G}\frac{dP}{dt}
\label{a1}
\end{equation}
\noindent where $s$ is shear strain, $\eta$ is viscosity, $G$ is shear modulus and $P$ is shear stress.

(\ref{a1}) states that shear deformation in a liquid is the sum of the viscous and elastic deformations, given by the first and second right-hand side terms. As mentioned above, both deformations are treated in (\ref{a1}) on equal footing.

Frenkel proposed \cite{frenkel} to represent the Maxwell interpolation by introducing the operator $A$ as

\begin{equation}
A=1+\tau\frac{d}{dt}
\label{a2}
\end{equation}

Then, Eq. (\ref{a1}) can be written in the operator form as

\begin{equation}
\frac{ds}{dt}=\frac{1}{\eta}AP
\label{a3}
\end{equation}

In (\ref{a2})-(\ref{a3}), $\tau$ formally is Maxwell relaxation time $\frac{\eta}{G}$. At the microscopic level, Frenkel's theory approximately identifies this time with the time between consecutive diffusive jumps in the liquid \cite{frenkel}. This is supported by numerous experiments \cite{dyre} as well as modelling results \cite{egami}.

(\ref{a1}-\ref{a3}) enable two approaches to describe liquid viscoelasticity: generalizing $\eta$ to account for short-term elasticity and generalizing $G$ to allow for long-time hydrodynamic flow \cite{frenkel}.

In the first approach, (\ref{a1}) enables us to generalize viscosity as

\begin{equation}
\frac{1}{\eta}\rightarrow\frac{1}{\eta}\left(1+\tau\frac{d}{dt}\right)
\label{sub2}
\end{equation}

In the second approach, $G$ is generalized by noting that if $A^{-1}$ is the reciprocal operator to $A$, (\ref{a3}) can be written as $P=\eta A^{-1}\frac{ds}{dt}$. Because $\frac{d}{dt}=\frac{A-1}{\tau}$ from (\ref{a2}), $P=G(1-A^{-1})s$. Comparing this with the solid-like equation $P=Gs$, we see that the presence of hydrodynamic viscous flow is equivalent to the substitution of $G$ by the operator

\begin{equation}
M=G(1-A^{-1})
\label{operator}
\end{equation}

Adopting the hydrodynamic approach as a starting point of liquid description, we write the Navier-Stokes equation as

\begin{equation}
\nabla^2{\bf v}=\frac{1}{\eta}\left(\rho\frac{d{\bf v}}{dt}+\nabla p\right)
\label{navier2}
\end{equation}

\noindent where ${\bf v}$ is velocity, $\rho$ is density and the full derivative is $\frac{d}{dt}=\frac{\partial}{\partial t}+{\bf v\nabla}$ and generalize $\eta$ according to (\ref{sub2}):

\begin{equation}
\eta\nabla^2{\bf v}=\left(1+\tau\frac{d}{dt}\right)\left(\rho\frac{d{\bf v}}{dt}+\nabla p\right)
\label{gener}
\end{equation}

Having written (\ref{gener}), Frenkel did not analyze its implications. We solved (\ref{gener}) \cite{ropp} and considered the absence of external forces, $p=0$ and the slowly-flowing fluid so that
$\frac{d}{dt}=\frac{\partial}{\partial t}$. Then, Eq. (\ref{gener}) reads

\begin{equation}
\eta\frac{\partial^2v}{\partial x^2}=\rho\tau\frac{\partial^2v}{\partial t^2}+\rho\frac{\partial v}{\partial t}
\label{gener2}
\end{equation}

\noindent where $v$ is the velocity component perpendicular to $x$.

In contrast to the Navier-Stokes equation, Eq. (\ref{gener2}) contains the second time derivative of $v$ and hence allows for propagating waves. Using $\eta=G\tau=\rho c^2\tau$, where $c$ is the shear wave velocity, we re-write Eq. (\ref{gener2}) as

\begin{equation}
c^2\frac{\partial^2v}{\partial x^2}=\frac{\partial^2v}{\partial t^2}+\frac{1}{\tau}\frac{\partial v}{\partial t}
\label{gener3}
\end{equation}

Seeking the solution of (\ref{gener3}) as $v=v_0\exp\left(i(kx-\omega t)\right)$ gives

\begin{equation}
\omega^2+\omega\frac{i}{\tau}-c^2k^2=0
\label{quadratic}
\end{equation}

We will encounter Eq. (\ref{quadratic}) throughout this review and in several other areas where GMS operate.

Eq. (\ref{quadratic}) yields complex frequency

\begin{equation}
\omega=-\frac{i}{2\tau}\pm\sqrt{c^2k^2-\frac{1}{4\tau^2}}
\label{solution}
\end{equation}

If $ck<\frac{1}{2\tau}$, $\omega$ in (\ref{quadratic}) does not have a real part and propagating modes. For $ck>\frac{1}{2\tau}$, the real part of $\omega$ is

\begin{equation}
\omega=\sqrt{c^2k^2-\frac{1}{4\tau^2}}
\label{omega}
\end{equation}

\noindent and the solution of (\ref{gener3}) is

\begin{equation}
v\propto\exp\left(-\frac{t}{2\tau}\right)\exp(i\omega t)
\label{omega0}
\end{equation}

According to Eq. (\ref{omega}), the gap in $k$-space emerges in the liquid transverse spectrum: in order for $\omega$ in (\ref{omega}) to be real, $k>k_g$ should hold, where

\begin{equation}
k_g=\frac{1}{2c\tau}
\label{kgap}
\end{equation}

\begin{figure}
\begin{center}
{\scalebox{0.8}{\includegraphics{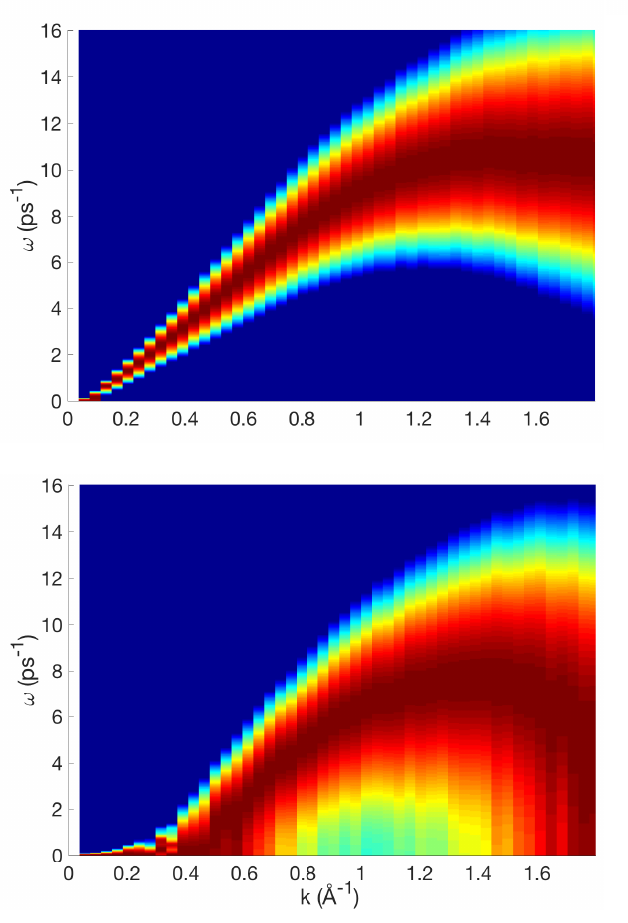}}}
\end{center}
\caption{Intensity maps showing the spectra of transverse currents calculated in molecular dynamics simulations as the Fourier transform of the real part of transverse current correlation functions. The intensity maps are shown for a model Ar liquid at 205 K (top) and 450 K (bottom) and 10 kbar pressure. The maximal intensity corresponds to the middle points of dark red areas and reduces away from them. The emergence of the gap in $k$-space is seen at high temperature. Adapted from Ref. \cite{prl}.
}
\label{3}
\end{figure}

More recently \cite{prl}, detailed evidence for the $k$-gap was presented on the basis of molecular dynamics simulations. According to (\ref{kgap}), the gap in $k$-space increases with temperature because $\tau$ decreases. In agreement with this prediction, Figure \ref{3} shows the $k$-gap emerging in the liquid at high temperature.

\begin{figure}
\begin{center}
{\scalebox{0.32}{\includegraphics{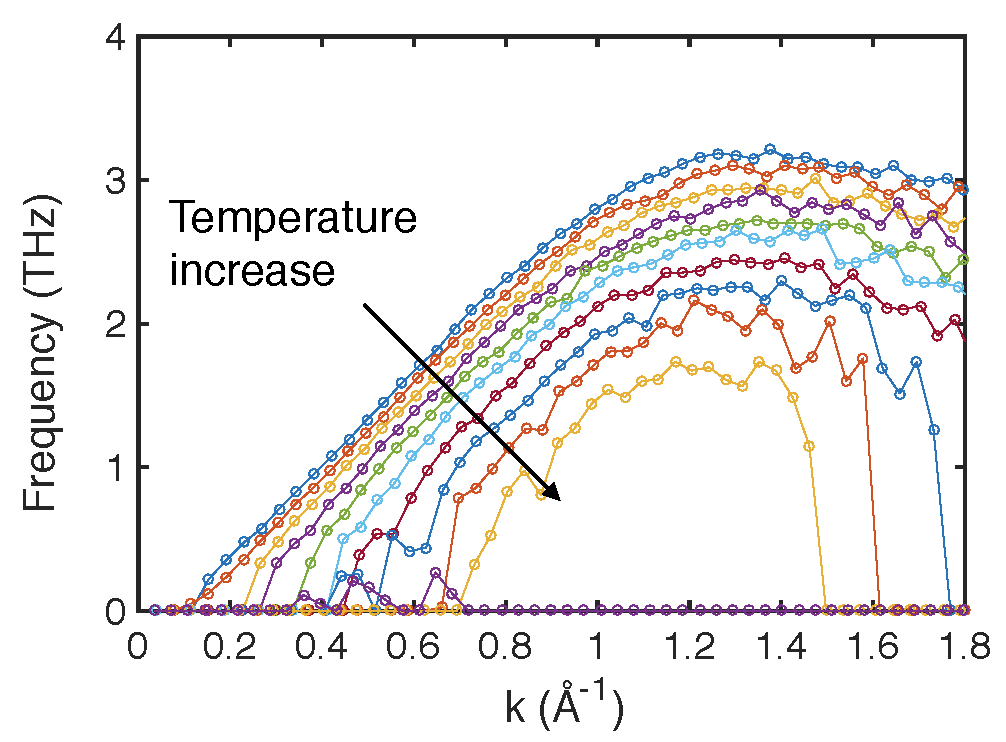}}}
\end{center}
\caption{Phonon dispersion curves of supercritical Ar calculated in molecular dynamics simulations \cite{prl}. The curves are shown at 200-500 K and 550 K with the temperature increment of 30 K. The deviation from linearity (curving over) of dispersion curves at large $k$ is related to probing the effects comparable to interatomic separations (in the solid this correspond to curving over of $\omega\propto\sin ck$ at large $k$). This effect is not accounted for in the theory leading to the gap (\ref{kgap}) because the theory is formulated in the continuous medium \cite{ropp} and therefore describes the $k$-gap in the linear part of the dispersion law. We show $k$ in the range slightly extending the first pseudo-zone boundary (FPZB) at low temperature. As volume and interatomic separation increase with temperature, FPZB shrinks, resulting in the decrease of $\omega\propto\sin ck$ at large $k$ beyond the FPZB. This effect is unrelated to the $k$-gap.}
\label{2}
\end{figure}

Maxima of intensity maps at frequency $\omega$ correspond to a propagating mode at that frequency and gives a point ($k,\omega$) on the dispersion curve. The dispersion curves are plotted in Figure \ref{2} and show a detailed evolution of the gap with temperature.

We observe that $k_g$ increases with temperature. This is consistent with (\ref{kgap}) predicting that the gap increases because $\tau$ decreases with temperature. In further agreement with
(\ref{kgap}), $k_g$ calculated from the dispersion curves increases approximately as $\frac{1}{\tau}$ \cite{prl}.

The emergence of $k_g$ has also been ascertained in hard disk models used to understand the rigidity transition in glasses and liquid-glass transition \cite{naumis}. Here, the $k$-gap increases at low packing fractions. Because rigidity is related to propagating shear modes, the value of $k_g$ was interestingly proposed as an order parameter quantifying the rigidity transition \cite{naumis}.

Microscopically, the gap in $k$-space can be related to a finite propagation length of shear waves in a liquid: if $\tau$ is the time during which the shear stress relaxes, $c\tau$ gives the shear wave propagation length, or liquid elasticity length, $d_{\rm el}$ \cite{jpcm}:

\begin{equation}
d_{\rm el}=c\tau
\label{del}
\end{equation}

The allowed wavelengths in the system can not be longer than the wave propagation length. Therefore, the condition $k>k_g=\frac{1}{2c\tau}$ (\ref{kgap}) approximately corresponds to propagating waves with wavelengths shorter than the propagation length.

From the point of view of elasticity, the $k$-gap suggests that we can consider a liquid as collection of dynamical regions of characteristic size $c\tau$ where the solid-like ability to support shear waves operates.

It is useful to mention the propagation range of shear waves $d$ in two regimes, hydrodynamic regime where $\omega\tau<1$ and solid-like elastic regime where $\omega\tau>1$. $d$ can be derived from complex shear modulus emerging from Maxwell interpolation (\ref{a1}). The result is that $d\approx\lambda\cdot\omega\tau$ in the solid-like elastic regime, where $\lambda$ is the wavelength \cite{ropp}. Hence $d=c\tau$, the same propagation length discussed in the previous paragraph. It is useful to write this as

\begin{equation}
\frac{d}{\lambda}=\omega\tau
\end{equation}

\noindent showing that the propagation length is much longer than the wavelength in the regime $\omega\tau\gg 1$. We will come back to this ratio when we discuss the attenuation of the electromagnetic waves and associated gaps in their spectrum.

In the hydrodynamic regime where $\omega\tau<1$, $d=\frac{\lambda}{2\pi}$ \cite{frenkel}, showing that the propagation length is comparable to the wavelength and implying a substantial attenuation of the low-frequency waves.

It is interesting to ask why the gap develops in $k$-space in (\ref{omega})-(\ref{kgap}) but not in the frequency space as envisaged in (\ref{omegaF}) by Frenkel originally? The answer lies in a difference between a local nature of a relaxation event (particle jump) and an extended character of a wave. Indeed, (\ref{omegaF}) gives the condition at which a local environment of a jumping atom is solid-like. This condition was applied to predict propagating transverse modes in liquids, but we now understand the condition to be too strong. A propagating wave does not require all particles it encounters during its propagation to obey (\ref{omegaF}) and be solid-like. Instead, the wave propagation is affected only by particles jumping at a distance $d=c\tau$ where the wave front has reached, disrupting the wave continuity and dissipating the wave. If $\tau$ is average time between particle rearrangements, this distance is given by $c\tau$, setting the maximal wavelength and minimal $k$, or the $k$-gap.

We note that Eq. (\ref{gener3}) has the form of the telegrapher's equation discussed in the context diffusion and related processes \cite{gaveau} as well as earlier consideration of transmission of electromagnetic waves (see, e.g., Ref. \cite{masoliver} for a brief review). Perhaps surprisingly, GMS were not discussed as a solution to the telegrapher's equation.

\subsection{Further properties of the $k$-gap}

We make several further observations regarding the emergence of $k$-gap in the liquid transverse spectrum. First, the $k$-gap emerges not only to liquids below the critical point but also to supercritical fluids as long as the system is below the Frenkel line \cite{prl}. We will discuss the Frenkel line in more detail below. Here, it suffices to note that the Frenkel line separates the low-temperature liquid-like supercritical state where particle dynamics combines oscillatory and diffusive components of motion and where transverse modes exist from the high-temperature gas-like state where particle dynamics is purely diffusive and where no transverse modes operate \cite{ropp,fl1,fl3,fl2}.

Second, the gap in momentum space in liquids emerges in the transverse spectrum rather the longitudinal one. The longitudinal mode remains gapless at low $k$ because the bulk modulus always has a non-zero static component $B_0$, giving rise to the hydrodynamic sound wave with velocity $\sqrt{\frac{B_0}{\rho}}$ and
small $k$ corresponding to long wavelengths extending to system size. This wave is present in a continuum (hydrodynamic) approximation of different media with non-zero static bulk modulus, including solids and gases.

Our third remark is related to the crossover between propagating and non-propagating modes related to the presence of imaginary and real terms in (\ref{solution}). The imaginary part defines decay
time or decay rate $\Gamma$ of an excitation. According to (\ref{solution}), the decay time and decay rate
are $2\tau$ and $\Gamma=\frac{1}{2\tau}$. If, as is often assumed, the crossover between propagating and non-propagating modes is given by the equality between the decay time and inverse frequency
(period of the wave), the propagating modes correspond
to $k>\frac{1}{\sqrt{2}}\frac{1}{c\tau}$ from (\ref{solution}), or $k>k_g\sqrt{2}$.

Our next observation is related to the velocity of propagating waves with the $k$-gap. We note that the dispersion relation with the mass (energy) gap (\ref{mgap}) gives zero group velocity
$\frac{d\omega}{dk}$ in the limit $k\rightarrow 0$ and a diverging phase velocity $\frac{\omega}{k}$ in the same limit. This is interestingly {\it reversed} for the dispersion relation with the $k$-gap. Indeed, at the smallest $k$-point, $k=k_g$, (\ref{omega}) gives $0$ for the phase velocity and a diverging
group velocity $\propto\frac{1}{\sqrt{c^2k^2-\frac{1}{4\tau^2}}}$. This might appear to contradict the necessity
for the wave group velocity to be subluminal. However, there is no contradiction if we note that the ratio of the mode frequency (\ref{omega}) and decay rate $\Gamma=\frac{1}{2\tau}$ is
$\sqrt{4c^2\tau^2k^2-1}$. This tends to $0$ as $k\rightarrow k_g$, implying non-propagating excitations.

Finally, we comment on the experimental evidence of the $k$-gap. There are currently no transverse dispersion relations with the $k$-gap directly obtained from inelastic neutron or X-ray scattering experiments. However, there are indirect pieces of evidence supporting the existence of $k$-gap. The first piece of evidence comes from the fast sound or positive sound dispersion (PSD), the increase of the measured speed of sound over its hydrodynamic value \cite{ropp}. As first noted
by Frenkel \cite{frenkel}, a non-zero shear modulus of liquids implies that the propagation velocity crosses over from its hydrodynamic value $v=\sqrt{\frac{B}{\rho}}$ to the solid-like elastic value $v=\sqrt{\frac{B+\frac{4}{3}G}{\rho}}$, where $B$ and $G$ are bulk and shear moduli, respectively
\cite{landelast,dyre}. According to the discussion in the previous section, shear modes become propagating $k>k_g$, implying PSD at these $k$-points. This further implies that PSD should disappear with temperature starting from small $k$ because the $k$-gap increases with temperature.
This is confirmed experimentally \cite{morkel}: inelastic X-ray experiments in liquid Na show that PSD is present in a wide range of $k$ at low temperature. As temperature increases, PSD disappears starting from small $k$, in agreement with the $k$-gap picture. At high temperature, PSD is present at large $k$ only.

Another piece of evidence comes from low-frequency shear elasticity of liquids at small scale \cite{noirez1,noirez2}. According to (\ref{omega}), the frequency at which a liquid supports shear stress can be arbitrarily small provided $k$ is close to $k_g$. This implies that small systems are able to support shear stress at low frequency. This has been ascertained experimentally \cite{noirez1,noirez2}. This important result was to some extent surprising, given the widely held view that, according to (\ref{omegaF}), liquids were thought to be able to support shear stress at high frequency only \cite{frenkel,dyre}.

\subsection{Essential ingredients of gapped momentum states and symmetry of liquid desription}\label{sss}

We have seen that one needs two essential ingredients for the $k$-gap to emerge in the wave spectrum. First, we need a wave-like component in equations enabling wave propagation. Second, we need a dissipative effect, the process that disrupts the wave continuity and dissipates it over a certain distance,
thus destroying waves with long wavelengths and setting the gap in $k$-space.

There are two important features of the two ingredients above, which are to some extent related to each other. First, neither solid-like propagating nor dissipative viscous terms are assumed to be small in our equations. Consequently, a theory of $k$-gap can not proceed by starting with either term and treating the other term with perturbation theory.

The second feature is more fundamental and is related to finding a correct starting point for liquid description altogether and $k$-gap in particular. In our derivation of the $k$-gap above, we have started with the hydrodynamic approach describable by the Navier-Stokes equation and generalized it to endow the liquid with solid-like elastic response. This approach agrees with the spirit of generalized hydrodynamics mentioned earlier and discussed in section \ref{hydrodyn} in more detail. This approach is consistent with our everyday experience that common liquids flow and hence are hydrodynamic systems. Yet Maxwell interpolation (\ref{a1}) gives no preference to either hydrodynamic or solid-like elastic terms to serve as a starting point of liquid description. Instead, Maxwell interpolation treats the viscous hydrodynamic term and solid-like elastic term on equal footing. Does this imply that propagating solid-like transverse modes with the gap in $k$-space can be derived starting from solid-state equations instead of hydrodynamic ones?

The answer is positive: it can be shown that the central equation (\ref{gener3}) predicting the gap in $k$-space can be derived by adopting the solid-state description as a starting point
\cite{pre-l}. We start with the wave equation describing a non-decayed propagation of transverse waves in the solid:

\begin{equation}
G\frac{\partial^2v}{\partial x^2}=\rho\frac{\partial^2v}{\partial t^2}
\label{2ap}
\end{equation}

We generalize (\ref{2ap}) by substituting $G$ by $M$ (\ref{operator}):

\begin{equation}
G(1-A^{-1})\frac{\partial^2v}{\partial x^2}=\rho\frac{\partial^2v}{\partial t^2}
\label{2ap1}
\end{equation}

We act on both sides of (\ref{2ap1}) with operator $A$ in (\ref{a2}):

\begin{equation}
G\left(A-1\right)\frac{\partial^2v}{\partial x^2}=\rho A\frac{\partial^2v}{\partial t^2}
\label{2ap2}
\end{equation}

According to (\ref{a2}), $A-1$ in (\ref{2ap2}) is $\tau\frac{\partial d}{\partial t}$. Using $G=\rho c^2$ as earlier and rearranging the right side of (\ref{2ap2}) gives

\begin{equation}
c^2\tau\frac{\partial}{\partial t}\frac{\partial^2v}{\partial x^2}=\frac{\partial}{\partial t}\left(\frac{\partial v}{\partial t}+\tau\frac{\partial^2 v}{\partial t^2}\right)
\label{2ap3}
\end{equation}

Integrating over time and setting the integration constant to 0 gives the equation identical to (\ref{gener3}) predicting solid-like shear modes with the gap in $k$-space.

That liquid properties and the gap in momentum space can be derived in approaches starting from either hydrodynamic or solid state equations helps us understand the physical origin of GMS. It also
implies that liquids with their emerging GMS occupy a symmetrical place between the hydrodynamic and solid-like approaches from the point of view of physical description.

\subsection{Implications for liquid thermodynamics}

As mentioned earlier, a general theory of liquid thermodynamics at the level comparable to solids and gases was deemed impossible \cite{landau}. However, understanding propagating modes in liquids enables us to
calculate their energy and, subsequently, the total liquid energy. In other words, liquid thermodynamics can be discussed on the basis of collective modes as is done in the solid state theory.
Indeed, the $k$-gap in the transverse spectrum implies that the energy of transverse modes can be calculated as

\begin{equation}
E_t=\int\limits_{k_{\rm gap}}^{k_{\rm D}}E(k,T)\,g(k)\,dk
\label{ene1}
\end{equation}
\noindent where $g(k)=\frac{6N}{k_{\rm D}^3}k^2$ is the density of states in the Debye model, $k_{\rm D}$ is Debye wave vector, $N$ is the number of molecules and $k_{\rm gap}=\frac{1}{c\tau}$ (here $\tau$ is understood to be the full period of the particle's jump motion equal to twice Frenkel’s $\tau$). Taking $E=T$ ($k_{\rm B}=1$) in the classical case and integrating gives

\begin{equation}
E_t=2NT\left(1-\left(\frac{\omega_{\rm F}}{\omega_{\rm D}}\right)^3\right)
\label{ene2}
\end{equation}

\noindent where $\omega_{\rm D}=ck_{\rm D}$ is Debye frequency.

Adding the energy of remaining longitudinal mode and the kinetic energy of diffusing atoms to $E_t$ in (\ref{ene2}) gives the total liquid energy as \cite{ropp}:

\begin{equation}
E=NT\left(3-\left(\frac{\omega_{\rm F}}{\omega_{\rm D}}\right)^3\right)
\label{ene3}
\end{equation}

At low temperature when $\omega_{\rm F}\ll{\omega_{\rm D}}$, Eq. (\ref{ene3}) predicts $c_v=\frac{1}{N}\frac{dE}{dT}=3$ as in solids. At high temperature, the range in $k$-space where the transverse modes propagate reduces according to (\ref{kgap}). When $\tau$ reaches its limiting value of $\tau_{\rm D}$ at high temperature ($\omega_{\rm F}$ reaches its limiting value $\omega_{\rm D}$), $k_g$ increases to the zone boundary (in disordered systems, the zone boundary is introduced in the Debye model \cite{landau}). At this point, all transverse modes disappear from the liquid spectrum. According to (\ref{ene3}), setting $\omega_{\rm F}=\omega_{\rm D}$ gives $c_v=2$, in agreement with experimental results \cite{ropp}. If $\omega_{\rm F}$ is known from experiments or
molecular dynamics simulations, the agreement with experimental and modelling data can be made quantitative in the entire temperature range where $c_v$ reduces from 3 at low temperature to 2 at high \cite{ropp,ling-pre}.

\subsection{A relationship between $\omega$ and $k$-gap}

\begin{figure}
\begin{center}
{\scalebox{0.6}{\includegraphics{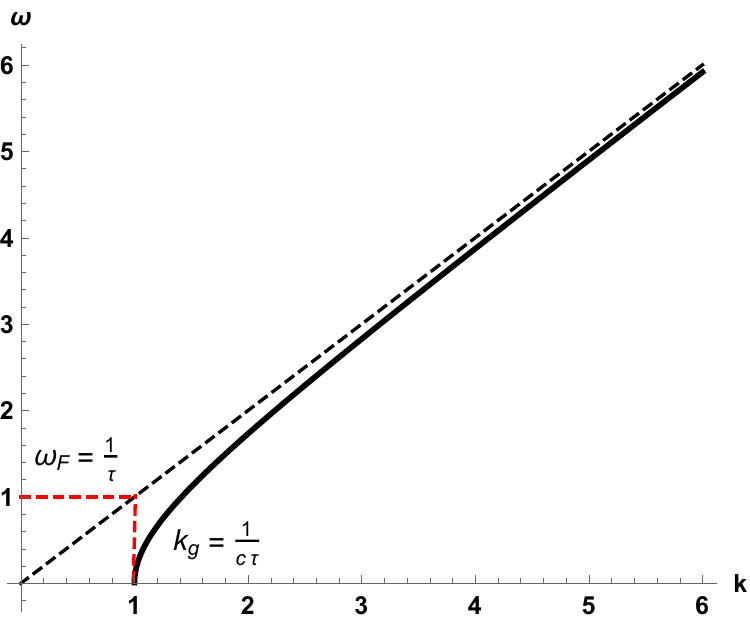}}}
\end{center}
\caption{Dispersion relation with the $k$-gap and its approximation in the Debye model by the straight line starting at $k=k_g$ and $\omega=\omega_{\rm F}=\frac{1}{\tau}$.}
\label{omegakgap}
\end{figure}

It is important to return to the original Frenkel's assumption that liquids support transverse modes above the frequency $\frac{1}{\tau}$ as discussed in Eq. \ref{omegaF}. This implies the gap in $\omega$ space. However, the gap turns out to be $k$-space as discussed above, rather that in $\omega$. Interestingly, it can be seen that operating in terms of $\omega$-gap gives a good approximation from the point of view of system thermodynamic properties. Indeed, the density of states $g(\omega)\propto\frac{1}{\frac{d\omega}{dk}}$ becomes small at $k\rightarrow k_g$ because $\frac{d\omega}{dk}$ with $\omega$ in (\ref{omega}) diverges at $k\rightarrow k_g$. This means that the states close to $k_g$ can be neglected in the integral representing different properties such as the energy. Moreover, in the Debye model widely used in isotropic disordered systems, $\omega=ck$ \cite{landau}. In this model, the frequency corresponding to the $k_g\approx\frac{1}{c\tau}$ is $\frac{1}{\tau}$, which is the Frenkel frequency in Eq. \ref{omegaF}. Figure \ref{omegakgap} illustrates this point: neglecting states close to $k_g$ approximates the dispersion relation by the straight line in the Debye model starting at $k=k_g$ and $\omega=\omega_{\rm F}=\frac{1}{\tau}$.

In the classical case, calculating $E_t$ in the picture with $\omega$-gap and $k$-gap gives identical results. Indeed, assuming the frequency gap $\omega_{\rm F}$ gives $E_t$ as

\begin{equation}
E_t=\int\limits_{\omega=\omega_{\rm F}}^{\omega_{\rm D}}E(\omega,T)\,g(\omega)\,d\omega
\label{eom1}
\end{equation}

\noindent where $g(\omega)=\frac{6N}{\omega_D^3}\omega^2$ is the density of states of transverse modes.

Putting $E=T$ gives the same result as (\ref{ene2}). We therefore find that the frequency (energy) gap in liquids {\it effectively} emerges in the approximate description of the phonon states.

Fundamentally, the emergence of the $k$-gap and $\omega$-gap as its effective approximation is related to strong non-linear coupling in liquids alluded to earlier and discussed in the next section in more detail. In this sense, this insight may be interesting in the context of the emergence of energy gaps in other non-linear problems including the Yang-Mills theory.

\subsection{Gapped momentum states in Lagrangian formulation}\label{lagr-gap}

From the point of view of field-theoretical description of liquid dynamics, it is interesting to ask what kind of Lagrangian gives GMS. The challenge is to represent the viscous term
$\propto\frac{1}{\tau}$ in (\ref{gener3}) in the Lagrangian. Notably, when the existence of propagating transverse modes above $k_g$ were unknown in the past, the Lagrangian formulation of liquids involving dissipation was deemed impossible (see, e.g., \cite{leut}).

We consider the scalar field $\phi(x,t)$ describing velocities or displacements. The viscous energy can be written as the work $W$ done to move the liquid. If $s$ is the strain, $W\propto Fs$, where
$F$ is the viscous force $F\propto\eta\frac{ds}{dt}$. Hence, the Lagrangian should contain the term $s\frac{ds}{dt}$ or, in terms of the field $\phi$, the term

\begin{equation}
L\propto\phi\frac{\partial\phi}{\partial t}
\label{fdf}
\end{equation}

However, the term $\phi\frac{d\phi}{dt}$ disappears from the Euler-Lagrange equation

\begin{equation}
\frac{\partial L}{\partial\phi}=\frac{\partial}{\partial t}\frac{\partial L}{\partial\frac{\partial\phi}{\partial t}}+\frac{\partial}{\partial x}\frac{\partial L}{\partial\frac{\partial\phi}{\partial
x}}
\label{lagr}
\end{equation}

\noindent because $\frac{\partial L}{\partial\phi}=\frac{\partial}{\partial t}\frac{\partial L}{\partial\frac{\partial\phi}{\partial t}}=\frac{\partial\phi}{\partial t}$. Another way to see this is
note that the viscous term $\phi\frac{d\phi}{dt}\propto\frac{d}{dt}\phi^2$.

To circumvent this problem, we consider {\it two} fields $\phi_1$ and $\phi_2$, i.e. we invoke two scalar field theory used in a different context \cite{zee}. We note that a two-coordinate
description of a localised damped harmonic oscillator was discussed earlier \cite{bateman,dekker}. In section \ref{tre}, we will discuss the Keldysh-Schwinger approach to dissipative effects and will see how two fields naturally emerge in that formulation, describing an open system of interest and its environment (bath). We note that in the area of liquids and disordered solids, theories of interaction between the system and its environment have been used to explain several important effects involving dissipation (see, e.g., Ref. \cite{zaccone1,zaccone2}).

In terms of two fields $\phi_1$ and $\phi_2$, the dissipative term can be written as a combination of (\ref{fdf}) as

\begin{equation}
L\propto\phi_1\frac{\partial\phi_2}{\partial t}-\phi_2\frac{\partial\phi_1}{\partial t}
\label{twof}
\end{equation}

\noindent and the Lagrangian becomes

\begin{equation}
L=\frac{\partial\phi_1}{\partial t}\frac{\partial\phi_2}{\partial t}-c^2\frac{\partial\phi_1}{\partial x}\frac{\partial\phi_2}{\partial x}+\frac{1}{2\tau}\left(\phi_1\frac{\partial\phi_2}{\partial
t}-\phi_2\frac{\partial\phi_1}{\partial t}\right)
\label{l1}
\end{equation}

Setting $\tau\rightarrow\infty$ in (\ref{l1}) gives the form of the complex scalar field theory and corresponds to no particle jumps and, therefore, solid dynamics. $\tau\rightarrow\infty$ in (\ref{l1}) gives the Lagrangian describing waves in solids as expected.

We note that (\ref{l1}) follows from the two-field Lagrangian

\begin{equation}
\begin{split}
&L=\frac{1}{2}\left(\left(\frac{\partial\psi_1}{\partial t}\right)^2-\left(\frac{\partial\psi_1}{\partial x}\right)^2+\left(\frac{\partial\psi_2}{\partial
t}\right)^2-\left(\frac{\partial\psi_2}{\partial x}\right)^2\right)+\\
&\frac{i}{2\tau}\left(\psi_2\frac{\partial\psi_1}{\partial t}-\psi_1\frac{\partial\psi_2}{\partial t}\right)
\label{l11}
\end{split}
\end{equation}

\noindent using the standard transformation employed in the complex field theory: $\phi_1=\frac{1}{\sqrt{2}}(\psi_1+i\psi_2)$ and $\phi_2=\frac{1}{\sqrt{2}}(\psi_1-i\psi_2)$. 

The free part in \ref{l11} has a standard two-field scalar field theory form. The advantage of using (\ref{l1}) in terms of $\phi_1$ and $\phi_2$ is that the equations of motion for $\phi_1$ and $\phi_2$ decouple as we see below. This is not an issue, however: one can use (\ref{l11}) to obtain the system of coupled equations for $\psi_1$ and $\psi_2$ and decouple them using the same transformation between $\phi$ and $\psi$, resulting in the same equations for $\phi$ as those following from (\ref{l1}). Note that the imaginary term in (\ref{l11}) may be related to dissipation
\cite{bender,bender1,sudarshan1,sudarshan2}, however the Hamiltonian
corresponding to (\ref{l11}) does not have an explicit imaginary term: $H=\frac{1}{2}\left(\left(\frac{\partial\psi_1}{\partial t}\right)^2+
\left(\frac{\partial\psi_1}{\partial x}\right)^2+\left(\frac{\partial\psi_2}{\partial t}\right)^2+\left(\frac{\partial\psi_2}{\partial x}\right)^2\right)$, where terms with $\tau$ cancel out. We will comment on the Hamiltonian of the composite system and the energy below.

The Lagrangian (\ref{l11}) is non-Hermitian and, notably, ${\mathcal PT}$-symmetric as follows from its invariance under changing the sign of time and swapping two fields \cite{bender1}. The implications of this will be discussed elsewhere.

Applying (\ref{lagr}) to the two fields in (\ref{l1}) gives two de-coupled equations

\begin{equation}
\begin{aligned}
c^2\frac{\partial^2\phi_1}{\partial x^2}=\frac{\partial^2\phi_1}{\partial t^2}+\frac{1}{\tau}\frac{\partial\phi_1}{\partial t}\\
c^2\frac{\partial^2\phi_2}{\partial x^2}=\frac{\partial^2\phi_2}{\partial t^2}-\frac{1}{\tau}\frac{\partial\phi_2}{\partial t}
\end{aligned}
\label{2eq}
\end{equation}

\noindent with the solution

\begin{equation}
\begin{aligned}
\phi_1=\phi_0\exp\left(-\frac{t}{2\tau}\right)\cos(kx-\omega t)\\
\phi_2=\phi_0\exp\left(\frac{t}{2\tau}\right)\cos(kx-\omega t)\\
\omega=\sqrt{c^2k^2-\frac{1}{4\tau^2}}
\end{aligned}
\label{2phi}
\end{equation}

\noindent with the same dispersion relation as in (\ref{omega}) and where, for simplicity, we assumed zero phase shifts in $\phi_1$ and $\phi_2$.

We consider the dissipative process over time scale comparable to $\tau$ because the phonon with the $k$-gap dissipates after time comparable to $\tau$ (see (\ref{solution}) and (\ref{omega0})). Both $\phi_1$ and $\phi_2$ change appreciably over this time scale.

We observe that the first equation in (\ref{2eq}) is identical to (\ref{gener3}), resulting in the first solution in (\ref{2phi}) as in (\ref{omega0}). The second solution increases with time. $\phi_1$ and $\phi_2$ in (\ref{2phi}) can be viewed as energy exchange between waves $\phi_1$ and $\phi_2$: $\phi_1$ and $\phi_2$ appreciably reduce and grow over time $\tau$, respectively. This process is not dissimilar from phonon scattering in crystals due to defects or anharmonicity where a plane-wave phonon ($\phi_1$) decays into other phonons (represented by $\phi_2$) and acquires a finite lifetime $\tau$ as a result.

We note that $\phi_2$ can be viewed as the wave propagating back in time and space with respect to $\phi_1$ because $\phi_2(x,t)=\phi_1(-x,-t)$, implying that a Lagrangian formulation of a
non-reversible dissipative process involves two waves moving in the opposite space-time directions, resulting in this sense in the reversibility of the Lagrangian description. As we will see in section \ref{tre}, the two-field description is analogous to the Keldysh-Schwinger approach where the integration is extended to a complementary plane and integration contour is closed as a result.

The total energy of the composite system does not have exponential terms $\exp\left(\pm\frac{t}{2\tau}\right)$ due to their cancellation. Indeed, the Hamiltonian is $H=\pi_1\frac{\partial\phi_1}{\partial t}+\pi_2\frac{\partial\phi_2}{\partial t}-L$, where $\pi_1=\frac{\partial\phi_2}{\partial t}-\frac{\phi_2}{2\tau}$ and $\pi_2=\frac{\partial\phi_1}{\partial t}+\frac{\phi_1}{2\tau}$ from (\ref{l1}). This gives $H=\frac{\partial\phi_1}{\partial t}\frac{\partial\phi_2}{\partial t}+c^2\frac{\partial\phi_1}{\partial x}\frac{\partial\phi_2}{\partial x}$. Using the solutions $\phi_1$ and $\phi_2$ above gives the system energy $E=\phi_0^2\left(2c^2k^2\sin^2(kx-\omega t)-\frac{1}{4\tau^2}\right)$. Averaged over time, $E=\phi_0^2\left(c^2k^2-\frac{1}{4\tau^2}\right)=\phi_0^2\omega^2$ and is constant and positive.

\begin{figure}
{\scalebox{0.4}{\includegraphics{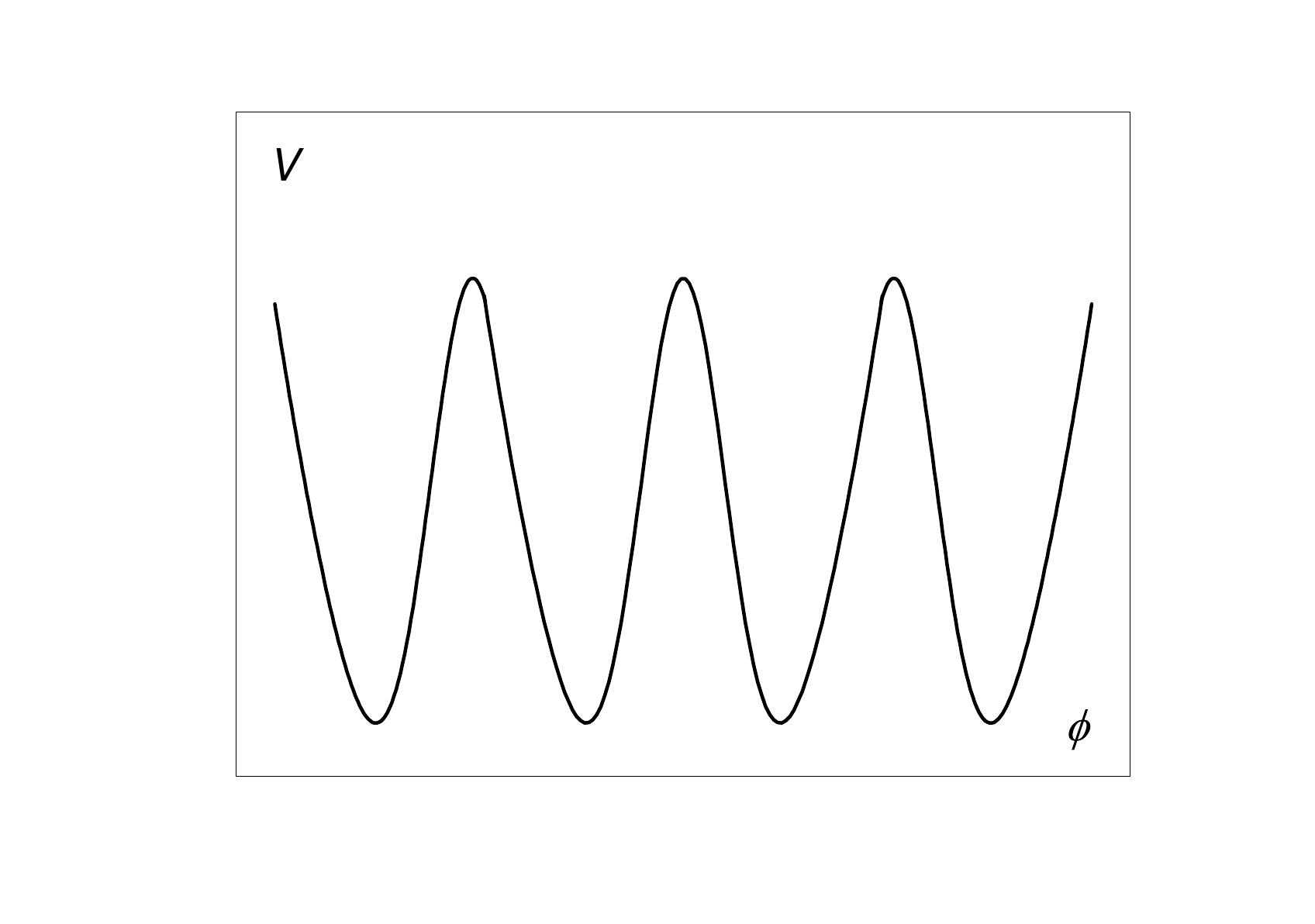}}}
\caption{Schematic illustration of an interaction potential.}
\label{potential}
\end{figure}

From the point of view of field theory, the last term in (\ref{l1}) describes dissipative hydrodynamic motion and represents a way to treat strongly anharmonic self-interaction of the field. Indeed, if this interaction has a double-well (or multi-well) form, the field can move from one minimum to another in addition to oscillating in a single well \cite{prd}. This interaction potential is illustrated in Fig. \ref{potential}. This motion is analogous to diffusive particle jumps in the liquid responsible for the viscous $\propto\frac{1}{\tau}$ term in (\ref{gener3}). Therefore, the dissipative $\propto\frac{1}{\tau}$ term in (\ref{l1}) and (\ref{2eq}) describes the hopping motion of the field (via thermal activation or tunneling \cite{coleman}) between different wells with frequency $\frac{1}{\tau}$. We refer to this term as dissipative, although we note that no energy dissipation takes place in the system as discussed above. Rather, the dissipation concerns the propagation of plane waves in the anharmonic field of Lagrangian (\ref{l1}). The dissipation varies as $\propto\frac{1}{\tau}$: large $\tau$ corresponds to rare transitions of the field between different potential minima and non-dissipative wave propagation due to the first two terms in (\ref{l1}).

\subsection{Interplay between the dissipative and mass terms}

From the point of view of field theory, it is interesting to see how the gap in $k$ space in (\ref{l1}) can vary and to elaborate on general effects of the dissipative term (\ref{twof}). We start with adding the mass term to (\ref{l1}), $-m^2\phi_1\phi_2$:

\begin{equation}
\begin{split}
&L=\frac{\partial\phi_1}{\partial t}\frac{\partial\phi_2}{\partial t}-c^2\frac{\partial\phi_1}{\partial x}\frac{\partial\phi_2}{\partial x}+\\
&\frac{1}{2\tau}\left(\phi_1\frac{\partial\phi_2}{\partial t}-\phi_2\frac{\partial\phi_1}{\partial t}\right)-m^2\phi_1\phi_2
\end{split}
\label{lagr1}
\end{equation}
\noindent where $m$ is bare mass.

Seeking the solution in the form of a plane wave as before we find the real part of $\omega$ corresponding to propagating waves for both $\phi_1$ and $\phi_2$ as

\begin{equation}
\omega=\sqrt{c^2k^2+m^2-\frac{1}{4\tau^2}}
\label{newomega}
\end{equation}

$\omega$ in (\ref{newomega}) behaves differently depending on the sign of $m^2-\frac{1}{4\tau^2}$. If $m>\frac{1}{2\tau}$, (\ref{newomega}) gives the mass (energy) gap as

\begin{equation}
\omega(k=0)=\sqrt{m^2-\frac{1}{4\tau^2}}
\label{mass-em}
\end{equation}

We observe that the dissipative term reduces the mass (energy gap) from its bare value $m$.

When $m<\frac{1}{2\tau}$, (\ref{newomega}) predicts the gap in $k$-space. Indeed, under this condition the expression under the square root in (\ref{newomega}) is negative unless $k>k_g$, where

\begin{equation}
k_g=\sqrt{\frac{1}{4c^2\tau^2}-\frac{m^2}{c^2}}
\label{newgap}
\end{equation}

Comparing with (\ref{kgap}) we see that the bare mass reduces the gap in $k$-space.

The mass gap and $k$-gap both close when

\begin{equation}
m=\frac{1}{2\tau}
\label{compete1}
\end{equation}

\noindent i.e. when the bare mass $m$ becomes close to the field hopping frequency. In this case, (\ref{newomega}) gives the photon-like dispersion relation $\omega=ck$, corresponding to the first two terms of (\ref{lagr1}) only.

These results bring us back to Fig. 1b and our earlier discussion of how a DP with energy (mass) gap can continuously transform into the DP with the gap in momentum space. Eqs.
(\ref{newomega})-(\ref{newgap}) show how the transformation between  the mass-gapped and momentum-gapped DR proceeds as $\omega_{\rm F}=\frac{1}{\tau}$ increases and how the gapless DP emerges in the
process.

In section \ref{sechol}, we will see that the same interplay between the dissipative and mass terms and the associated transformation between the mass-gapped and momentum-gapped DR originates in the holographic models. It also emerges in the Keldysh-Schwinger formalism discussed in section \ref{tre}.

\subsection{How wide can the gap get?}

According to (\ref{kgap}), $k_g$ increases with temperature because $\tau$ decreases. How large can $k_g$ be and how wide can the $k$-gap become?

In condensed matter systems and liquids in particular, there is an upper limit to $k_g$ set by the interatomic distance $a$ playing the role of the UV cutoff. Recall that $\tau$ approximately corresponds to the average time between diffusive jumps in a liquid. Hence its shortest value can not exceed an elementary, Debye, vibration period, $\tau_{\rm D}$. Using $\tau\approx\tau_{\rm D}$ in (\ref{kgap}) and noting that $c\tau_{\rm D}\approx a$ gives $k_g\approx\frac{1}{a}$. This is close to the largest, Debye, $k$ set by the zone boundary in the spherical isotropic approximation, $k_{\rm D}$:

\begin{equation}
k_g\approx k_{\rm D}
\label{kmax}
\end{equation}

(\ref{kmax}) sets the largest width of the $k$-gap.

If, as discussed in section \ref{lagr-gap}, $\tau$ in (\ref{kgap}) is related to the motion in the multi-well potential in Fig. \ref{potential}, another, and larger, UV cutoff emerges. This cutoff is related to the height of the activation barrier of the potential, $U$. Indeed, the potential provides no restoring force for energies larger than $U$, hence the starting harmonic point of perturbation theories involving creation and annihilation operators does not apply for energies larger than $U$ \cite{scirep2019}.

\begin{figure}
\begin{center}
{\scalebox{0.45}{\includegraphics{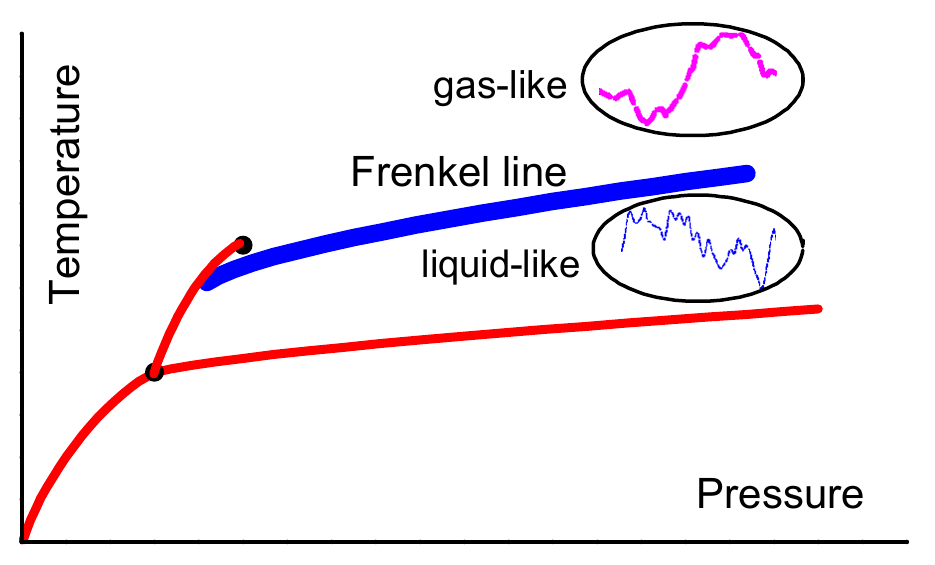}}}
\end{center}
\caption{Phase diagram of matter showing the triple and critical points and the Frenkel line in the supercritical region. Particle dynamics includes both oscillatory and diffusive components below the line, and is purely diffusive above the line. Below the line, the system is able to support rigidity and transverse modes. Above the line, particle motion is purely diffusive, and the ability to support rigidity and transverse modes is lost. Crossing the Frenkel line from below corresponds to the transition between the ``rigid'' liquid to the ``non-rigid'' gas-like fluid.}
\label{frenline}
\end{figure}

Interestingly, $k_g\approx k_{\rm D}$ not only sets the largest value of the $k$-gap but is also related to an important crossover in the behavior of both subcritical and supercritical fluids. Supercritical fluids in particular started to be widely deployed in many important industrial processes once their high dissolving and extracting properties were appreciated \cite{sup1,sup2}. Theoretically, little was known about the supercritical state, apart from the general assertion that supercritical fluids can be thought of as high-density gases or high-temperature fluids whose properties change smoothly with temperature or pressure and without qualitative changes of properties. This assertion followed from the known absence of a phase transition above the critical point. A recent discussion challenging this understanding introduced the Frenkel line (FL) separating two supercritical states \cite{fl1,fl3,fl2}. The FL is illustrated in Fig. \ref{frenline}.

The main idea of the FL lies in considering how particle dynamics changes in response to pressure and temperature. Recall that particle dynamics in the liquid can be separated into solid-like oscillatory and gas-like diffusive components. This separation applies equally to supercritical fluids as it does to subcritical liquids: increasing temperature reduces $\tau$, and each particle spends less time oscillating and more time jumping; increasing pressure reverses this and results in the increase of time spent oscillating relative to jumping. Increasing temperature at constant pressure (or decreasing pressure at constant temperature) eventually results in the disappearance of the solid-like oscillatory motion of particles; all that remains is the diffusive gas-like motion. This disappearance represents the qualitative change in particle dynamics and gives the point on the FL in Figure \ref{frenline}. The change of particle dynamics at the FL gives a practical criterion to calculate the line based on the disappearance of minima of the velocity autocorrelation function (VAF) \cite{fl3}. Notably, the FL exists at arbitrarily high pressure and temperature (as long as chemical bonding is unaltered), as does the melting line. At low temperature the FL touches the boiling line at around 0.8$T_c$, where $T_c$ is the critical temperature (note that the system does not have cohesive liquid-like states at temperatures above approximately 0.8$T_c$, hence crossing the boiling line above this temperature can be viewed as a gas-gas transition \cite{fl3}).

Experimentally, the operation of the FL was ascertained in supercritical Ne \cite{ne}, CH$_4$ \cite{ch4} and CO$_2$ \cite{co2}.

The significance of the Frenkel line for the purpose of our current discussion of GMS is that the line marks the maximal value of $k_g$ at which point all transverse modes disappear from the system spectrum. Indeed, it is readily seen on general grounds that the presence or absence of solid-like oscillatory particle motion implies the presence or absence of solid-like shear modes in the system. This can be seen more specifically on the basis of $k_g$: increasing $k_g$ to its maximal value $k_{\rm D}$ implies no $k$ at which transverse modes can propagate, i.e. complete disappearance of these modes from the system's spectrum.

Above the FL, only the longitudinal mode propagates. This is confirmed by calculations of current correlation functions above and below the FL \cite{fomin}. Since the ability to support transverse waves is associated with solid-like rigidity, the FL corresponds to the crossover from the ``rigid'' liquid to the ``non-rigid'' gas-like fluid where no transverse modes exist, implying the qualitative change of the excitation spectrum \cite{fl1,fl3,fl2}.

\begin{figure}
\begin{center}
\includegraphics[width=1\linewidth]{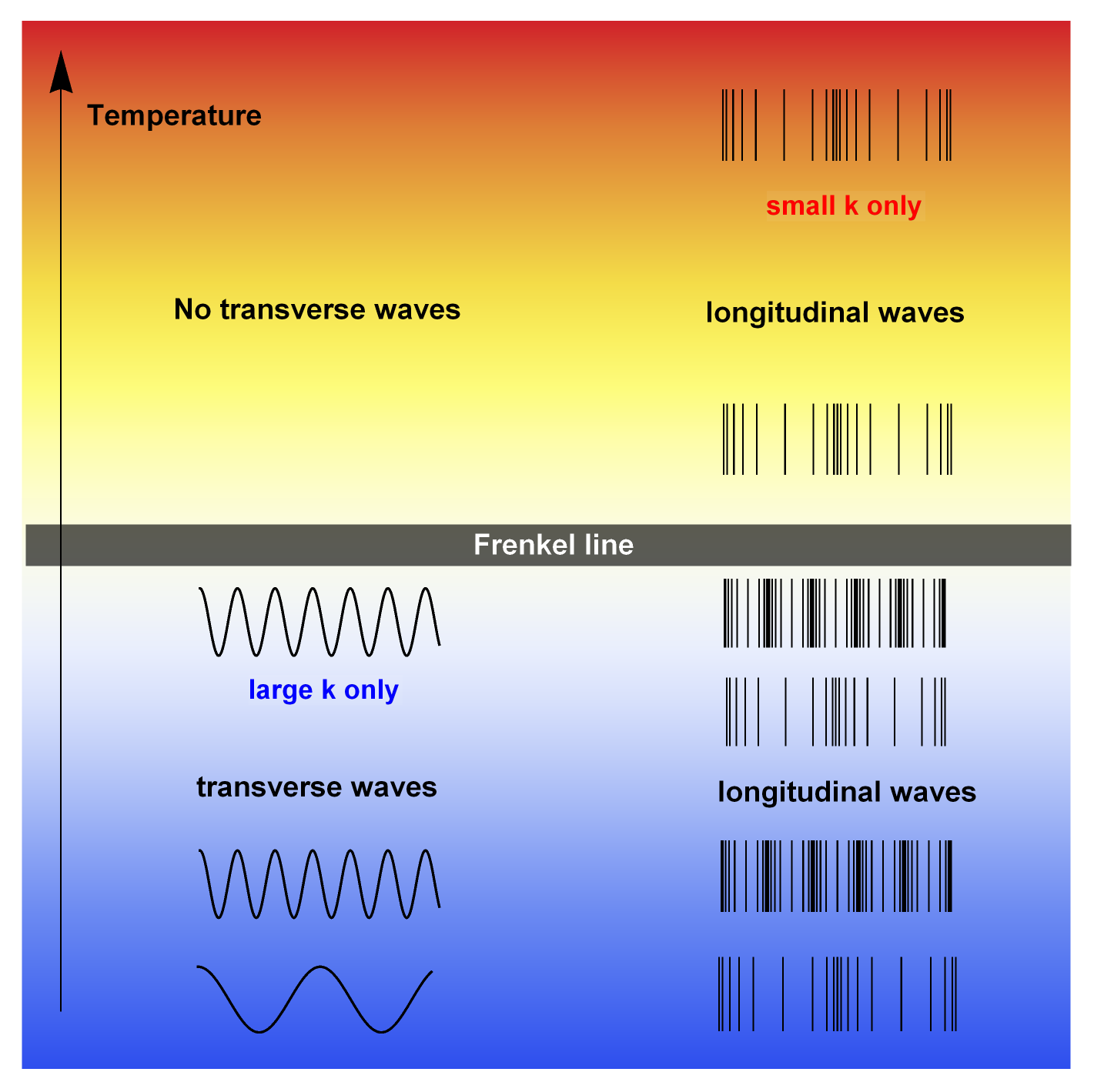}
\end{center}
\caption{Evolution of transverse and longitudinal waves in disordered matter, from viscous liquids and glasses at low temperature to gases at high. The variation of colour from deep blue at the bottom to light red at the top corresponds to temperature increase. The Figure shows that transverse waves (left) start disappearing with temperature starting with long wavelength modes (small $k$) and completely disappear at the Frenkel line where $k_g$ increases to its maximal value comparable to the zone boundary. The longitudinal waves (right) propagate up to the Frenkel line but start disappearing above the line starting with the shortest wavelength, with only long-wavelength longitudinal modes propagating at high temperature.}
\label{ricky}
\end{figure}

Figure \ref{ricky} illustrates the above discussion and shows the evolution of collective modes in liquid and supercritical states with temperature. The Figure shows that at low temperature, liquids and supercritical fluids have the same set of collective modes as in solids: one longitudinal mode and two transverse modes. On temperature increase, the number of transverse modes propagating above $k_g$ decreases. At the FL where particles lose the oscillatory component of motion and start moving diffusively as in a gas, the two transverse modes disappear. Above the FL, only the longitudinal mode remains and has wavelengths larger than the particle mean free path \cite{ropp}; at high temperature (low density) this represents the long-wavelength sound.

We note that the disappearance of the shear modes from the system's spectrum implies a specific value of system's specific heat: $c_v=2k_{\rm B}$. This $c_v$ is the sum of the kinetic term $\frac{3}{2}k_{\rm B}$ and the term due to the potential energy of the remaining longitudinal mode, $\frac{1}{2}k_{\rm B}$ (here a quasi-harmonic approximation is implied where the energy per mode is $k_{\rm B}T$). $c_v=2k_{\rm B}$ serves as a thermodynamic criterion of the Frenkel line and gives the line coinciding with the dynamical VAF criterion \cite{fl3}. Above the FL, the longitudinal mode remains propagating but only with the wavelengths larger than the particle mean free path, and its energy progressively decreases with temperature until it becomes close to the ideal gas \cite{ropp}. This implies a crossover of $c_v$ at $c_v$ close to $2k_{\rm B}$ because temperature dependence of $c_v$ are different below and above the FL. This crossover was ascertained on the basis of molecular dynamics simulations \cite{c-cross}, and the possibility of a phase transition at the Frenkel line was raised.

\begin{figure}
\begin{center}
{\scalebox{0.7}{\includegraphics{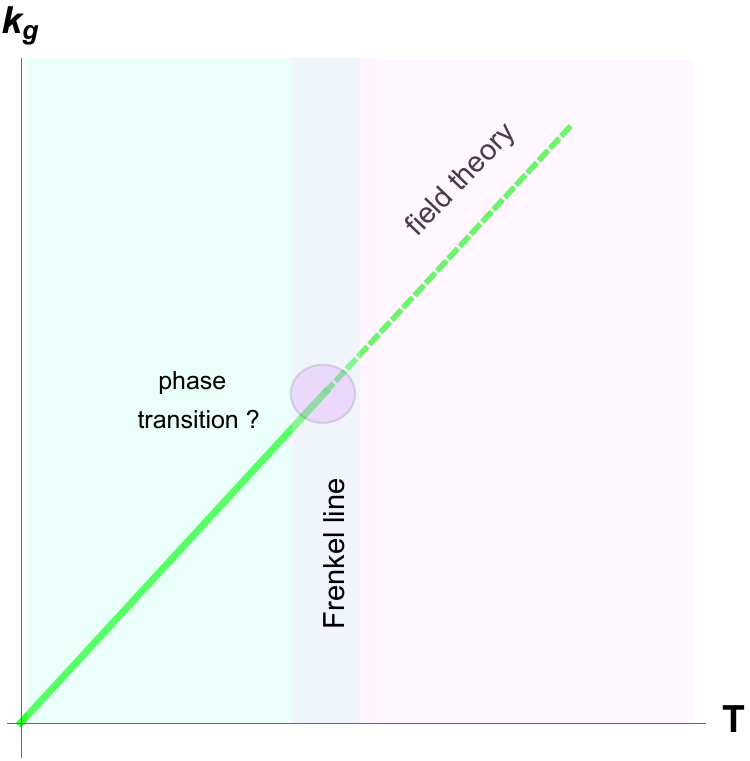}}}
\end{center}
\caption{$k_g$ increases in condensed matter systems up to the Frenkel line where $k_g$ reaches its maximal value (solid line). In field theories with no UV cutoff, $k_g$ increases without bound. Schematic illustration.}
\label{gms-max}
\end{figure}

In summary to this section, we observe that $k_g$ reaching its maximal value corresponds to a physically significant situation where (a) shear modes disappear and the system's spectrum changes qualitatively; (b) a special line (Frenkel line) exists on the phase diagram where dynamical and thermodynamic properties qualitatively change. This takes place because condensed matter systems have a natural UV cutoff related to inter-atomic spacing. In this sense, the operation of GMS in condensed matter systems is different from field theories discussed later in this review: if no UV cutoff is imposed (at, e.g., Planck scale) in these theories, $k_g$ may increase without bound. Fig. \ref{gms-max} illustrates this point: $k_g$ reaches its maximal value at the Frenkel line in condensed matter systems but keeps increasing in field theories with no UV cutoff.

\section{Keldysh-Schwinger approach to dissipation: two-field description}\label{tre}

\subsection{Formulation in terms of two fields}
In previous sections, we discussed the emergence of gapped momentum states due to dissipation in classical systems. In this section, we will discuss the generalization of this picture to the quantum case. We will see that the Keldysh-Schwinger approach similarly involves doubling of the number of degrees of freedom.

In this section, we discuss the emergence of gapped momentum states in the Keldysh-Schwinger approach \cite{Keldysh,Schwinger} to dissipation. This approach describes quantum-mechanical evolution of a non-equilibrium system. In this description, two fields emerge describing a central dissipative system and its external environment, in line with a more general proposal in a field theory that a dissipative behavior is associated with doubling the number of fields \cite{Janssen,Dominicis}. The two fields exchange energy, with the total energy remaining constant. This is similar to the two-field description of the dissipative Lagrangian we encountered earlier.

The Keldysh-Schwinger technique describes the non-equilibrium dynamics of a quantum mechanical system with a large number of particles \cite{Keldysh,Schwinger}. In several texts
\cite{DF,KO,KAMENEV,Milton}, this description is formulated in terms of path integrals. Here, we give our own brief interpretation of this formulation, based on the continuum integration formalism, which we consider to be most concise and transparent and hence suitable for a compact review.

Let us mention the important connection between the Keldysh-Schwinger formalism and the formulation of dissipative hydrodynamics from an action principle \cite{Grozdanov:2013dba,Glorioso:2018wxw,Jensen:2018hse,Haehl:2018lcu} which further links this section with the rest of the paper.

The probability of a transition between the states of a system is represented by a functional integral:
\begin{gather*}
\langle q, t''| q', t'\rangle=\langle q_{t''}|\mathcal{\hat U}| q_{t'}\rangle = \int \mathfrak{D}q\exp \left[\dfr i{\hbar}\int\limits_{t'}^{t''}\mathcal{L}(q)\mathrm{d}t \right]
\end{gather*}
where $\mathcal{L}(q)=p\dot q-H(q)$ is the Lagrangian of the system, $H$ is the Hamiltonian, $\int \mathfrak{D}q$ is the functional integration (see Appendix I), $\mathcal{\hat U}$ is the evolution operator, $q$ is the microscopic state parameter
(a particle position for example), $p=\dot q/m$ is momentum, the symbols $ |q,\,t\rangle $ denote states in the Heisenberg representation, $ |q,\,t\rangle = e^{- iHt / \hbar} | q \rangle $ and $ | q \rangle $ is the
dimensionless eigenstate of the operator $ \hat q $ in the Schr\"{o}dinger representation. Below we will consider the case of slow changes of a microscopic state parameter and length scales larger
than the lattice constant.

For a many-particle system $\{q\}=\{q_1,\,q_2,\,\dots,\,q_N\}$ the Hamiltonian of a quasi-equilibrium system at time $t$ is
\begin{gather*}
H\{q\}=\sum\limits_{a=1}^N\frac{p_a^2}{2\mu}+U(q_1,\,q_2,\dots,\,q_N)
\end{gather*}
where $U$ is the potential energy and $\mu$ is the particle mass.

Let us represent the potential energy as:
\begin{gather*}
U\{q\}=\sum\limits_{a,b}^N k_{ab}q_aq_b+\dots
\end{gather*}
write $2q_aq_b=q_a^2+q_b^2-(q_a-q_b)^2$ and assume that $k_{ab}$ bonds only nearest points in the lattice~\cite{Zee}. Then
\begin{gather*}
U\{q\}=\dfr{l^2k_l}2\sum\limits_{i=1}^n\sum\limits_{a}^N\dfr{(q_{a+i}-q_a)^2}{l^2}+k_l\sum\limits_{a}^Nq_a^2+\dots
\end{gather*}
where $l$ is the lattice constant and $n$ is the number of nearest points.

In the continuous limit
\begin{gather*}
U(q)=\dfr{\varepsilon}2 (\nabla q)^2+v(q)
\end{gather*}
where $\varepsilon $ is the parameter characterising the smoothness of $q$ function and $v(q)$ is the function which depends on $q$. Considering system's microscopic states at all space points $\bf
r$, $q_a(t)\to \phi_{\bf r}(t)$, we have
\begin{align*}
&\langle \phi, t''| \phi', t'\rangle=\langle \phi_{t''}|\mathcal{\hat U}| \phi_{t'}\rangle =\\&=
\int \mathfrak{D}\phi   \exp \left[\dfr i{\hbar V}\int\limits_{t'}^{t''}\mathrm{d}t \left(\int\mathrm{d}V\left(\dfr12\mu\dot \phi_{\bf r} ^2-\dfr{\varepsilon}2 (\nabla\phi_{\bf
r})^2\right)-v\{\phi_{\bf r}\}\right)\right]
\end{align*}
where $\phi $ is the vector of infinite dimensions with components are $\phi_{\bf r}$, $V$ is the system volume and $\int\mathrm{d}V$ is the integration over this volume.

The initial and final states are assumed to be equilibrium and are related to the ground state by $ | \phi_{eq}, t \rangle = \sqrt{\rho_0} | 0 \rangle $, where $ \rho_0 (T) $
is the equilibrium density of states. Therefore, $ \langle \phi, \infty | \phi', - \infty \rangle = \langle \phi_{eq}, \infty | \phi_{eq}, - \infty \rangle = \rho_0 \langle 0 | 0 \rangle = \rho_0 $, i.e. it is a certain constant that depends on temperature. In this case, a mean value $A$
\begin{equation*}
\langle A(\phi )\rangle =\dfr {1}{\langle \phi_{eq}, \infty| \phi_{eq}, -\infty\rangle}\int \mathfrak{D}\phi A(\phi )\exp \left[\dfr i{\hbar}\int\limits_{-\infty }^{\infty }\mathrm{d}t \,\mathcal{L}(\phi )\right]
\end{equation*}
is well-defined.

In order to make connection to equilibrium statistical mechanics, we assume $ i(t-t') = \hbar \beta $, and the Hamiltonian of the system is unchanged: $ H(\phi (t)) = H $. Then we arrive at the standard (equilibrium) statistical mechanics:
\begin{gather*}
Z =\langle \phi_{i\beta }| \phi_{-i\beta} \rangle= \mathcal{N}\int \mathfrak{D}\phi \exp \left[-\beta H(\phi ) \right]=1.
\end{gather*}

Let us consider a quantum many-body system governed by a time-dependent Hamiltonian $\hat H(t)$. The time evolution of the system is given by the evolution operator $\mathcal{\hat U}$: $|\phi,
t+\delta t\rangle=\mathcal{\hat U}_{\delta t}|\phi,\,t\rangle$. This operator evolves according to the Heisenberg equation of motion $\hbar\partial_t\mathcal{\hat U}_{\delta t}=i\left[\mathcal{\hat U}_{\delta t},\,\hat
H\right]$, which is formally solved as
\begin{gather*}
\mathcal{\hat U}_{\delta t}=\exp \left[-\dfr i{\hbar}\delta t\hat H\right].
\end{gather*}

Let us divide the time interval $(t',\,t'')$ into the infinitesimal parts $\delta t$, then the probability of transition from $\phi_{t'}$ state to $\phi_{t''}$ state is
\begin{align*}
&\langle \phi_{t''}|\mathcal{\hat U}|\phi_{t'}\rangle= \int\mathfrak{D}\phi_{t''-\delta t}\dots\int\mathfrak{D}\phi_{t'+\delta t}\langle \phi_{t''}|\mathcal{\hat U}_{\delta t}| \phi_{t''-\delta t}
\rangle \times\\
&\langle \phi_{t''-\delta t}|\mathcal{\hat U}_{\delta t}| \phi_{t''-2\delta t} \rangle \dots \langle \phi_{t'+2\delta t}|\mathcal{\hat U}_{\delta t}| \phi_{t'+\delta t} \rangle \langle \phi_{t'+\delta
t}|\mathcal{\hat U}_{\delta t}| \phi_{t'} \rangle ,
\end{align*}
where $\mathcal{\hat U}_{\delta t}$ is the evolution operator during the time interval $\delta t$. The $| \phi_{j-1}\rangle$, and $| \phi_{j}\rangle$ states are not coherent. Therefore, the evolution
operator elements are given by:
\begin{align*}
&\langle \phi_j|\mathcal{\hat U}_{\delta t}| \phi_{j-1}\rangle \equiv
\langle \phi_j|e^{-i\hat H\hbar^{-1}\delta t}| \phi_{j-1}\rangle \\& \approx
\langle \phi_j|\hat 1-{i\hbar^{-1}\hat H\delta t}| \phi_{j-1}\rangle =
\langle \phi_j| \phi_{j-1}\rangle (1-{i\hbar^{-1}H_j\delta t})\\ &=
\langle \phi_j| \hat 1+\delta t\partial_t|\phi_{j}\rangle (1-{i\hbar^{-1}H_j\delta t}) \\ & =
\langle \phi_j|\phi_{j}\rangle e^{\delta t\phi_j^*\partial_t\phi_{j}}e^{-i\hbar^{-1}H_j\delta t}=e^{\phi_j^*\partial_t\phi_{j}\delta t-i\hbar^{-1}H_j\delta t}
\end{align*}
(see Appendix II), and in the continuous limit we have
\begin{gather*}
\langle \phi_{t''}|\mathcal{\hat U}|\phi_{t'}\rangle=
 \int \mathfrak{D}\phi\exp \left[i\int\limits_{t'}^{t''}\mathrm{d}t\,\left(\hbar^{-1}\mathcal{L}(\phi)+i\phi\partial_t\phi\right) \right],
\end{gather*}
where $\phi\partial_t\phi$ denotes $V^{-1}\int\mathrm{d}V
\,\phi_{\bf r}\partial_t\phi_{\bf r}$.

Introducing the imaginary time $t\to -it$, this expression can be represented as
\begin{gather*}
\langle \phi_{t''}| \mathcal{\hat U}|\phi_{t'}\rangle=
\int \mathfrak{D}\phi \exp \left[-\int\limits_{t'}^{t''}\mathrm{d}t\,\phi G^{-1}\phi \right],
\end{gather*}
where

\begin{equation}
G^{-1} = \hbar^{-1}\mu\partial_t^2-\hbar^{-1}\varepsilon \nabla^2 + \partial_t
\label{green}
\end{equation}
\noindent is the operator inverse to the Green function operator \cite{KAMENEV}.

The quadratic part of (\ref{green}) corresponds to the wave equation $\mu\partial_t^2\phi-\varepsilon \nabla^2\phi =0$, thus $\mu$ and $\varepsilon $ are related by the wave velocity $c$:
$\varepsilon =c^{2}\mu$. The last term in (\ref{green}) describes dissipation.

So far, the time scale in the system described by (\ref{green}) is given by the quantum-mechanical time scale related to the Planck constant. (\ref{green}) is a convenient place to introduce system's relaxation time $\tau$, similarly to its introduction in the Frenkel's theory of liquids discussed earlier. Similarly to Frenkel's approach, the introduction of $\tau$ here does not involve a first-principles calculation: as discussed earlier, such a calculation in the case of liquids is exponentially complex and therefore is not tractable. Rather, $\tau$ is introduced on physical grounds and from observation that a transit between different potential minima on a potential energy landscape involves a characteristic time $\tau$. As discussed earlier, this time in liquids is related to the system's viscosity via the Maxwell interpolation. We therefore write

\begin{equation}
G^{-1} = \tau\partial_t^2-c^2\tau \nabla^2 + \partial_t
\label{green1}
\end{equation}

\noindent where the first two terms describe the propagating wave and the last term describes dissipation as in (\ref{green}).

We consider a non-equilibrium system interacting with a thermal reservoir and coming to thermal equilibrium at $t\to \infty$. We assume that at $t=0$, the system is out of equilibrium and is in the
state $ | \phi_0 \rangle $, and evolves to its final equilibrium state, $\langle \phi_{\infty} |\equiv\langle \phi_{eq} | $, characterized by the equilibrium density of states $\rho _{eq}$. The
transition probability is:
\begin{align*}
&\langle \phi_{eq}, \infty |\phi_0, 0 \rangle = \langle \phi_{eq}|\mathcal{\hat U}_{\infty} |\phi_0 \rangle = \\&=
\int \mathfrak{D}\phi_{eq}\mathfrak{D}\phi
\exp \left[\int\limits^{\infty}_{0}\mathrm{d}t\,(\hbar^{-1}\mathcal{L}(\phi)-\phi\partial_t\phi) -\hbar^{-1}\tau E_{eq}\phi_{eq}\phi_{\infty}\right]
\end{align*}

One can show that this probability does not depend on shifting the time $t=\infty$ (see Appendix III). However, this probability depends on the initial state of the system and, accordingly, on the
choice of the initial time. The averaging operation is not defined in this case and, as a consequence, the statistical theory can not be formulated.

\begin{figure}
   \centering
   \includegraphics[scale=0.47]{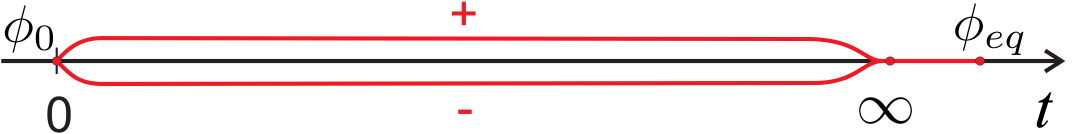}
   \caption{At $t=0$ the present states $\phi_0$ of both fields are equivalent and, therefore, are coherent: $\langle \phi^+_0 |\phi^-_0\rangle =\langle \phi_0 |\phi_0\rangle=1$. At $t=\infty$, the    system reaches the equilibrium state $\phi_{\infty}\equiv\phi_{eq}$ which is also equivalent and coherent for both fields $\phi^+_{\infty}\equiv\phi^-_{\infty}$.  Therefore, the contour is closed for $t=0$ and $t=\infty $. At $t>\infty$, the system is in equilibrium, hence two contour branches are equivalent and can be represented by one line.}
   \label{F1}
\end{figure}

To get around this problem, we use the following approach: consider a copy of our system, with the same transition probability. We denote the field in the initial system as $\phi^+$ and in the
copy as $\phi^-$. Recall that this is the same field, hence $\langle \phi_0^-, 0 | \phi_0^+, 0 \rangle = 1 $. Using the two fields, we close the integration contour in $ t = \infty $ point (see
Fig.\,\ref{F1}) and write
\begin{equation*}
1\equiv \langle \phi_0^- , 0|\phi_0^+ , 0 \rangle=
\int \mathfrak{D}\phi_{eq}\langle \phi_0^- , 0| \phi_{eq}, \infty \rangle
\langle \phi_{eq}, \infty |\phi_0^+ , 0 \rangle
\end{equation*}

Now the integration over the contour yields unity, and the averaging operation in the system with two fields is well defined because it does not depend on the initial state of the system. We now
represent the preceding expression in the form of the functional integral:
\begin{align*}
&\langle\phi_0^-, 0 | \phi_0^+, 0\rangle =
\mathcal{N}\int \mathfrak{D}\phi_{eq}\int \mathfrak{D}\phi^+ \mathfrak{D}\phi^- \times\\ &
\exp \left[\int\limits^{\infty}_{0}\mathrm{d}t\,\left(\hbar^{-1}\mathcal{L}(\phi^-)-\phi^-\partial_t\phi^--\hbar^{-1}\mathcal{L}(\phi^+)+\phi^+\partial_t\phi^+\right) -\right. \\& \left.
\hbar^{-1}\tau E_{eq}\left(\phi^-(\infty)-\phi^+(\infty)\right)\phi_{eq}-\hbar^{-1}\tau E_{eq}\phi_{eq}^2
\right],
\end{align*}
in which we ``glued'' the two branches of the contour at the point $ t = \infty $, since $ \phi_{eq} $ state does not depend on the choice of the contour. After the integration over $ \phi_{eq} $
(see Appendix I), we obtain
\begin{align*}
&\langle\phi_0^-, 0 | \phi_0^+, 0\rangle =
\mathcal{N}'\int \mathfrak{D}\phi^+ \mathfrak{D}\phi^-  \times\\&
\exp \left[\int\limits^{\infty}_{0}\mathrm{d}t\,\left(\hbar^{-1}\mathcal{L}(\phi^-)-\phi^-\partial_t\phi^--\hbar^{-1}\mathcal{L}(\phi^+)+\phi^+\partial_t\phi^+\right) \right.-\\ &\left.-\hbar^{-1}\tau
E_{eq}\left(\phi^-(\infty)-\phi^+(\infty)\right)^2
\right]
\end{align*}

It is convenient to operate in terms of frequency representation. Because $E_{eq}\phi_{eq}^2=\tau\int\limits_{-\infty}^{\infty}\hbar \omega \,\rho(\omega )\,\phi ^2_{\omega}\mathrm{d}\omega $, where
$\rho (\omega)=\dfr 12\,\mbox{coth}\left(\dfr{\hbar\omega}{2kT}\right)$ is the equilibrium states density (see Appendix IV), we obtain
\begin{align*}
& \langle \phi^-_0, 0 | \phi^+_0, 0 \rangle= \mathcal{N}'\int \mathfrak{D}\phi^+\mathfrak{D}\phi^- \exp \left[\tau^2\int\limits_{-\infty}^{\infty}\mathrm{d}\omega\left\{ \dfr
1{\hbar}\mathcal{L}_{\omega}(\phi^-) -\right.\right.\\ & \left.\left.
\dfr 1{\hbar}\mathcal{L}_{\omega}(\phi^+)-i\omega(\phi^-\phi^- -\phi^+\phi^+)-\omega \rho(\omega )(\phi^+ - \phi^-)^2\right\} \right]
\end{align*}

We can now perform the famous Keldysh rotation and introduce new fields: $\phi ^{cl}=\sqrt{1/2}(\phi^+ +\phi ^-)$, $\phi ^{q}=\sqrt{1/2}(\phi^+ -\phi ^-)$ ($\phi ^{+}=\sqrt{1/2}(\phi^{cl} +\phi
^{q})$, $\phi ^{-}=\sqrt{1/2}(\phi^{cl} -\phi ^{q})$), which are called ``classical'' and ``quantum'' fields, respectively. Using this rotation, the theory is represented in compact and convenient
form. In momentum representation, we have:
\begin{align*}
&1=\langle \phi^-_0, 0 | \phi^+_0, 0 \rangle= \mathcal{N}'\int \mathfrak{D}\phi^{cl}\mathfrak{D}\phi^{q} \times\\&
\exp \left[- V_{\bf k}^{-1}\tau^2\int\limits_{-\infty}^{\infty}\int\mathrm{d}{\bf k}\mathrm{d}\omega\left\{ \phi^{cl}_{\bf k}G^{-1}_{\bf k}\phi^q_{-\bf k} -\phi^q_{\bf k}G^{-1}_{\bf k}\phi^{cl}_{-\bf
k}+ \right.\right.  \left.\left.
2\omega \rho(\omega )\phi^q_{\bf k}\phi^q_{-\bf k}\right\} \right]=\\&=
\mathcal{N}'\int \mathfrak{D}\phi^{cl} \mathfrak{D}\phi^{q}\exp \left[- V_{\bf k}^{-1}\tau^2\int\limits_{-\infty}^{\infty}\int\mathrm{d}{\bf k}\mathrm{d}\omega\,\bar\phi_{\bf k}\hat G^{-1}_{\bf
k}\bar\phi_{-\bf k}\right]
\end{align*}
where $V_{\bf k}=l^{-d}$ is the system volume in momentum space, $\bar \phi=\left\{\phi^{cl},\,\phi^q\right\}$ and $\hat G^{-1}$ is the inverse operator to the Green functions operator
\begin{equation}
\hat G=\left[ \begin{array}{cc}\displaystyle\dfr{\tau^{-1}\omega \,\mbox{coth}\left({\hbar\omega}/{2kT}\right)}{\Box^2+ \omega^2} & \displaystyle\dfr{\tau^{-1}}{\Box -i \omega} \\[12pt]
\displaystyle\dfr{\tau^{-1}}{\Box +i \omega} & \displaystyle 0 \end{array}\right]
\label{gr}
\end{equation}
where $\Box =\tau(c^2{\bf k}^2 -\omega^2)$ is the  d'Alembert operator describing evolution of a conservative elastic system. The elements of matrix $\hat G$ are called as ``advanced'', ``retarded'',
and ``Keldysh'' Green functions.

We observe that the above approach to dissipation represents a quantum approach to the problem and is equivalent to a quantum field theory where the number of fields is doubled. The doubling of the number of degrees of freedom is equivalent to the two-field description discussed in section \ref{lagr-gap}.

\subsection{Gapped momentum states}

The solutions of equations in this theory are found from the asymptotic of the retarded correlation function $G^R(\omega)$ in (\ref{gr}):
\begin{gather*}
G^R(\omega)=\displaystyle\dfr{\tau^{-1}}{\Box -i \omega}=\dfr{\tau^{-1}}{\tau c^2{\bf k}^2-\tau\omega^2 -i \omega}
\end{gather*}

In terms of propagating plane waves, the real part of $\omega$ corresponding to propagating waves is
\begin{gather*}
    \mbox{Re}\,\omega=\sqrt{c^{2}{\bf k}^2-\frac{1}{4\tau^2}}
\end{gather*}
from which the gap in $k$- space emerges
\begin{gather*}
k_g=\dfr{1}{2c\tau}
\end{gather*}
and is the same as in Eq. (\ref{kgap}).

If a mass term is present, $G^R(\omega)$ reads

\begin{gather*}
G^R(\omega)=\displaystyle\dfr{\tau^{-1}}{\Box -i \omega}=\dfr{\tau^{-1}}{\tau c^{2}{\bf k}^2-\tau\omega^2+\tau m^2 -i \omega}.
\end{gather*}
where $m$ is the mass, yielding the dispersion relation

\begin{gather*}
    \mbox{Re}\,\omega=\sqrt{c^{2}{\bf k}^2+m^2-\frac{1}{4\tau^2}}
\end{gather*}

\noindent as in Eq. (\ref{newomega}), resulting in the same interplay between the dissipative and mass terms. This and other equations discussed earlier generalize GMS to the quantum case.

\section{Strongly-coupled plasma}

In this and following sections, we continue discussing different systems where gapped momentum states emerge. In this section, we discuss GMS in strongly-coupled plasma.

Plasma is often described as a different state of matter since its properties are largely defined by the charged character of constituent particles. One might expect that dispersion relations and GMS in particular should depend on the charged nature of particles in plasma in some way. Interestingly, it transpires that GMS in plasma can be rationalized in terms of the same viscoelastic picture discussed in the area of liquids. Particle charges are important in setting the screened-Coloumb interactions and other properties, however GMS appears to be a generic effect emerging for a large variety of interatomic interactions including those operating in plasma.

Strongly-coupled plasma (SCP) can be defined as a system of charged particles where interactions are strong enough to give an approximate equality between potential and kinetic energies.
Investigation of SCP is currently one of the hottest fundamental branches of physics which lies at the interface between different fields: plasma physics, condensed matter physics, atomic and molecular physics \cite{pl1,pl2,pl3}. It is widely recognised that theoretical understanding strongly-coupled plasma faces a fundamental challenge because strong inter-particle interaction precludes using conventional methods of theoretical physics such as perturbation theory \cite{pl1,pl2}. This is the same problem that we encounter in theoretical description of liquids discussed in section \ref{liquid-section}. As a result, most advances in understanding SCP have been coming from experiments, while no theoretical guidance exists for better planning and performing new experiments involving SCP \cite{pl2}.

\begin{figure}
\begin{center}
{\scalebox{0.35}{\includegraphics{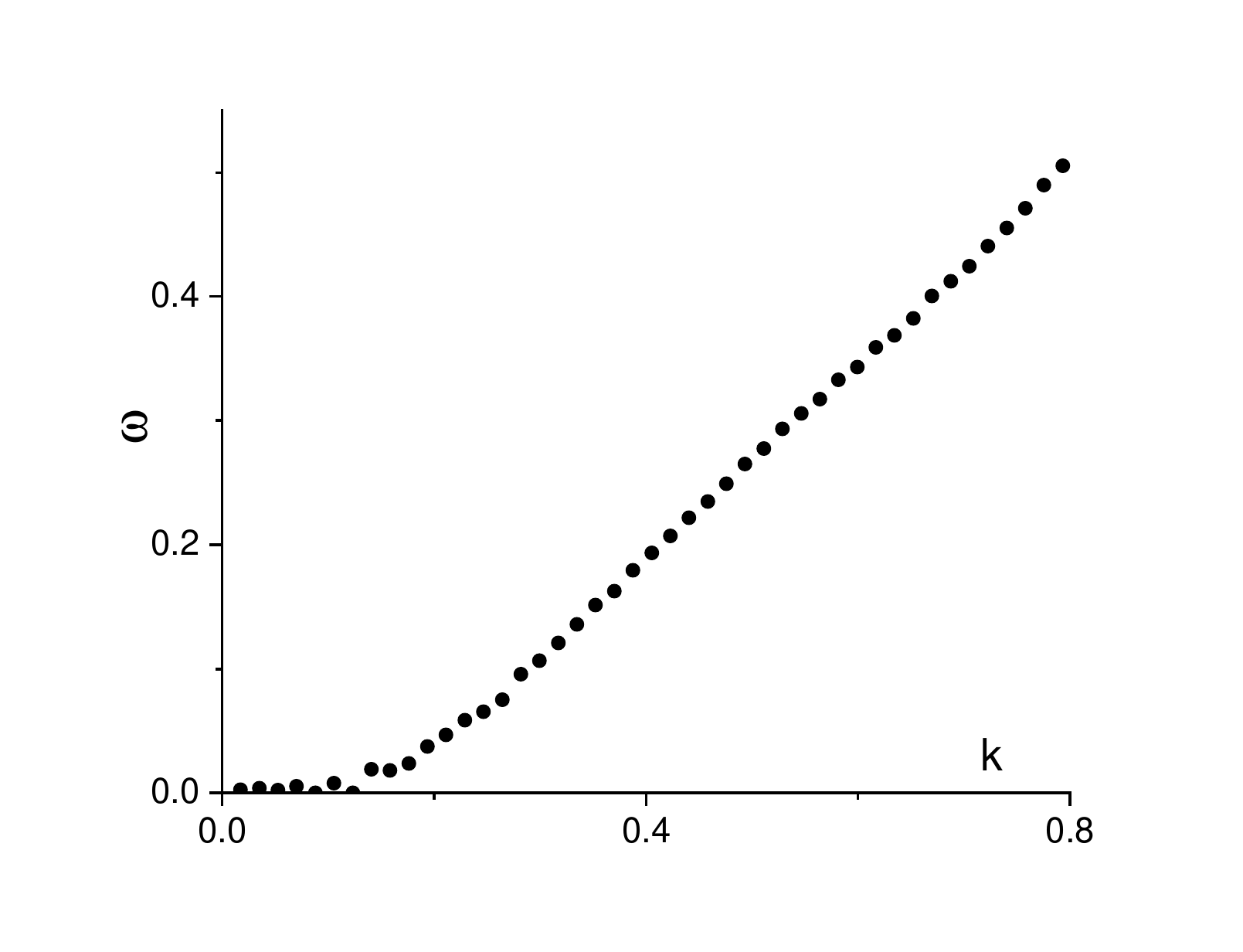}}}
\end{center}
\caption{Transverse dispersion relation calculated in the simulation of the strongly-coupled two-dimensional plasma, showing the gap in $k$-space. $\omega$ and $k$ are shown in reduced units. The data are from Ref. \cite{pg6}.}
\label{gap-plasma}
\end{figure}

There has been extensive research into collective modes in plasma (see, e.g., Refs. \cite{pl2,pmode1,pg-theory1} for review and references therein). We do not attempt to review this large field, but instead remark that, similarly to liquids, the $k$-gap seen in molecular dynamics simulations of plasma models (see, e.g. Refs \cite{pg1,pg-md-theory,pg3,pg4,pg5,pg6}. Fig. \ref{gap-plasma} shows one such example. The gap tends to reduce with the plasma coupling parameter. Qualitatively, this is consistent with the liquid result (\ref{kgap}) if stronger coupling is related to larger
$\tau$.

\begin{figure}
\begin{center}
{\scalebox{0.35}{\includegraphics{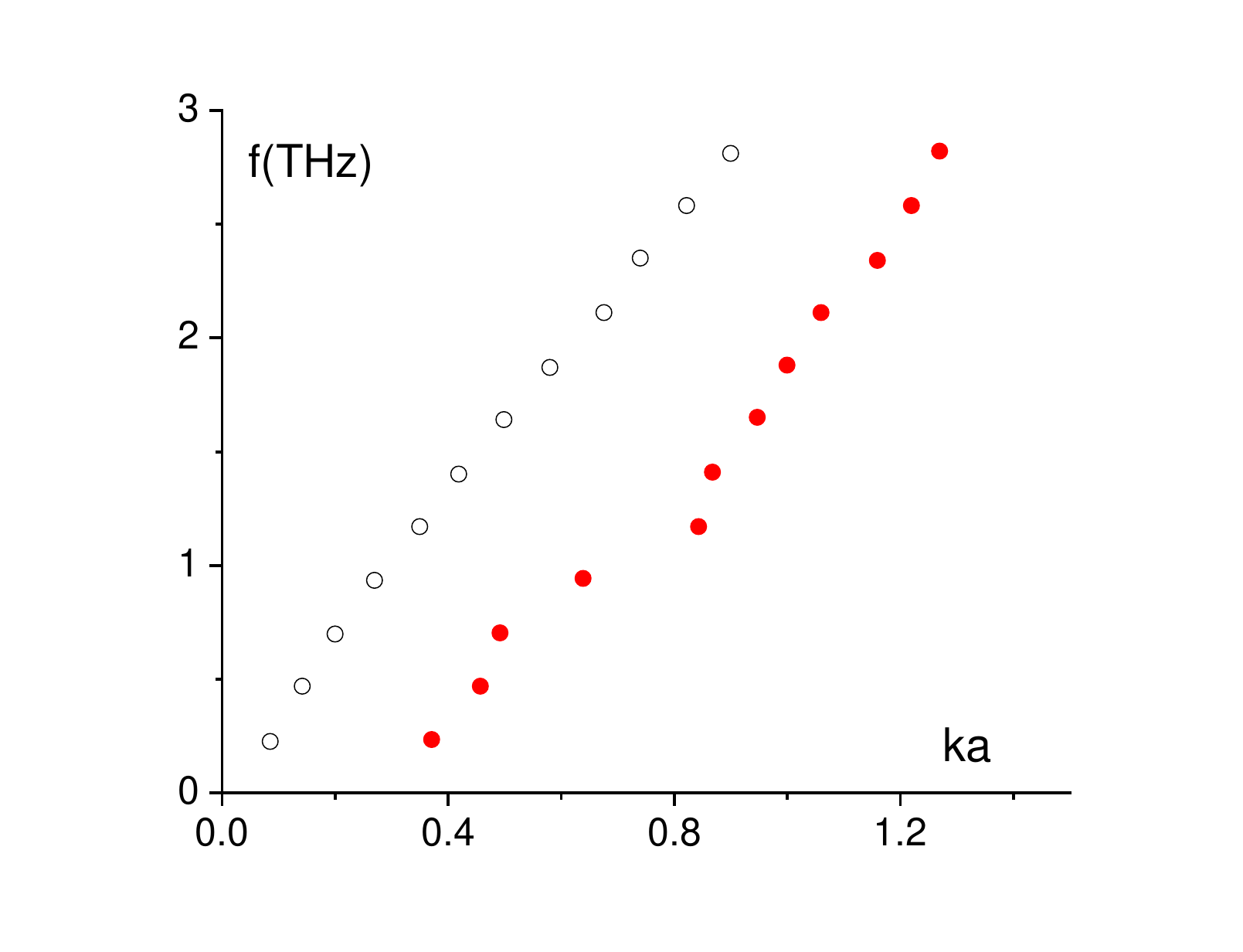}}}
\end{center}
\caption{Experimental transverse dispersion relations in dusty plasma in the solid crystalline state (open black circles) and the liquid state (red bullets). $a=0.325$ mm. The emergence of the gap in $k$-space is seen in the liquid state. The data are from Ref. \cite{pg-exp}.}
\label{plasma-exp}
\end{figure}

An experimental confirmation of gapped momentum states was reported in Ref. \cite{pg-exp}. This study used dusty plasma where particles were imaged by camera. This was possible because, in contrast to liquids, this system has long inter-particle distances and low frequencies: typical inter-particle distance characteristic frequencies are of the order of 1 mm and 10 s$^{-1}$, respectively. Dispersion relations were derived from calculated transverse current correlation functions using the imaged trajectories of dusty plasma particles. Figure \ref{plasma-exp} shows a gapless transverse dispersion relation in the solid crystalline state of dusty plasma and the emergence of the gap in the liquid state.

Theoretically, the gapped momentum states in plasma are rationalized using generalized hydrodynamics, the approach that extends the hydrodynamic description to larger $k$ and $\omega$ (see, e.g., \cite{pg-md-theory,pg-theory1,pg-theory2}) as in liquids mentioned earlier and discussed in section \ref{hydrodyn} in more detail. Interestingly, the authors of Ref. \cite{pg-theory1} trace the generalized hydrodynamics approach used in plasma back to the viscoelastic theory of liquids developed by Frenkel \cite{frenkel}.

\section{Electromagnetic waves}\label{EMsec}

\subsection{$k$-gap and skin effect}

In this section, we discuss the emergence of GMS in electromagnetic waves.

Equations governing the propagation of electromagnetic (EM) waves in a conductor involve combining Maxwell equations with constituent relations between the field and the current. If the latter is taken in the form of the Ohm's law ${\bf J}=\sigma {\bf E}$, where $J$ is the current density and $\sigma$ is conductivity, the wave equation for the electric field $E$ becomes
\cite{griffiths,dresselhaus}:

\begin{equation}
\nabla^2{\bf E}=\mu\,\epsilon\,\frac{\partial^2{\bf E}}{\partial t^2}+\mu\,\sigma\,\frac{\partial{\bf E}}{\partial t}
\label{em1}
\end{equation}
\noindent

The magnetic component follows the same equation.

Seeking the solution in the form ${\bf E}={\bf E_0}e^{i(kx-\omega t)}$ gives

\begin{equation}
k^2=\mu\,\epsilon\,\omega^2+i\,\mu\,\sigma\,\omega
\label{em2}
\end{equation}

We re-write (\ref{em2}) as

\begin{equation}
\omega^2+i\frac{\sigma}{\epsilon}\omega-c^2k^2=0
\label{em3}
\end{equation}

\noindent where $c^2=\frac{1}{\mu\epsilon}$ is the propagation speed and observe that the form of (\ref{em3}) is identical to (\ref{quadratic}) and that Eq. (\ref{em1}) is identical to Eq. (\ref{gener3}), from which the $k$-gap emerges.

A comparison (\ref{em3}) to (\ref{quadratic}) identifies the relaxation time $\tau$ with $\frac{\epsilon}{\sigma}$, giving the same equation as (\ref{quadratic}):

\begin{equation}
\omega^2+\frac{i}{\tau}\omega-c^2k^2=0
\label{em31}
\end{equation}

\noindent and resulting in the $k$-gap for EM waves as

\begin{equation}
k_g=\frac{1}{2c\tau}=\frac{\sigma}{2c\epsilon}
\label{em4}
\end{equation}

Interestingly, the $k$-gap (\ref{em4}) is not commonly discussed in the well-researched area of EM waves. In order to understand this, let us discuss the relationship between $k_g$ and the skin effect.

The discussions of skin depth involve solving (\ref{em2}) for $k$ \cite{griffiths,dresselhaus} rather than $\omega$ as we did for transverse modes in liquids. We write the solution as

\begin{equation}
k=k_1+ik_s
\label{em5}
\end{equation}
\noindent

\begin{equation}
k_1=\frac{\omega}{c\sqrt{2}}\left(\sqrt{1+\left(\frac{1}{\omega\tau}\right)^2}+1\right)^\frac{1}{2}
\label{em5}
\end{equation}
\noindent and

\begin{equation}
k_s=\frac{\omega}{c\sqrt{2}}\left(\sqrt{1+\left(\frac{1}{\omega\tau}\right)^2}-1\right)^\frac{1}{2}
\label{em6}
\end{equation}

\noindent so that ${\bf E}={\bf E_0}e^{-k_sz}e^{i(kx-\omega t)}$ and the skin depth $d_s=\frac{1}{k_s}$.

To aid our further analysis, it is convenient to write the ratio $\frac{d_s}{\lambda}$, an indicator of to what extent the penetration depth is related to the oscillatory behavior. Noting that
$\frac{d_s}{\lambda}=\frac{k_1}{k_s}$ and using (\ref{em5}) and (\ref{em6}) gives

\begin{equation}
\left(\frac{d}{\lambda}\right)^2=1+\,2\,(\omega\,\tau)^2\,\left(1+\sqrt{1+\frac{1}{(\omega\tau)^2}}\right)
\label{em8}
\end{equation}

We observe that in the regime $\omega\tau\gg 1$, $\frac{d}{\lambda}\gg 1$, implying many wavelengths in a propagation length. In the opposite ``hydrodynamic'' regime $\omega\tau\ll 1$, $\frac{d}{\lambda}$ tends to 1, implying strong attenuation. Thus the two regimes are identical to the solid-like elastic and hydrodynamic regimes of shear wave propagation envisaged by Frenkel and
discussed in section \ref{uno}.

In the regime $\omega\tau\gg 1$, $k_1$ in (\ref{em5}) gives $k_1=\frac{\omega}{c}$ as expected for propagating waves, and $k_s$ in (\ref{em6}) becomes $k_s=\frac{1}{2c\tau}$, the same as $k_g$ in
(\ref{em4}):

\begin{equation}
k_s=k_g
\label{em9}
\end{equation}

This implies that in the weakly-attenuated regime, the skin depth includes the full range of wavelengths at which EM waves propagate, from $d_s=\frac{1}{k_s}$ to the shortest wavelength. In this sense, specifying the propagation length (skin depth) implies the $k$-gap because it imposes the smallest $k$ of propagating waves in the system. Note that the opposite does not apply: specifying the $k$-gap implies the allowed values of $k$ of propagating waves but not the propagation length.

It therefore appears that phonon propagation in liquids and propagation of EM waves in conductors was historically discussed in different terms: the first phenomenon was quantified in terms of the $k$-gap, whereas the second one was discussed in terms of the skin depth. Although the equation governing the two effects is the same, solving it for $\omega$ gives the $k$-gap, whereas solving it for $k$ gives the skin depth. One can reverse this state of affairs and discuss the propagation of EM waves in conductors in terms of the $k$-gap (not commonly done) and to discuss the propagation of phonons in liquids in terms of the ``skin depth'', or propagation length, with the proviso that phonons in liquids are internal excitations as compared to an external electromagnetic field penetrating a conductor. Recall that we have previously referred to the propagation length of transverse phonons in liquids as the liquid elasticity length $d_{\rm el}$ in Eq. \ref{del}, which has the same physical meaning as the skin depth for EM waves.

In the opposite regime $\omega\tau\ll 1$, $k_1$ and $k_s$ become $\frac{1}{c\tau\sqrt{2}}{\sqrt{\omega\tau}}$, resulting in the skin depth varying as $\propto\frac{1}{\sqrt{\omega}}$ as commonly discussed.


\subsection{Interplay between the gaps in $k$- and $\omega$-space}

Returning to our original Figure 1 showing three possible dispersion relations (frequency gap, $k$-gap and gapless line), it is interesting to see how the gaps in $k$-space and $\omega$ space move in
response to parameter change. To see this, it is convenient to use the Drude model for conductivity:

\begin{equation}
\sigma=\frac{n\,e^2\,\tau_c}{m(1-i\,\omega\,\tau_c)}
\label{em10}
\end{equation}

\noindent where $n$ is carrier density, $e$ and $m$ are carrier charge and mass, respectively, and $\tau_c$ is related to the relaxation time of charge carriers.

Using (\ref{em10}) in (\ref{em3}) gives

\begin{equation}
\omega^2+\frac{i\,\omega\,\tau_c}{1-i\,\omega\,\tau_c}\,\omega_p^2-c^2k^2=0
\label{em11}
\end{equation}
\noindent where $\omega_p$ is plasma frequency, $\omega_p^2=\frac{ne^2}{m\epsilon}$.

In the $\omega\tau_c\gg 1$ limit, Eq. (\ref{em11}) gives

\begin{equation}
\omega^2=\omega_p^2+c^2k^2
\label{em12}
\end{equation}

(\ref{em12}) is a dispersion relation with the gap in $\omega$-space and represents a well-known result that EM waves in a conductor propagate above the plasma frequency.

For $\omega\tau_c\ll 1$, Eq. (\ref{em11}) gives $\omega^2+i\omega_p^2\omega\tau_c-c^2k^2=0$, the equation that yields the $k$-gap. Comparing it with (\ref{em31}) gives the $k$-gap as

\begin{equation}
k_g=\frac{\omega_p^2\,\tau_c}{2c}
\label{em13}
\end{equation}

The gapless line $\omega=ck$ results from Eq. (\ref{em11}) in the absence of charge careers and $\omega_p=0$.

\begin{figure}
{\scalebox{0.65}{\includegraphics{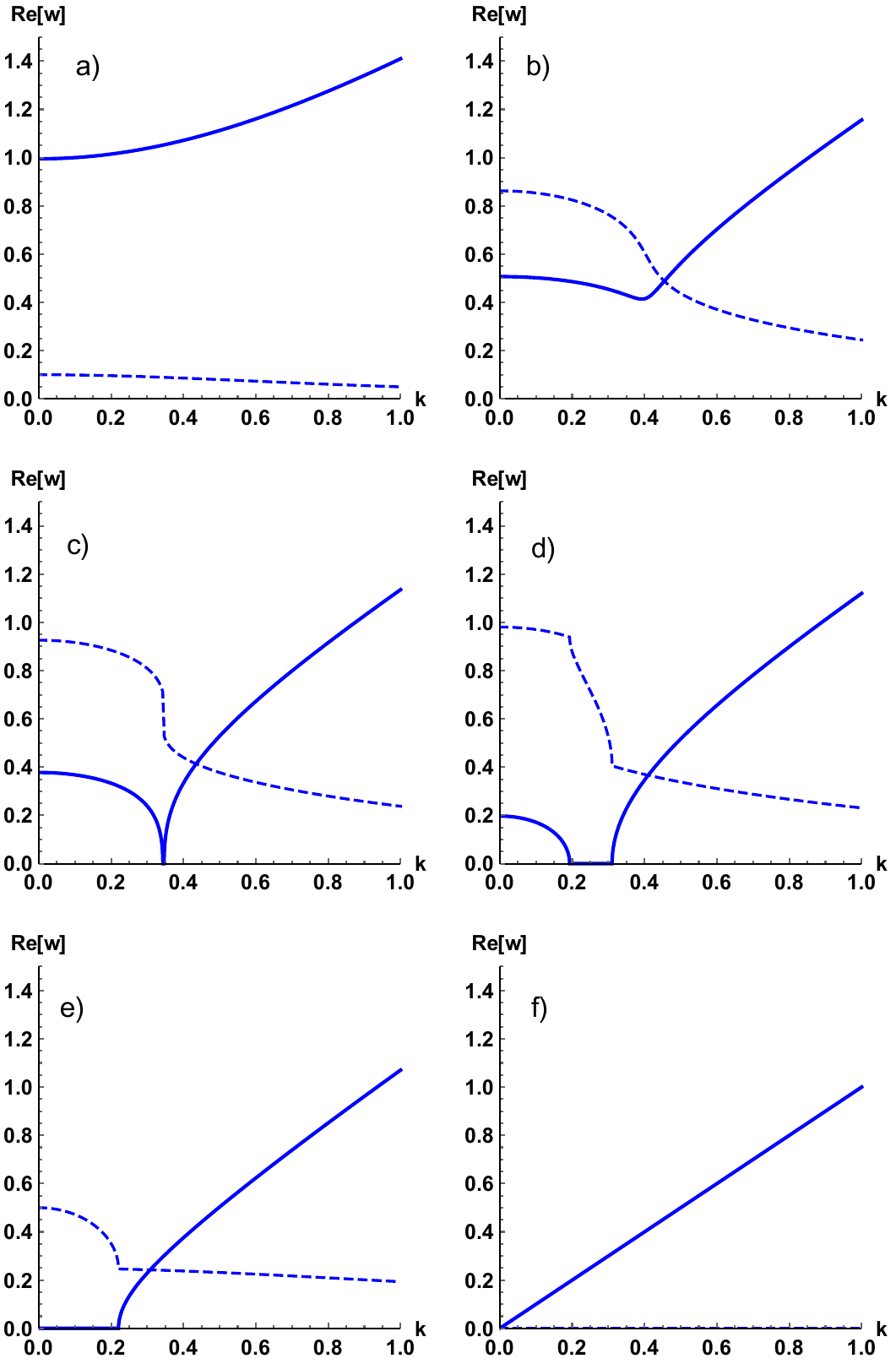}}}
\caption{Dispersion relations resulting from solving (\ref{em11}) for different $\tau_c$: (a) $\tau_c=5$, (b) $\tau_c=0.58$, (c) $\tau_c=0.54$, (d) $\tau_c=0.51$, (e) $\tau_c=0.4$ and (f) $\tau_c=10^{-4}$. The line shows a non-negative real solution for $\omega$, and the dashed line shows the imaginary part. $\omega_p=1$ and $c=1$ are assumed.
}
\label{em-crossover}
\end{figure}

The limiting cases above illustrate the crossovers between $\omega$-gapped and $k$-gapped dispersion relations. The full picture of this crossover requires solving the cubic equation (\ref{em11}) for different values of $\omega\tau_c$. To illustrate the crossover, we assume $\omega_p=1$ and $c=1$, vary $\tau_c$ and show a non-negative real solution of $\omega$ for different $\tau_c$ in Fig. \ref{em-crossover}.

Fig. \ref{em-crossover} shows an interesting and non-trivial behavior. For large $\tau_c$, we observe the dispersion relation with $\omega$ gap in Fig. \ref{em-crossover}a, in agreement with (\ref{em12}). At smaller $\tau_c$, the dispersion relation develops a checkmark-type feature in Fig. \ref{em-crossover}b, which touches the $x$-axis on further decrease of $\tau_c$ in Fig. \ref{em-crossover}c and subsequently develops a gap at intermediate values of $k$ in Fig. \ref{em-crossover}d. This is followed by the disappearance of the low-$k$ part with negative dispersion and the emergence of the $k$-gap in Fig. \ref{em-crossover}e at $\tau_c=0.4$. The observed value of the $k$-gap in Fig. \ref{em-crossover}e is close to 0.2 predicted by Eq. (\ref{em13}) for $\tau_c=0.4$ (recall $\omega_p=1$ and $c=1$).

The $k$-gap further decreases with $\tau_c$ and finally closes at small $\tau_c$ in Fig. \ref{em-crossover}f, resulting in a gapless line in agreement with Eq. (\ref{em13}). We note that $\tau_c\rightarrow 0$ gives the gapless line as expected from Eq. (\ref{em11}) where small $\tau_c$ is equivalent to the second term becoming small, corresponding to the absence of dissipation.

The imaginary part of the solution is plotted in Fig. \ref{em-crossover} as the dashed line. We observe that the real part of $\omega$ exceeds the imaginary part in the entire rage of $k$ in Fig. \ref{em-crossover}a and Fig. \ref{em-crossover}f. In Fig. \ref{em-crossover}b-\ref{em-crossover}e, this is the case for large $k$ only. In particular, modes with $k$-points close to the $k$-gap are non-propagating.

We note that the limiting cases of large and small $\tau_c$ ($\omega\tau_c\gg 1$ and $\omega\tau_c\ll 1$) give the dispersion curves with the frequency gap (Fig. \ref{em-crossover}a) and $k$-gap (Fig. \ref{em-crossover}e). The non-trivial behavior in Fig. \ref{em-crossover}b-d takes place in a fairly narrow range of $\tau_c$, corresponding to the intermediate regime. It would be interesting to investigate to what extent this behavior may be characteristic in real systems with the right combination of system properties and external parameters.

Interestingly, the same behavior of dispersion curves is found for plasmon modes studied in holographic models \cite{Gran:2018vdn}. Its currently unclear how the two theories are related and to what extent the underlying equations are similar. This is the subject of ongoing work and serves as one of the points for this review: discussing similar results from different fields, with the view of deeper understanding the underlying physics.

\section{Sine-Gordon model}

As discussed in previous sections, GMS emerge as a result of dissipation of transverse waves due to a relaxation process with a characteristic time $\tau$ and the appearance of a finite range of wave propagation. This process can be attributed to an anharmonic or nonlinear potential: recall that the $k$-gap is zero in a linear problem with the harmonic potential where a plane wave is an eigenstate and where $\tau\rightarrow\infty$ in Eq. (\ref{gener3}. For particles (fields) to have the liquid-like ability to move between different minima due to thermal activation in addition to solid-like oscillations in single minima, the anharmonicity (non-linearity) has the form shown in Figure \ref{potential}. For other forms of non-linearity and its effect on wave propagation, see, for example, Refs. \cite{kosevich,rbook1,rbook2}.

As discussed in the introduction, obtaining this picture from first principles involves the complexity of solving the problem of coupled non-linear oscillators. However, it is interesting to what extent the problem can be reduced to a single-particle model in an effective potential of the form in Figure \ref{potential}. A simple model of a potential shown in Figure \ref{potential} is a periodic function such as $\sin$. This results in the model described by the non-linear Sine-Gordon equation (SGE):

\begin{equation}
\frac{\partial^2\phi}{\partial t^2}-\frac{\partial^2\phi}{\partial x^2}+\sin\phi=0
\label{sg}
\end{equation}

The SGE has been used to discuss a variety of systems and effects, including dislocations in crystals, Josephson junctions, waves in ferromagnetic materials \cite{whitham,kosevich} as well as the Beresinskii-Kosterlitz-Thouless transition.

\begin{figure}
\begin{center}
{\scalebox{0.58}{\includegraphics{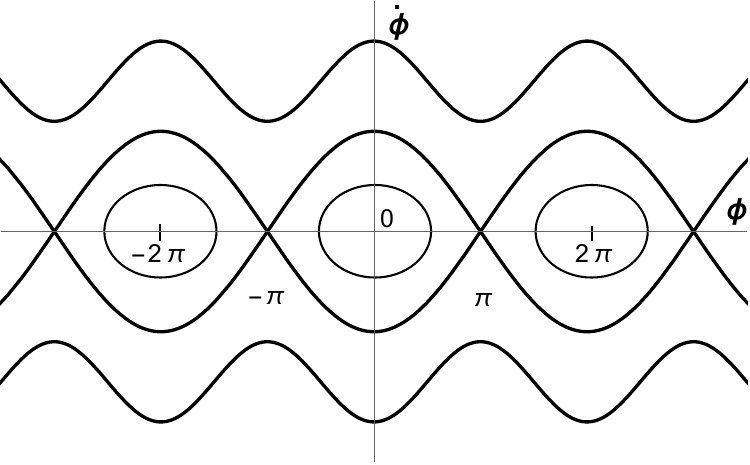}}}

 \vspace{1cm}

{\scalebox{0.5}{\includegraphics{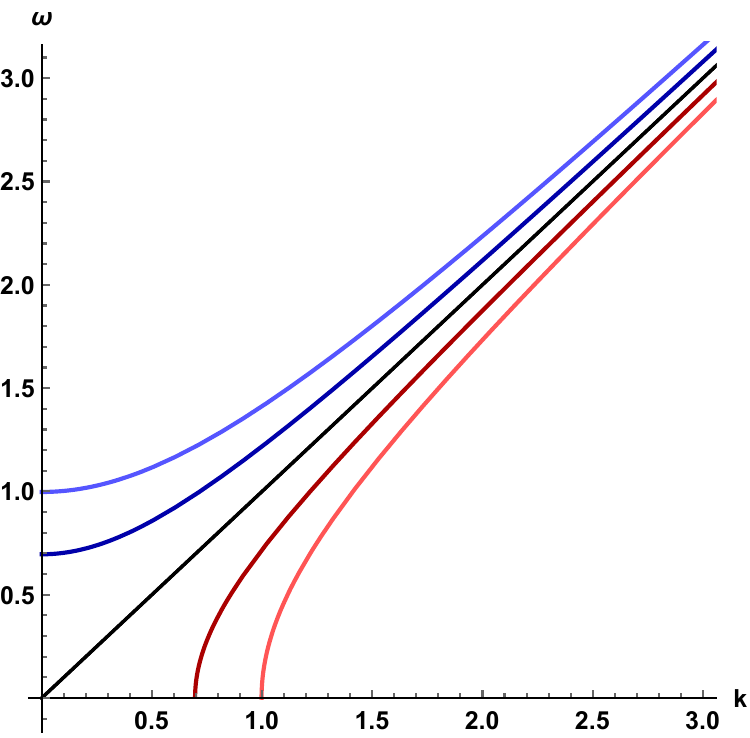}}}
\end{center}
\caption{Phase map ($\phi,\dot{\phi}$) of the Sine-Gordon Equation showing closed and open trajectories separated by a soliton solution (top).
Dispersion relations for nonlinear stationary waves of the Sine-Gordon equation (bottom). Fast and slow waves have the gap in energy and momentum, respectively. Schematic illustration adapted from
Ref. \cite{kosevich}.}
\label{sgfig}
\end{figure}

We observe that the SGE also describes solid-like oscillatory particle dynamics as well as gas-like motion. Indeed, the SGE yields two dynamical regimes: oscillations in a single potential well at
low energy of the system and the motion above the potential barrier at high energy. On the ($\phi,\dot{\phi}$) phase map, the first regime corresponds to closed trajectories where $\dot{\phi}$ and
$\phi$ are bound. The second regime corresponds to periodic $\dot{\phi}$ as a function of $\phi$, where $\phi$ is unbound. This is illustrated in Figure \ref{sgfig}. The two regimes are separated by
a separatrix which is a {\it soliton} \cite{kosevich}.

We therefore find that depending on the energy given by initial conditions, the SGE predicts solid-like and gas-like motion, but not the liquid-like motion where the oscillatory component in one
potential minimum is followed by jumps between different minima. This is not surprising because this motion requires thermal fluctuations enabling activated jumps over the potential barrier, and
these are absent in the SGE model involving no thermal bath. In this sense, the SGE describes the solid-like and gas-like dynamics and the solid-gas sublimation transition between the two states
given by the soliton solution. As far as we know, the SGE was not previously discussed in the context of these processes.

Notably, the non-linearity of the SGE can result in gapped momentum states. Considering the waves of stationary profile where variables $x$ and $t$ depend on $x-vt$, where $v$ is the
propagation speed of the nonlinear waves, Eq. (\ref{sg}) can be re-written as

\begin{equation}
(v^2-1)\frac{\partial^2\phi}{\partial t^2}-\frac{\partial^2\phi}{\partial x^2}+\sin\phi=0
\label{sg1}
\end{equation}

General solutions of (\ref{sg1}) are given in terms of Jacobi elliptic functions and are different depending on whether $v>1$ or $v<1$, corresponding to fast and slow waves. Considering the solutions
corresponding to closed trajectories in Figure \ref{sgfig}, it is found that dispersion relations for fast nonlinear waves are

\begin{equation}
\omega^2=k^2+\left(\frac{\pi}{2K(\chi)}\right)^2
\label{sg2}
\end{equation}

\noindent where $K(\chi)$ is the complete elliptic integral of the first kind and $\chi$ is the nonlinearity parameter that depends on the system energy and increases from 0 to 1 as nonlinearity
increases \cite{kosevich}. For slow waves, the dispersion relation is

\begin{equation}
\omega^2=k^2-\left(\frac{\pi}{2K(\chi)}\right)^2
\label{sg3}
\end{equation}

\noindent and implies the gap in momentum space.

These two dispersion relations are shown in Figure \ref{sgfig}, illustrating the energy gap for fast waves and momentum gap for slow waves. This graph is very similar to Figure 1 discussed in the Introduction.

In our earlier discussion of GMS in liquids, we have seen that increasing dissipation promotes the gap in $k$-space. If the mass term is present, increasing dissipation first reduces the mass gap, eventually zeroes it and subsequently opens up the gap in $k$-space. Interestingly, nonlinearity acts differently
for fast and slow waves in the Sine-Gordon model: increasing nonlinearity reduces the energy gap for fast waves and $k$-gap for slow waves. More specifically, the dispersion for fast waves becomes the same as for the Klein-Gordon equation obtainable from (\ref{sg}) by $\sin\phi\rightarrow\phi$, with the maximal frequency gap.
Increasing nonlinearity reduces the frequency gap and yields $\omega=k$ in the strongly nonlinear case. For slow waves, weak non-linearity gives large $k$-gap in Figure \ref{sgfig}. Increasing nonlinearity reduces the $k$-gap and yields $\omega=k$ in the strongly nonlinear case \cite{kosevich}.

We recall that our previous discussion of liquids identified dissipation and relaxation processes as essential ingredients of GMS. On the other hand, GMS and their evolution in the Sine-Gorodn model emerge solely from the nonlinearity of the SGE equation rather than from an explicit presence of dissipation and relaxation.

\section{Generalized Hydrodynamics, Quasihydrodynamics and applications}\label{due}

Hydrodynamic description is an effective field theory framework to describe the flow of fluids and gases. It describes low-energy and low-frequency degrees of freedom of a system using a gradient expansion. At the same time, we can think about hydrodynamics as a theoretical framework describing a set of conserved currents and relative charges. Its validity relies on small frequency/momentum expansion $\omega/T,k/T \ll 1$ where $T$ is the characteristic thermal scale of the system. In section \ref{uno}, we  have shown how to start with the hydrodynamic Navier-Stokes equation and modify it to in order to obtain the $k$-gap. In this section, we will present other areas of hydrodynamic framework where the $k$-gap arises naturally. In particular, we will discuss the relation between hydrodynamics and the $k$-gap dispersion relation in terms of the recently proposed quasihydrodynamic theory \cite{Grozdanov:2018fic}.
\begin{figure}[hbtp]
\centering
\includegraphics[width=7cm]{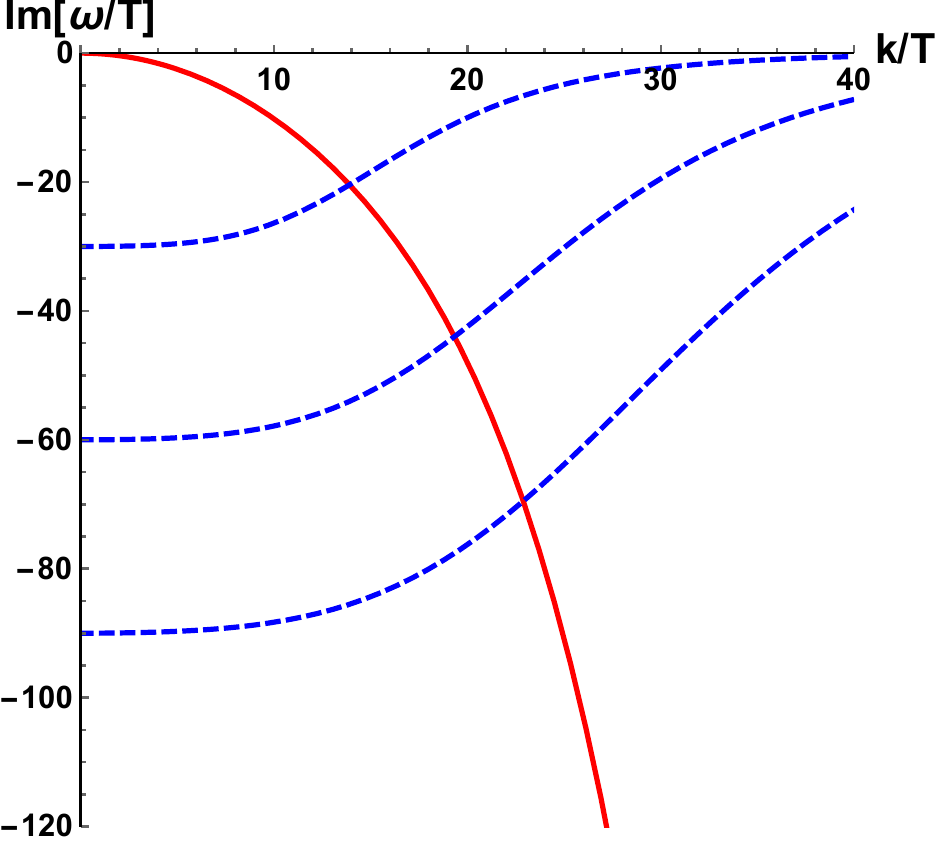}

 \vspace{1cm}

\includegraphics[width=7cm]{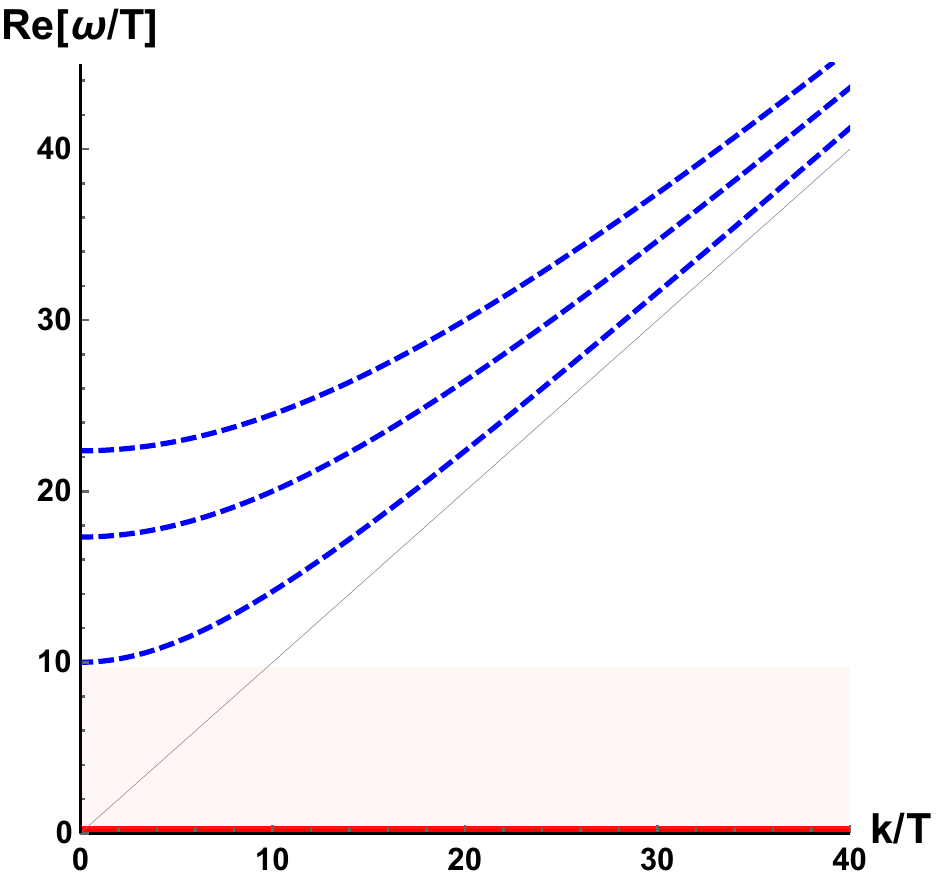}
\caption{The spectrum of excitations in dissipative relativistic hydrodynamics showing the hydrodynamic shear diffusive mode $\omega=-i D k^2$ (red) and additional non-hydrodynamic modes (blue). The crossing around $k/T \sim 15$ determines the breakdown of hydrodynamics. The shaded region emphasizes the absence of low-frequency propagating modes which is the main difference with the $k$-gap dispersion relation.}
\label{figrel}
\end{figure}

Before proceeding, we briefly review what hydrodynamic mode we would naturally expect to find in the transverse spectrum, and for convenience we do this within relativistic hydrodynamics \cite{Kovtun:2012rj}. In the absence of charge or additional conserved current, relativistic hydrodynamics arises from the conservation of the stress tensor $\partial_\mu T^{\mu\nu}=0$, and its transverse sector is codified in the $T_{ij}$ component, where $T_{ij}$ is the helicity-2 part of the stress tensor (in two spatial dimension this would simply be the $T_{xy}$ component). The only hydrodynamic transverse mode, at low frequency, is the shear diffusion mode:
\begin{equation}
\omega\,=\,-\,i\,D\,k^2\,+\,\dots
\end{equation}
which is the same mode as that coming from Fick's law or Brownian motion. A diffusive mode is natural to expect in the presence of a conserved current. In this specific case, the diffusion constant $D=\eta/(\epsilon+p)$ is fixed by the hydrodynamic coefficient known as shear viscosity $\eta$ which can be obtained via Kubo's formula from the Green function of the $T_{ij}$ operator. It follows that in the hydrodynamic limit $\omega/T,k/T \ll 1$, no $k$-gap emerges.

Beyond the hydrodynamic limit, different situations can appear. In the transverse sector, the most common behaviour is encountered in dissipative relativistic hydrodynamics. In this scenario (see for example \cite{Festuccia:2008zx,Fuini:2016qsc}), the hydrodynamic diffusive mode $\omega=-i D k^2$ crosses the second non-hydrodynamic mode at a specific momentum $\bar{k}/T \gg 1$ (see Fig.\ref{figrel}). That point is usually referred to as the breakdown of hydrodynamics \cite{Grozdanov:2019kge}. For larger momenta, the second non-hydrodynamic mode becomes the least damped one and its imaginary part approaches zero at infinite momentum. At large momenta, we have a propagating shear wave. Importantly, we note that no propagating mode appears at low frequency. This last point represents the main and fundamental difference with the $k$-gap dispersion relation. Heuristically, it is due to the fact that the secondary non-hydro mode is not a pure imaginary pole (as in the $k$-gap case) but has a finite real part. The latter provides a natural frequency cutoff for the appearance of a propagating shear mode as shown in Fig.\ref{figrel}.

\subsection{Generalized Hydrodynamics}\label{hydrodyn}

In a broad sense, generalized hydrodynamics seeks to start with hydrodynamic equations for liquid properties and subsequently add non-hydrodynamic effects including those at large $k$ and $\omega$. In section \ref{uno}, we have shown how GMS can be obtained in the Maxwell-Frenkel approach which starts from Navier-Stokes equations and generalizes it to include solid-like elastic response. This is probably the earliest example of how generalized hydrodynamics works in a broad sense. Lets recall that GMS can also be obtained starting from non-hydrodynamic solid-like elastic equations and generalizing them by adding hydrodynamic flow effects, implying a symmetry of liquid description with regard to the starting point of the theory (see section \ref{sss}).

As a specific term, ``generalized hydrodynamics'' refers to a number of proposals seeking to extend the hydrodynamic equations into the domain of large $k$ and $\omega$, where the emphasis is often on current correlation functions. This is achieved using a number of different phenomenological approaches \cite{boon,hansen,march1,baluca}. Generalized hydrodynamics is a fairly large field which we discuss briefly here, with the aim to offer readers a feel for methods used.

The hydrodynamic description starts with viewing the liquid as a continuous homogeneous medium and constraining it with continuity equation and conservation laws such as energy and momentum conservation. Accounting for thermal conductivity and viscous dissipation using the Navier-Stokes equation, the system of equations can be linearized and solved. This gives several dissipative modes, from which the evaluation of the density-density correlation function gives the structure factor $S(k,\omega)$ in the Landau-Placzek form involving several Lorentzians \cite{boon}:

\begin{equation}
\begin{aligned}
&S(k,\omega)\propto\frac{\gamma-1}{\gamma}\frac{2\chi k^2}{\omega^2+\left(\chi k^2\right)^2}+\\
&\frac{1}{\gamma}\left(\frac{\Gamma k^2}{(\omega+ck)^2+\left(\Gamma k^2\right)^2}+\frac{\Gamma k^2}{(\omega-ck)^2+\left(\Gamma k^2\right)^2}\right)
\end{aligned}
\label{boon1}
\end{equation}

\noindent where $\chi$ is thermal diffusivity, $\gamma=\frac{C_p}{C_v}$ and dissipation $\Gamma$ depends on $\chi$, $\gamma$, viscosity and density.

The first term corresponds to the central Rayleigh peak and thermal diffusivity mode. The second two terms correspond to the Brillouin-Mandelstam peaks, and describe acoustic modes with the adiabatic speed of sound $c$. The ratio between the intensity of the Rayleigh peak, $I_{\rm R}$, and the Brillouin-Mandelstam peak, $I_{\rm BM}$, is the Landau-Placzek ratio: $\frac{I_{\rm R}}{I_{\rm BM}}=\gamma-1$. Applied originally to light scattering experiments, Eq. (\ref{boon1}) is also viewed as a convenient fit to high-energy experiments probing non-hydrodynamic processes where the fit that may include several Lorentzians or their modifications.

Generalizing hydrodynamic equations and extending them to large $k$ and $\omega$ is often done in terms of correlation functions. Solving the hydrodynamic Navier-Stokes equation for the transverse current correlation function $J_t(k,t)$, $\frac{\partial}{\partial t} J_t(k,t)=-\nu k^2J_t(k,t)$, where $\nu$ is kinematic viscosity, gives for the Fourier transform $J_t(k,\omega)$ a Lorentzian form similar to (\ref{boon1}):

\begin{equation}
J_t(k,\omega)=2v_0^2\frac{\nu k^2}{\omega^2+\left(\nu k^2\right)^2}
\label{boonh}
\end{equation}

\noindent where $\nu$ is kinematic viscosity and $v_0^2=J_t(k,t=0)$.

The generalization of the hydrodynamic transverse current correlation function (\ref{boonh}) is done in terms of the memory function $K_t(k,t)$ defined in the equation for $J_t(k,\omega)$ as

\begin{equation}
\frac{\partial}{\partial t}J_t(k,\omega)=-k^2\int\limits_0^tK_t(k,t-t^\prime)J_t(k,t^\prime)dt^\prime
\label{mem}
\end{equation}

\noindent where $K_t(k,t-t^\prime)$ is the shear viscosity function or the memory function for $J_t(k,\omega)$ which describes its time dependence (``memory'').

Introducing $\tilde{J_t}(k,s)$ as the Laplace transform $J_t(k,\omega)=2\mathrm{Re}[\tilde{J_t}(k,s)]_{s=i\omega}$ and taking the Laplace transform of (\ref{mem}) gives

\begin{equation}
\tilde{J_t}(k,s)=v_0^2\frac{1}{s+k^2\tilde{K}_t(k,s)}
\label{tilde}
\end{equation}

The generalization introduces the dependence $k$ and $\omega$ by writing $\tilde{K}_t(k,s)$ as the sum of real and imaginary parts $[\tilde{K}_t(k,s)]_{s=i\omega}=K_t^\prime(k,\omega)+iK_t^{\prime\prime}(k,\omega)$. Then,

\begin{equation}
J_t(k,\omega)=2v_0^2\frac{k^2K_t^\prime(k,\omega)}{\left(\omega+k^2K_t^{\prime\prime}(k,\omega)\right)^2+\left(k^2K_t^{\prime}(k,\omega)\right)^2}
\label{genera}
\end{equation}
\noindent giving the generalized hydrodynamic description of the transverse current correlation function with a resonance spectrum.

Further analysis depends on the form of $K_t(k,t)$, which is often postulated to have an exponential time decay with $k$-dependent $\tau$ as decay time:

\begin{equation}
K_t(k,t)=K_t(k,0)\exp\left(-\frac{t}{\tau(k)}\right)
\label{assu}
\end{equation}

In generalized hydrodynamics, Eq. (\ref{assu}) is used not only for $K$ but also for several types of correlation and memory functions. These often include modifications such as including more exponentials with different decay times in order to improve the fit to experimental or simulation data.

Taking the Laplace transform of (\ref{assu}):

\begin{equation}
\begin{aligned}
&K_t^\prime(k,\omega)=K_t(k,0)\frac{\tau(k)}{1+\omega^2\tau^2(k)}\\
&K_t^{\prime\prime}(k,\omega)=-K_t(k,0)\frac{\omega\tau^2(k)}{1+\omega^2\tau^2(k)}
\end{aligned}
\label{boon2}
\end{equation}

\noindent and using $K_t^\prime(k,\omega)$ and $K_t^{\prime\prime}(k,\omega)$ in (\ref{genera}) gives \cite{boon}:

\begin{equation}
J_t(k,\omega)\propto\frac{1}{\left(\omega^2-\left(k^2K_t(k,0)-\frac{1}{2\tau^2(k)}\right)\right)^2+\left(k^2K_t(k,0)-\frac{1}{4\tau^2(k)}\right)\frac{1}{\tau^2(k)}}
\label{boon3}
\end{equation}

The condition for the resonant frequency in (\ref{boon3}) to be real is $k>\frac{1}{\sqrt{2K_t(k,0)}\tau(k)}$ and results in the $k$-gap in the liquid transverse spectrum.

\subsection{Quasihydrodynamics}\label{quasihydro}
Hydrodynamics is governed by the equations of motion that reflect the conservation of a finite set of currents $\mathcal{J}_a$ associated with a certain collection of global symmetries and conserved charges $\rho_a$. A more interesting and quite frequent situation appears when the system possesses at least one operator $\mathcal{O}$ (typically the momentum operator) which is not strictly conserved but is dissipated at a slow rate fixed by a relaxation time $\tau$ such that $\langle \mathcal{O}(t) \mathcal{O}(0)\rangle \sim e^{-t/\tau}$. Assuming that the dissipation rate is small compared to the typical time scale of the system, \textit{i.e.} $\tau T \gg 1$, we can still describe the system using the following equations:
\begin{align}
& \partial_t \langle \rho_a\rangle \,+\,\partial_i\,\mathcal{J}_a^i\,=\,0\,,\quad\partial_t \langle \mathcal{O}\rangle \,+\,\partial_i\,\mathcal{J}_\mathcal{O}^i\,=\,-\,\frac{\langle \mathcal{O}\rangle}{\tau}
\end{align}
which are thought as a deformation of the hydrodynamic framework in the presence of a non-conserved current.

\begin{figure}
\centering
\includegraphics[width=7cm]{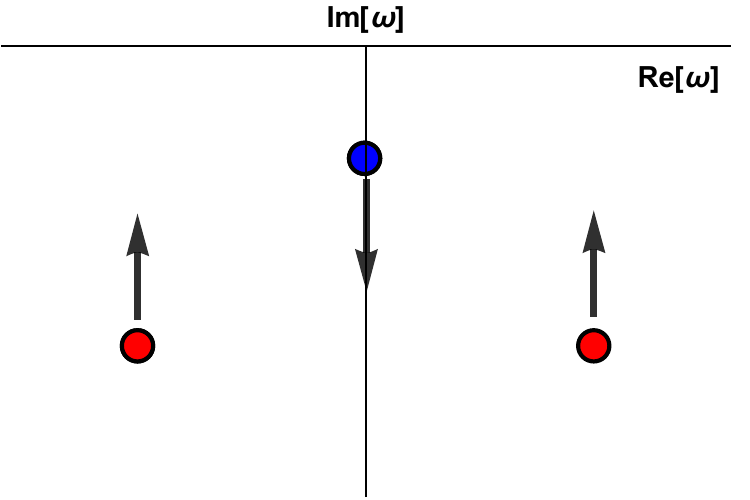}

 \vspace{1cm}

\includegraphics[width=7cm]{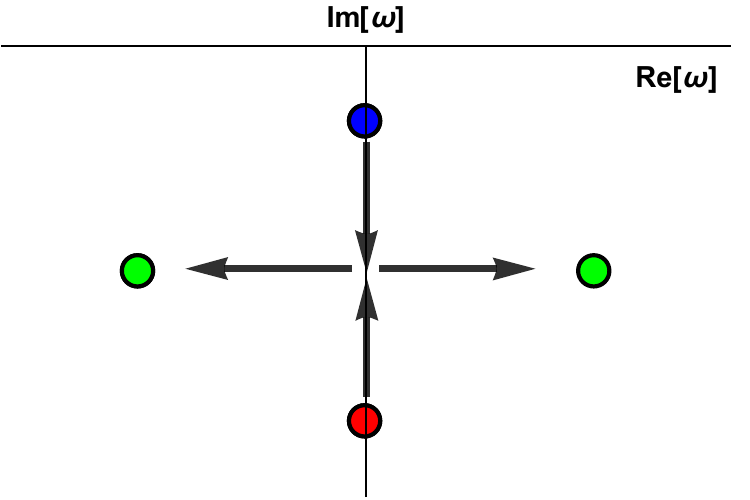}
\caption{Two typical behaviours on the complex frequency plane. The top figure shows pole-crossing: the hydro pole (blue) and the first non-hydro poles (red) cross each other. The bottom figure shows pole-collision: the hydro pole (blue) and the first non-hydro pole (red) collide on the imaginary axes and produce two specular poles with finite $Re[\omega]$.}
\label{hydropoles}
\end{figure}

The problem of formulating hydrodynamics in the presence of broken symmetries is the subject of Ref. \cite{forster1975hydrodynamic}. More recently, using the assumptions above, the authors of Ref. \cite{Grozdanov:2018fic} find that the appearance of the $k$-gap is universal and is due to a pole-collision between the hydro diffusive pole $\omega=-i D k^2 +\dots$ and purely imaginary non-hydro pole $\omega=-i/\tau +\dots$ which corresponds to the non-conserved operator $\mathcal{O}$. However, this is not the only option that a hydrodynamic system can support. Let us describe two typical situations which can take place on increasing the momentum of the modes in the transverse sector (see Fig. \ref{hydropoles}). One possibility is that the diffusive mode and a pair of non-hydro off-axis modes cross each other, in which case a diffusion-to-sound crossover takes place but no $k$-gap emerges. A different effect appears when the diffusive mode collides with a purely imaginary pole and produces a pair of propagating modes with a finite real part. In this case, the $k$-gap emerges. We also note that the diffusion-to-sound crossover in the transverse stress tensor correlators can be also derived using the relaxation time approximation for the Boltzmann equation in kinetic theory (see Ref. \cite{Romatschke:2009im} for a review).
The main idea of Ref. \cite{Grozdanov:2018fic} is that the presence of the $k$-gap and the corresponding emergent propagating mode is intimately connected with the existence of an \textit{approximately conserved} global symmetry. In simple cases the identification of such a symmetry appears to be straightforward, but generally it is far from obvious and is often related to a specific treatment of hidden global symmetries of the system.

To be more specific about the idea of \textit{quasihydrodynamics}, let us take a simple example presented in Ref. \cite{forster1975hydrodynamic} and consider an operator $\mathfrak{O}$ whose dynamics is well described by a hydrodynamic diffusion equation:
\begin{equation}
\partial_t\,\mathfrak{O}\,=\,D\,\partial_x^2\,\mathfrak{O}
\end{equation}
whose conserved current is $J=-D \partial_x \mathfrak{O}$. Its dispersion relation is a simple diffusive one: $\omega=-iD k^2$. Now let us assume the presence of an additional quasihydrodinamic field $\mathfrak{O}_2$, with corresponding relaxation time $\tau$, such that the system of equations is modified into:
\begin{equation}
\partial_t \mathfrak{O}\,+\,\partial_x \mathfrak{O}_2\,=\,0,\quad \partial_t \mathfrak{O}_2\,+\,
D\partial_x \mathfrak{O}\,=\,-\,\frac{1}{\tau}\,\mathfrak{O}_2
\end{equation}

Considering $\tau=\infty$, we recover a purely diffusive mode. Writing the solution in the Fourier space, we find our usual equation for the $k$-gap (see Eq. \ref{quadratic}):

\begin{equation}
\omega^2\,+\,\frac{i}{\tau}\omega\,-\,D\,k^2\,=\,0\quad
\end{equation}

and its solution:

\begin{equation}
\omega\,=\,-\,\frac{i}{2\,\tau}\,\pm\,\sqrt{\frac{D}{\tau}k^2\,-\,\frac{1}{4 \tau^2}}
\label{kk}
\end{equation}

As a result, the $k$-gap appears as is shown Fig. \ref{disp}. More specifically because of the non conservation of one of the operators, one mode will acquire a zero momentum damping $\omega=-i \tau^{-1}$. Increasing the momentum, this mode will collide with the other diffusive mode, corresponding to a conserved quantity, and will produce the $k$-gap (see Fig. \ref{hydropoles}). At low momenta, the system is governed by the usual diffusive mode. At higher momenta, namely at $k_g^2=1/(4 D \tau)$, a collision of the type of Fig.\ref{hydropoles} takes place and a propagating mode appears in the transverse spectrum. At yet larger momenta, we have a propagating transverse mode with the asymptotic speed $c^2 \equiv \frac{D}{\tau}$.\\

\begin{figure}
\centering
\includegraphics[width=8cm]{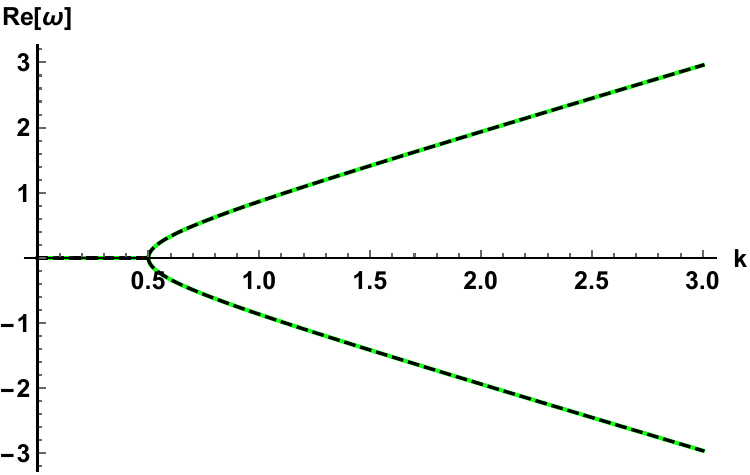}

 \vspace{1cm}

\includegraphics[width=8cm]{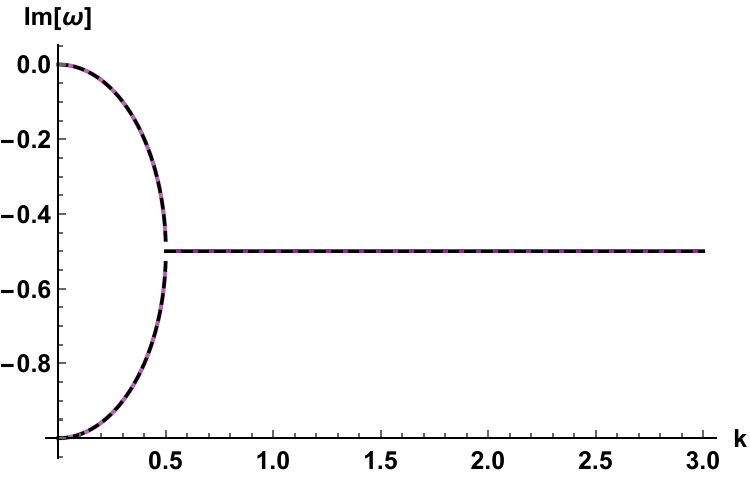}
\caption{A typical dispersion relation $\omega(k)$ of the type \eqref{kk}. Here we take $D=\tau=1$.}
\label{disp}
\end{figure}

Interestingly, this effect can have potential experimental implications and has been discussed to possibly appear in two-dimensional electronic systems such as graphene or Weyl semimetals \cite{2018PhRvB..97k5449L}.

\subsection{Global formulation of EM and dual photon}
As an application of the quasihydrodynamic method, we discuss the global symmetry formulation of electromagnetism coupled to matter introduced in Ref. \cite{Grozdanov:2016tdf} and further discussed in a series of papers \cite{Grozdanov:2017kyl,Hofman:2017vwr,Armas:2018ibg,Grozdanov:2018fic,Glorioso:2018kcp,Armas:2018atq,Armas:2018zbe}.

We start with Maxwell equations in four dimensions:
\begin{equation}
\frac{1}{e^2}\,\partial_\mu\,F^{\mu\nu}\,=\,J_{electric}^\nu \label{maxw1}
\end{equation}
where $e$ is the gauge coupling, $F=dA$ is the electromagnetic field strength or Maxwell tensor and $J^\nu_{electric}$ is the current induced by a finite electric charge density in the medium. Eq. \eqref{maxw1} is related to $U(1)$ local gauge symmetry which implies the conservation of the electric charge. We can now use the field strength to define a two-form $\mathcal{J}^{\mu\nu}$:
\begin{equation}
\mathcal{J}^{\mu\nu}\,=\,\frac{1}{2}\,\epsilon^{\mu\nu\rho\sigma}\,F_{\rho\sigma}
\end{equation}

Using the original Bianchi identity for the field strength $F$, $d*F=0$, which implies the absence of magnetic monopoles, we find:
\begin{equation}
\partial_\mu\,\mathcal{J}^{\mu\nu}\,=\,0.\label{concon}
\end{equation}
or, in other words, we find that the two-form current $\mathcal{J}^{\mu\nu}$ is conserved. This is not related to the conservation of the electric charge, but rather informs us that magnetic field line cannot end because of the absence of magnetic monopoles. The symmetry behind the conservation of such a current expressed in \eqref{concon} is due to a generalized global symmetry. The associated conserved charge can be written as:
\begin{equation}
Q\,=\,\int_S\,*\,\mathcal{J}
\end{equation}
and counts the number of magnetic lines across a codimension-2 surface $S$ (see Fig. \ref{stri}).

\color{black}In the absence of dynamical magnetic monopoles, the Maxwell theory enjoys a $1-$form generalized global symmetry, with the conserved current $\mathcal{J}^{\mu\nu}$ counting the magnetic lines. The operator charged under $\mathcal{J}$ is the t'Hooft line $W(C)$. If we now consider EM with no charged matter, it is easy to show that the t'Hooft line displays a perimeter law; thus the $1-$form symmetry associated to $\mathcal{J}$ is spontaneously broken \cite{Hofman:2018lfz}. Curiously, the associated Goldstone mode is the common photon. The phase in which this $1-$form symmetry is unbroken is the Higgs phase of EM, where the t'Hooft line exhibits an area law. We refer to Ref. \cite{Hofman:2018lfz} for more details.\color{black}

We can now notice that in addition to the conserved two-form current $\mathcal{J}^{\mu\nu}$, we also have its dual current $\tilde{\mathcal{J}}\,=\,*\mathcal{J}\,=\,F$ which counts the electric flux lines and it is simply the original field strength $F^{\mu\nu}$. The important point is that the dual electric form is now not conserved in the presence of charge matter, as Eq. \eqref{maxw1} clearly shows. This is the same as stating that electric lines end with an electric charge. As a consequence, in a media with finite electric charge density, there is only one exact global symmetry and one conserved two-form. In particular, we have:
\begin{equation}
\partial_\mu \tilde{\mathcal{J}}^{\mu\nu}\,=\,-\,\frac{1}{\tau}\,\tilde{\mathcal{J}}^{t\nu}\label{nocons}
\end{equation}
which states that the conservation of the $F^{\mu\nu}$ two-form is broken by a finite relaxation time $\tau$ produced by the electric charge distribution. To obtain \eqref{nocons}, we have expressed $j^\nu_{electric}$ in terms of the electric field $F^{t\nu}$. Eq. \eqref{nocons} is analogous to Ohm's law for the case of a dynamical gauge field $A^\mu$. In the limit $\tau=\infty$, we recover the standard dispersion relation for EM waves $\omega=ck$, but otherwise we have a typical situation described by the quasihydrodynamic approach where one of the currents is not conserved. As a result, the dispersion relation acquires a $k$-gap as discussed in the previous section.

We therefore find this consideration provides an intriguing way to view the emergence of the $k$-gap as being due to the softly broken global symmetry. We also note that the usual plasmon dispersion relation
\begin{equation}
\omega^2\,=\,\omega_P^2\,+\,c^2\,k^2
\end{equation}
can be explained in this picture using the Goldstone theorem. In particular, the plasmon represents the pseudo-goldstone boson for the global symmetry discussed, and the plasma frequency $\omega_P$ corresponds to the mass gap. It would be interesting to develop this picture further.

\subsection{Theory of elasticity and lattice defects}
The same arguments can be applied to a different area: the theory of elasticity with lattice defects. The idea is to express the theory of elasticity \cite{Lubensky} in two and three dimensions with defects \cite{NELSON1981108,PhysRevA.6.2401} in a dual formalism, which is partially inspired by Refs. \cite{Beekman:2016szb,Beekman:2017brx}. The framework was presented in Ref. \cite{Grozdanov:2018ewh}, and its interpretation in terms of quasihydrodynamics was discussed in Ref. \cite{Grozdanov:2018fic}.

Let us consider a set of scalar fields $\phi^I$ which define the displacements. The equation governing the dynamics of the system is the conservation of momentum:
\begin{equation}
\partial_\mu\,P^\mu_I\,=\,0
\end{equation}
where the momentum is defined via the relation:
\begin{equation}
P^\mu_I\,=\,C_{IJ}^{\mu\nu}\,\partial_\nu\,\phi^J
\end{equation}
where $C$ is the elastic tensor.

As in the previous section where the dynamics of the photon was arising from the spontaneous symmetry breaking, the theory of elasticity can be formulated in terms of the spontaneous breaking of the momentum conservation which gives rise to a set of Goldstone bosons, i.e. phonons, determining the mechanical response of the system.

In the absence of any defects (disclinations or dislocations) there exists an additional set of conserved higher spin currents:
\begin{equation}
J_I^{n_1\dots n_d}\,=\,\epsilon^{n_1\dots n_d \nu}\partial_\nu \phi^I
\end{equation}

The conservation of these currents plays the role of the topological Bianchi identity. The associated charge
\begin{equation}
\mathcal{N}\,=\,\int_{S^1}\,*\,J
\end{equation}
counts the number of domain walls of $\phi_I$ or defects lines.

To make an analogy with the electromagnetism discussed earlier, we observe that momentum conservation corresponds to the conservation of magnetic flux lines and the conservation of the topological currents corresponds to the conservation of the electric flux lines. As in the system discussed in the previous section, this last symmetry can be broken. In particular, a finite domain wall with density $\langle J^I\rangle \neq 0$, produced by the presence of lattice defects, breaks the conservation of the topological current:
\begin{equation}
\partial_\mu\,J_I^{\mu\nu}\,=\,-\,\frac{1}{\tau}\,J_I^{t \nu}
\end{equation}
and now acquires a finite relaxation time $\tau$. The usual argument of Ref. \cite{Grozdanov:2018fic} can be used, giving rise to the $k$-gap \cite{Grozdanov:2018ewh}. More precisely, in the presence of a finite density of defects, the currents overlap with the momentum operator and induce a propagating mode in the shear sector.

This framework is intriguingly close to our discussion of liquids in section \ref{liquid-section}. It is possible that the quasihydrodynamic framework could valuably add to a theoretical basis for the $k$-gap phenomenon.

\subsection{Israel-Stewart formalism}\label{is}
A similar earlier idea is related to the definition of relativistic hydrodynamics and the Israel-Stewart formalism \cite{1979AnPhy.118..341I}. Relativistic hydrodynamics \cite{Kovtun:2012rj} involves the conservation equation for the stress tensor $\nabla_\mu T^{\mu\nu}=0$, where the stress tensor at second order reads:
\begin{equation}
T^{\mu\nu}\,=\,\epsilon\,u^\mu\,u^\nu\,+\,p\,\Delta^{\mu\nu}\,-\,\eta\,\sigma^{\mu\nu}\,+\,\dots\label{str}
\end{equation}
where $\epsilon$ is the energy density, $u^\mu$ is velocity field, $p$ is pressure, $\eta$ is viscosity, $\Delta^{\mu\nu}=u_\mu u_\nu+g_{\mu\nu}$ and $\sigma_{\mu\nu}$ is shear stress tensor. Within the second-order hydrodynamics, the relevant transverse mode is again a diffusive mode whose diffusion constant is given by viscosity $D=\eta/(\epsilon+p)$. Understanding the operation of this mode has issues for several reasons. The group velocity of the diffusive mode $|\partial \omega/\partial k|  = 2 D k$ grows arbitrarily at large momentum because of its diffusive nature, and its propagation becomes superluminal and acausal. However, the superluminality takes place in a regime which is out of control of the second order and low-momentum approximation. Another related problem is that the corresponding Green function violates specific sum rules and therefore violates unitarity \cite{forster1975hydrodynamic}. There has been a lengthy discussion about the possible acausality problem of the linearized relativistic hydrodynamics which is mainly a problem for numerical simulations.

A possible resolution is the so-called Israel-Stewart mechanism. The idea is to extend the stress tensor \eqref{str} to next order, introducing new terms in the expansion. One of the coefficients that needs to be added plays the role of ''relaxation time'' and is usually denoted as $\tau_{\Pi}$. Once this extension is performed, the equation for the diffusive mode is modified as:
\begin{equation}
\omega^2\,+\,\frac{i}{\tau_{\Pi}}\,-\,\frac{\eta}{\epsilon\,+\,p}\,k^2\,=\,0
\end{equation}
from which the $k$-gap emerges. Note that the group velocity is now subluminal:
\begin{equation}
v\,=\,\lim_{k\rightarrow \infty}\left|\frac{\partial \omega}{\partial k}\right|\,=\,\sqrt{\frac{D}{\tau_{\Pi}}}\,\leq\,1\,
\end{equation}
and no casuality-related problem appears.

To relate this discussion to the previous section, we can think of the Israel-Stewart formalism as taking $\Pi^{\mu\nu}=-\eta \sigma^{\mu\nu}$ as an independent set of degrees of freedom whose conservation is broken by the $\tau_{\Pi}$ coefficient. According to the ideas of Ref. \cite{Grozdanov:2018fic}, this implies the dispersion relation with the $k$-gap.

We observe that dissipation is an essential ingredient for the $k$-gap to emerge in these models, similarly to our discussion in earlier sections. The link between dissipation and dispersion relations has also been discussed in the system of goldstone bosons in dissipative systems in \cite{Minami:2018oxl,Hayata:2018qgt}.
\begin{figure}
\centering
\includegraphics[width=7cm]{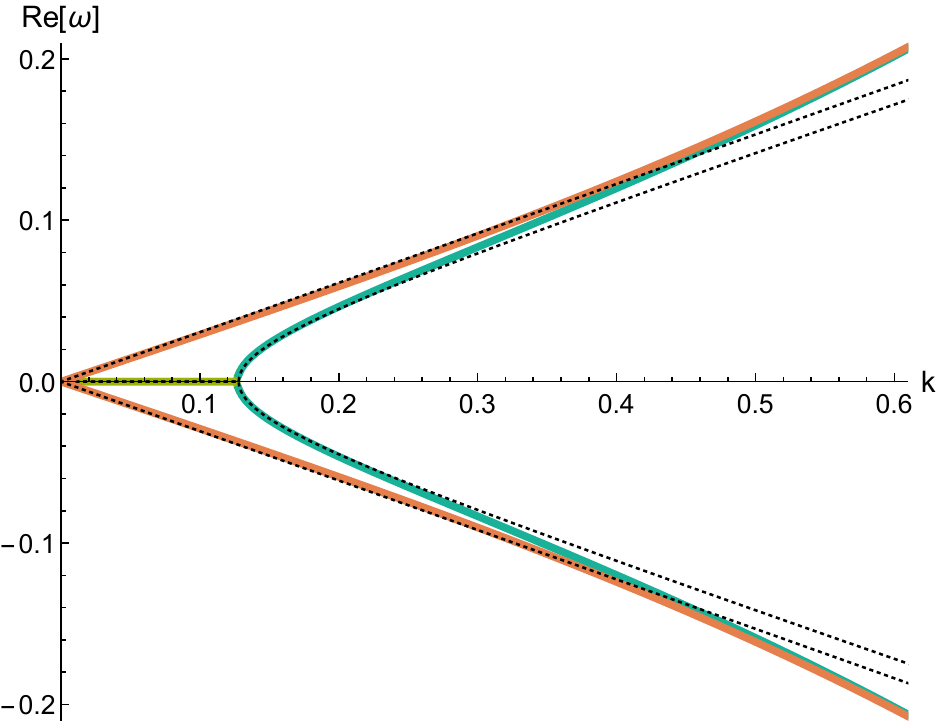}

 \vspace{1cm}

\includegraphics[width=7cm]{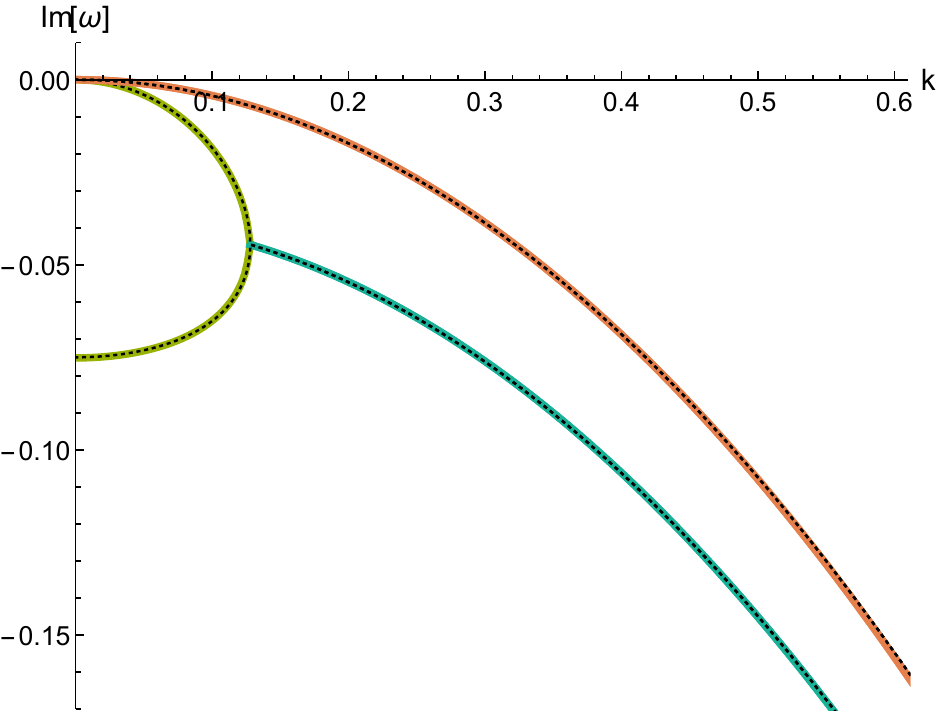}
\caption{The Chiral Magnetic Wave exhibiting a $k$-gap at a finite relaxation time for the axial charge. The plot is taken with permission from Ref. \cite{Jimenez-Alba:2014iia}.}
\label{chiralfig}
\end{figure}

\subsection{Anomalous transport}\label{anref}
Interestingly, the $k$-gap phenomenon also appears in a completely different type of hydrodynamics which is closely connected with anomalies and is relevant in the study of fermionic systems \cite{Landsteiner:2016led}. The description of specific condensed matter systems like graphene and Weyl semimetals is given in terms of massless relativistic fermions and shows peculiar phenomenological signatures in the transport properties. In these systems, a new collective excitation known as chiral magnetic waves (CMW) \cite{Kharzeev:2010gd} is expected to appear as a manifestation of the interplay between the chiral separation effect and the chiral magnetic effect. It is generated by the coupling of vector and axial density waves in the presence of an external magnetic field and is closely connected to the presence of the axial anomaly. The dispersion relation for this mode corresponds to a damped sound wave $\omega=v\,k\,-\,i\,D\,k^2$, where the velocity is proportional to the external magnetic field and the anomaly coefficient. Up to this point, vector and axial charges are taken to be conserved, but more generally, axial charge conservation can be broken by several effects. For example in Weyl semimetals dissipation is caused by the so-called "inter-valley" scattering between the electrons and impurities, while in Quark Gluon Plasma this takes place due to strong interactions. Here, we adopt a purely hydrodynamic point of view \cite{Stephanov:2014dma,Jimenez-Alba:2014iia} and write down the constitutive relations for the axial and vector currents:
\begin{equation}
j_A\,=\,\frac{\kappa\,B\,\rho_V}{\chi_V}\,-\,D\,\partial_x \rho_A\,,\quad j_V\,=\,\frac{\kappa\,B\,\rho_A}{\chi_A}\,-\,D\,\partial_x \rho_V
\end{equation}
in the presence of a background magnetic field $B$. We define the axial and vector charge densities and susceptibilities as $\rho_{A,V},\chi_{A,V}$. The parameter $\kappa$ is the coefficient of the U(1) anomaly. We write down the equation for the currents as:
\begin{equation}
\partial_\mu j_V^\mu\,=\,0\,,\quad  \partial_\mu j_A^\mu\,=\,-\,\frac{1}{\tau_A}\,\rho_A
\end{equation}
where relaxation time $\tau_A$ quantifies the dissipation of the axial charge due to various effects discussed above. We can write down the set of coupled equations in Fourier space as:
\begin{align}
&\omega \rho_V+\frac{k\,\kappa\,\rho_A\,B}{\chi_A}\,+\,i\,D\,k^2\,\rho_V=0\\
&\left(\omega+\frac{i}{\tau_A}\right) \rho_A+\frac{k\,\kappa\,\rho_V\,B}{\chi_V}\,+\,i\,D\,k^2\,\rho_A=0
\end{align}
We can now diagonalize the equations and obtain two independent modes with the following dispersion relation:
\begin{equation}
\omega\,=\,-\,\frac{i}{2\,\tau_A}\,\pm\,\sqrt{\frac{B^2\,k^2\,\kappa^2}{\chi_A\,\chi_V}\,-\,\frac{1}{4\,\tau_A^2}}\,-\,i\,D\,k^2\label{bb}
\end{equation}

Hence the chiral magnetic wave, in the presence of axial charge dissipation, develops a $k$-gap given by:
\begin{equation}
k_g^2\,\equiv\,\frac{\chi_A\,\chi_V}{4\,B^2\,\kappa^2\,\tau_A^2}
\end{equation}
which can be understood in terms of the theoretical setup discussed in the previous sections.
An example taken from \cite{Jimenez-Alba:2014iia} is shown in Fig. \ref{chiralfig}.

The experimental detection of chiral magnetic waves and their corresponding $k$-gap appears to be possible. Important studies related to the possible manifestation and detection of the dispersion relation in \eqref{bb} in Quark Gluon Plasma have been recently performed in Ref. \cite{Shovkovy:2018tks}.

\section{Holographic models}\label{quattro}
Holography, also known as AdS-CFT correspondence, is a duality between quantum field theories (QFTs) and gravitational theories in asymptotically Anti-De Sitter spacetime (AdS) which was conjectured in 1997 by J. Maldacena \cite{Maldacena:1997re}. Its formulation has been motivated by two main ideas.

First is the suggestion, known as the Holographic principle \cite{1995JMP....36.6377S}, that in quantum gravity the description of a volume of space and its dynamical degrees of freedom are encoded in a lower-dimensional ''boundary'' spacetime. As the famous example of the latter, we can mention the Area law for the black hole entropy by Hawking and Bekenstein \cite{PhysRevD.7.2333}. Emphasized by the term ''holographic'', the field theory we are interested in ''lives'' in a spacetime with one dimension less than the gravitational dual theory, and the extra dimension can be thought of as the energy scale of the field theory itself. In other words, the holographic extra dimension geometrizes the RG flow dynamics of the field theory at hand.

Second is the observation that the perturbative structure of a large-N quantum field theory can be unambiguously connected to String Theory \cite{HOOFT1974461}. The conjecture has been tested with several explicit string theories, however a final and mathematical proof is missing. For introductory-level reviews, see, e.g., Refs. \cite{hrev1,hrev2,hrev3,hrev4,Baggioli:2019lsz,Baggioli:2019rrs}.

The strength of the AdS-CFT correspondence is that it represents a strong-weak duality and is therefore able to tackle problems linked to a strongly coupled system using a weakly coupled and simpler gravitational description. In this way, it represents a new and valuable tool to study strongly-coupled or correlated field theories where perturbative methods are of no help \cite{Policastro:2002se,CasalderreySolana:2011us}. In the last decades, several applications of this idea to condensed matter have appeared \cite{Hartnoll:2016apf}, ranging from the study of strange metals \cite{Faulkner:2010da,Sachdev:2015efa} to understanding the mystery of high-temperature superconductivity \cite{Hartnoll:2008vx,Hartnoll:2008kx,Horowitz:2010gk}.

More recently, several holographic models exhibiting a $k$-gap dispersion relation have been discussed. In this section, we will review the major examples and their common features. We skip most of the technical details of the computations dealing with the holographic dictionary and focus on physical results.

\subsection{Viscoelasticity, relaxation and holographic k-gap}\label{sechol}
The simplest holographic model where the $k$-gap appears is probably the global symmetry model of \cite{Grozdanov:2018ewh}. This is a fairly simple setup with specific viscoelastic features which defined by the 4D bulk action:
\begin{equation}
S\,=\,\frac{1}{2\,\kappa_4^2}\int\,d^4x\,\sqrt{-g}\,\left(R\,+\,\frac{6}{L^2}\,-\,\frac{1}{12}\,H^I_{abc}H_I^{abc}\right)
\end{equation}
where $ H^I=dB^I$ is the field strength of a bulk two form $B^I_{ab}$. The theory admits a simple isotropic black brane solution:
\begin{equation}
ds^2\,=\,\frac{dr^2}{r^2\,f(r)}\,+\,r^2\,\left(-f(r)\,dt^2\,+\,dx^2+dy^2\right)
\end{equation}
where $r$ is the already mentioned holographic coordinate which parametrizes the extra dimension playing the role of the RG flow energy scale. The blackening factor, written in terms of the black hole mass M, reads:
\begin{equation}
f(r)\,=\,1\,-\,\frac{2\,M^2}{r^2}\,-\,\left(1\,-\,\frac{\rho^2}{2\,r_h^2}\right)\,\frac{r_h^3}{r^3}
\end{equation}
and vanishes at $r=r_h$, defining the position of the event horizon of the black hole solution. The black hole is sourced by the simple matter configuration:
\begin{equation}
H^1_{txr}\,=\,H^2_{tyr}\,=\,\rho
\end{equation}
where parameter $\rho$ is the number density of line defects immersed in the fluid and accounts for the solid properties of the dual system.

The temperature of the background is given as usual by the surface gravity at the horizon and reads:
\begin{equation}
T\,=\,\frac{r_h}{4\,\pi}\,\left(3\,-\,\frac{\rho^2}{2\,r_h^2}\right)
\end{equation}
For simplicity, we will consider the case with $\rho=0$ only. In this limit, the shear mode completely decouples from the rest of the excitations and can be therefore ignored. In particular, we will consider only the two-points function $\langle \mathcal{J}_{\mu\nu}^I \mathcal{J}_{\rho\sigma}^I\rangle$ where $\mathcal{J}_{\mu\nu}^I$ is the two-form current dual to the $B_{\mu\nu}$ bulk field. Because of the main features of the model, the poles of such correlator correspond to the vibrational degrees of freedom in the system. The aforementioned correlator will depend on the temperature of the system and on a renormalization scale $\mathcal{M}$ introduced to render the system UV-renormalizable. In the hydrodynamic limit $\omega/T,k/T \ll 1$ and assuming the UV cutoff to be large compared to the IR scale temperature, $\mathcal{M}/T \gg 1$, the equation governing the dynamics of the vibrational degrees of freedom can be written analytically as:
\begin{equation}
\omega\left(1\,-\,\frac{\omega}{\omega_g}\right)\,+\,i\,\left(\frac{\bar{\mathcal{M}}\,-\,1}{r_h}\right)\,k^2\,=\,0\label{Jmode}
\end{equation}
where $\bar{\mathcal{M}}\equiv \mathcal{M}/r_h$ is the dimensionless renormalization scale and:
\begin{equation}
\omega_g\,=\,-\,i\,\frac{r_h}{\bar{\mathcal{M}}\,-\,1\,+\,\frac{1}{2}\,\left(\log 3\,-\,\frac{\pi}{3\,\sqrt{3}}\right)}
\end{equation}
We immediately recognize that Eq. \eqref{Jmode} is the same as that giving rise to the $k$-gap analyzed in previous sections (see, e.g., Eq. \ref{quadratic} and \ref{em3}).

The following mechanism emerges in this picture. In the hydrodynamic limit, the relevant mode governing the low energy and late time dynamics is a simple diffusive mode. Increasing the momentum, the mode undergoes a collision with the first non-hydro and purely imaginary pole (as in Fig. \ref{hydropoles}), producing the $k$-gap effect. The model at hand contains an UV cutoff $\mathcal{M}$ which plays the role of the height of the activation barrier in the interatomic potential in liquids in Fig. \ref{potential}. For large values of this cutoff, the computations are particularly simple, and the relaxation time is given by $\tau=D/v^2$ where $v$ is not necessarily the speed of light but depends on the details of the theory and is given by the following expression:
\begin{equation}
v^2\,=\,\frac{\left(\bar{\mathcal{M}}-1\right)\,\omega_g}{r_h}
\end{equation}
and it is generically smaller than the speed of light.

It is interesting to note that in the limit of infinite cutoff $\mathcal{M}/T \rightarrow \infty$, the $k$-gap closes, and we are left with an undamped propagating shear wave as in solids. This effect can be thought as setting the activation barrier in Fig. \ref{potential} to infinity so that dynamics is confined to one potential minimum as in solids where plance waves are undamped.

\color{black}Interestingly, the emergence of the $k$-gap in this model cannot be understood from the presence of a non-conserved global symmetry which is related to the $2$-form bulk field used \cite{Grozdanov:2018fic}. In other words, the $k$-gap is not explained in the quasi-hydrodynamics framework \cite{Grozdanov:2018fic} but emerges due to the appearance of non-hydrodynamic modes, similarly to the approach of generalized hydrodynamics \cite{boon}.\color{black}

Another well-known and studied holographic model which gives rise to the $k-$gap is usually called the \textit{linear axions model} introduced in Ref. \cite{Andrade:2013gsa}. The model is a simple setup to introduce momentum relaxation into the holographic framework using certain specific massive gravity action in the bulk \cite{Vegh:2013sk,Blake:2013owa,Baggioli:2014roa}.
In particular, the model is written using the 4D bulk action:
\begin{equation}
S\,=\, M_P^2\int d^4x \sqrt{-g}
\left[\frac{R}2+\frac{3}{\ell^2}- \, m^2 \partial_\mu \phi^I \partial^\mu \phi^I\right]
\end{equation}
where $M_P$ is the Planck mass and the $\phi^I$ are two massless scalar fields which enjoy shift symmetry. The parameter $m$ parametrizes the mass for the graviton fluctuations. We can identify the scalar field $\phi^I$ with the St\"uckelberg fields of the massive gravity theory \cite{Rubakov:2008nh,Dubovsky:2004sg,Hinterbichler:2011tt,Alberte:2015isw}.
We study 4D AdS black brane geometries of the form:
\begin{equation}
\label{backg}
ds^2=\frac{\ell^2}{u^2} \left[\frac{du^2}{f(u)} -f(u)\,dt^2 + dx^2+dy^2\right] ~,
\end{equation}
where $u\in [0,u_h]$ is the radial holographic direction spanning from the boundary to the horizon, defined through $f(u_h)=0$, and $\ell$ is the AdS radius.
The $\phi^I$ fields admit a radially constant profile $\phi^I=x^I$ with $I=x,y$.
This is an exact solution of the system due to the shift symmetry.

In the dual picture, these fields represent marginal scalar operators which break the translational invariance because of their explicit dependence on the spatial coordinates. Despite the scalars profile breaking translational invariance, the geometry remains homogeneous due to residual diagonal symmetry which combines spacetime translations and internal shifts. The framework just described is close in spirit to effective field theories for the spontaneous breaking of Poincar\'e symmetry in flat spacetime elaborated in \cite{Nicolis:2013lma,Delacretaz:2014jka,Nicolis:2015sra}. In more detail, the blackening factor in this theory takes the form:
\begin{equation}
f(u)\,=\,1\,-\,m^2\,u^2\,\,-\,\frac{u^3}{u_h^3}\,+\,m^2\,\frac{u^3}{u_h}
\end{equation}
and the corresponding temperature is defined by:
\begin{equation}
T\,=\,\frac{3}{4\,\pi\,u_h}\,-\,\frac{m^2\,u_h}{4\,\pi}
\end{equation}

Studying the perturbations of the gravitons, i.e. gravitational waves on top of the above background, it is easy to see that the gravitational waves are massive and have a finite propagation length. The graviton mass breaks the diffeomorphism gauge invariance in the bulk, which corresponds to the conservation of the stress tensor of the dual theory. The momentum operator is no longer conserved in this theory. \color{black} At leading order, the broken Ward identity can be written as \cite{Davison:2013jba}
\begin{equation}
\partial_a\,T^{ai}\,=\,-\,\tau_{rel}^{-1}\,T^{ti}\,+\,\dots
\end{equation}
where the finite relaxation time $\tau_{rel}$ for the momentum operator \cite{Davison:2013jba} is given by:
\begin{equation}
\tau_{rel}^{-1}\,=\,\frac{m^2}{4\,\pi\,T}\,+\,\dots
\end{equation}
with $m$ being the graviton mass.
The finite graviton mass is related to the breaking of translations and is responsible for two important features: (a) finite DC conductivity \cite{Andrade:2013gsa} and (b) the violation of the Kovtun-Son-Starinets (KSS) viscosity to entropy ratio bound \cite{Hartnoll:2016tri,Alberte:2016xja,Burikham:2016roo}. The non-conservation of momentum has a direct consequence for the non-hydrodynamic shear diffusive mode, which is modified as:
\begin{equation}
\omega\,=\,-\,i\,\Gamma\,-\,i\,D\,k^2\,+\,\dots
\end{equation}
where the first contribution is of the Drude type with $\Gamma\equiv \tau_{rel}^{-1}$. The condition:
\begin{equation}
\Gamma/T\,\ll\,1
\end{equation}
corresponds to regime of weakly-broken symmetry and can be understood in the language of quasi-hydrodynamics of \cite{Grozdanov:2018fic}.\color{black}

\begin{figure}[htbp]
\centering
\includegraphics[width=7.5cm]{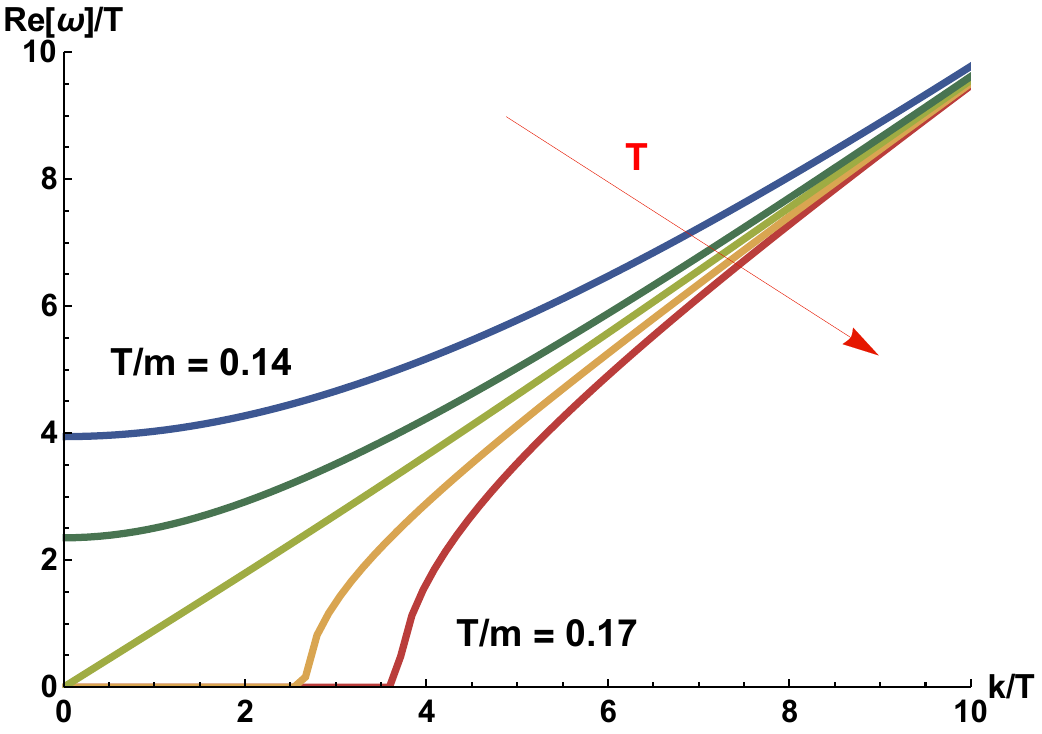}
\caption{Dispersion relation of the lowest QNM in the transverse sector, varying $T/m \in [0.14,0.17]$ for the model in Ref. \cite{Andrade:2013gsa}.}
\label{holkgap}
\end{figure}

Of particular interest is the quasinormal mode (QNM) spectrum at finite and large momentum shown in Fig. \ref{holkgap} for different values of the dimensionless parameter $T/m$. The real part of the dispersion relation shows a clear $k$-gap \cite{Baggioli:2018vfc,Baggioli:2018nnp} and is very well approximated by:

\begin{equation}
Re[\omega]=\sqrt{c^2 k^2+\mathfrak{m}^2-\frac{1}{4\tau^2}}
\label{omega-k-int}
\end{equation}

Eq. (\ref{omega-k-int}) emerges as a numerical solution as well as analytically for a specific value of $T/m$ at which the system enjoys an enhanced symmetry \cite{Davison:2014lua}. We note that (\ref{omega-k-int}) is the same as the dispersion relation (\ref{newomega}) we found when studying the Lagrangian formulation of liquid dynamics with the mass term included in section \ref{liquid-section}. Hence we expect to see the same interplay between the energy gap and $k$-gap as in liquids. Consistent with this expectation, the numerical solution shows that increasing temperature reduces the energy gap and subsequently results in the emergence of the $k$-gap as is illustrated in Fig. \ref{holkgap}.

This system can be treated analytically in the limit of $\tau T \gg 1$ and $\Gamma/T \ll 1$. More precisely, in the window corresponding to an intermediate range of the dimensionless graviton mass $m/T$, the dynamics of the system can be described at zero momentum by only two non-hydrodynamic modes:
\begin{equation}
\omega_1\,=\,-\,i\,\Gamma\,,\quad \quad \omega_2\,=\,-\,i\,\Gamma\,-\,\frac{1}{\tau}
\end{equation}
which at finite momentum collide as shown in Fig. \ref{hydropoles} and produce the $k$-gap shown in Fig. \ref{holkgap}.
Moreover, in this relativistic system, the relaxation time $\tau$ is related to the diffusion constant, and in the limit $\Gamma/T \ll 1, \tau T \gg 1$, reads:
\begin{equation}
\tau\,=\,\frac{D}{c^2}
\end{equation}
where $c$ is the speed of light and $D$ the diffusion constant of the underdamped shear mode $\omega=-i \Gamma-i D k^2+\dots$ On the contrary, for $m/T \gg 1$ and far from the quasyhydro limit, the formula above receives important corrections from the momentum dissipation terms. Similar but more complicated dispersion relations can be obtained in analogous models with higher-order corrections \cite{Alberte:2017cch}.

\begin{figure}[htbp]
\centering
\includegraphics[width=7cm]{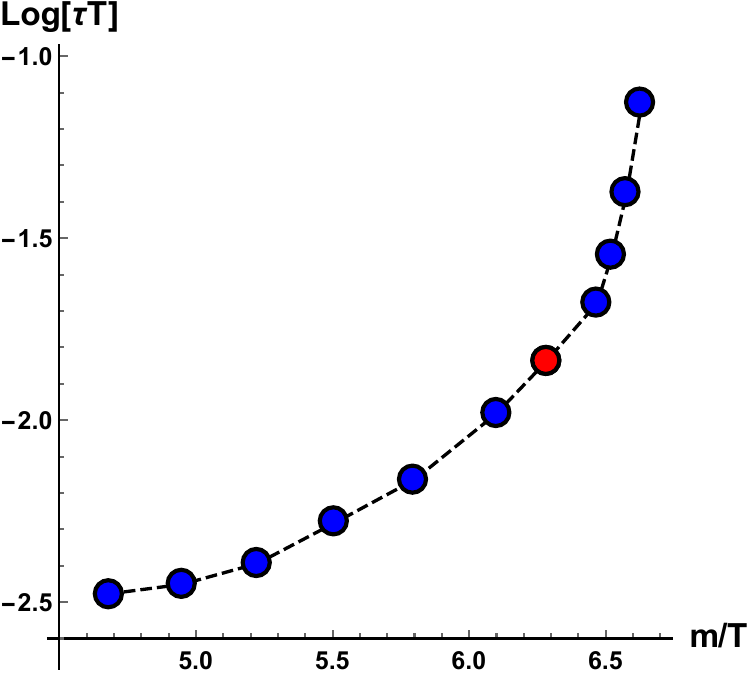}
\caption{The timescale $\tau$ governing the $k$-gap in function of temperature. The red dot is the analytic result at the ''self-dual'' temperature $T^\star$. The dashed line joins the numerical data. We note the similarity of this behaviour with the well-known Vogel-Fulcher-Tammann law for liquids.}
\label{law}
\end{figure}

We note that the relaxation time $\tau$ extracted from the holographic models does not coincide with the Maxwell relaxation time $\tau_{\rm M}=\eta/G_{\infty}$ where $\eta$ is the shear viscosity and $G_{\infty}$ the elastic modulus at infinite frequency. Nevertheless its temperature dependence is consistent with what is observed in liquids discussed in section \ref{liquid-section}. The numerical data in Fig. \ref{law} suggest that the behaviour of $\tau(T)$ is very similar to the well-known Vogel-Fulcher-Tammann law describing temperature dependence of relaxation time in liquids \cite{dyre}. Moreover, as shown in Fig. \ref{holkgap}, the $k$-gap moves to higher momenta with temperature increase. The last feature is in agreement with the trend seen in non-relativistic liquids discussed in section \ref{liquid-section}.

\subsection{$k$-gap ubiquity in other holographic models}
In the previous section, we have discussed two examples of how the $k$-gap emerges in holographic models. In addition to these examples, there are other holographic models discussed in different contexts which show the $k$-gap. Due to space limitations, we list and briefly describe them below.

1. \textit{Magnetohydrodynamics and emergent photon}. The use of global symmetries in the description of electromagnetism and in particular magnetohydrodynamics has received increasing attention in the holographic community in the recent years. Consistent progress has been made \cite{Hofman:2017vwr} and \cite{Grozdanov:2018fic} by using a bulk two form $J^{\mu\nu}$ enjoying a specific U(1) global symmetry. The Green function for such operator $\langle JJ\rangle$ exhibits a mode with a $k$-gap dispersion relation which has been interpreted as an emergent photon.

2. \textit{Gauss Bonnet}. Gauss Bonnet (GB) is a theory of gravity which deforms the Einstein-Hilbert action via specific higher-derivative terms encoding in an effective way ''quantum/stringy'' UV corrections to General Relativity. The $k$-gap dispersion relation has been observed in GB theories \cite{Grozdanov:2016fkt}. More recently, it was shown \cite{Grozdanov:2018fic} that the GB theory represents the gravity dual for the Israel-Stewart deformed relativistic hydrodynamics presented in section \ref{is}. It was shown that the GB coefficient $\lambda_{GB}$ controls the ''relaxation time'' $\tau_{\Pi}$ which is proportional to the speed of the emergent propagating mode at large momenta.

3. \textit{Holographic anomalous systems}. In Ref. \cite{Jimenez-Alba:2014iia}, the authors study a holographic model which exploits specific Chern-Simons terms in AdS$_5$ to encode the anomalies of the dual field theory in four dimensions. Anomalous transport and in particular the chiral magnetic waves are analyzed in the presence of an axial charg  relaxation time introduced via the Stueckelberg mechanism. It was shown that the chiral magnetic waves exhibit a $k$-gap, in agreement with the hydrodynamics approach and our discussion in previous sections.

4. \textit{Holographic P-wave superfluids}. Holographic superfluids/superconductors have been extensively studied in the last year, with the aim to understand the problem of high-temperature superconductivity. The authors of Ref. \cite{Arias:2014msa} studied the QNMs spectrum of a holographic P-wave superfluid and found the dispersion relation with the $k$-gap and diffusion-propagation crossover in the transverse sector.

5. \textit{Holographic plasmons}. The authors of Ref. \cite{Gran:2018vdn} studied plasmonic excitations in the Reissner-Nordstrom background and the poles of the density-density correlator $\langle \rho \rho \rangle$, finding the $k$-gap as well as more exotic dispersion relations that are similar in spirit to what was discussed in our section \ref{EMsec}. In Ref. \cite{Alberte:2017cch}, this effect was discussed in a different context.

6. \textit{Holographic fermions}. In Ref. \cite{Song:2018uds}, the authors studied Dirac bulk fermions with a finite mass $m$ and their spectral function. They obtained a crossover from a massive particle dispersion relation to the $k$-gap dispersion relation linked to the violation of the Breitenlohner-Freedman bound. It was proposed that this phenomenon can be related to the onset of an instability leading to the charge density waves background.

7. \textit{Top-Down constructions}. The $k$-gap phenomenon and the poles-collision mechanism has also been observed in the QNM transverse spectrum in different top-down and string theory-inspired models. In particular, the study in Ref. \cite{Myers:2008me} addressed holographic defects via D-brane constructions. The $k$-gap was also discussed studying the holographic dual of $\mathcal{N}=8$ super-Yang-Mills theory \cite{Miranda:2008vb}. More recently, the $k$-gap has been identified in genuine $R^4$ stringy corrections in $\mathcal{N}=4$ super-Yang-Mills theory \cite{Grozdanov:2016vgg} and in D-brane constructions related to anisotropic strongly-coupled liquids \cite{Itsios:2018hff}.

We conclude this section by observing that the $k$-gap phenomenon has been observed in disparate and apparently disconnected contexts. It is probably fitting to not refer to this effect as exotic as was considered in the examples above and to attempt to develop a more unified approach to this effect. In view of current interest in holographic models, it would be interesting to describe the $k$-gap phenomenon in terms of a general physical mechanism.

\section{Summary and outlook}

In summary, we have reviewed the emergence of gapped momentum states in several different areas of physics. Depending on the area, understanding GMS varies from a curiosity to a practical effect governing fundamental system properties or to a potentially general manifestation of underlying symmetries. We have seen how GMS emerge in liquids and supercritical fluids, strongly-coupled plasma, electromagnetic waves, Sine-Gordon model, quasihydrodynamics and holographic models. Throughout the review, we have been pointing out to essential physical ingredients required by GMS to emerge and making links between different areas of physics.

\begin{figure}
\includegraphics[width=\linewidth]{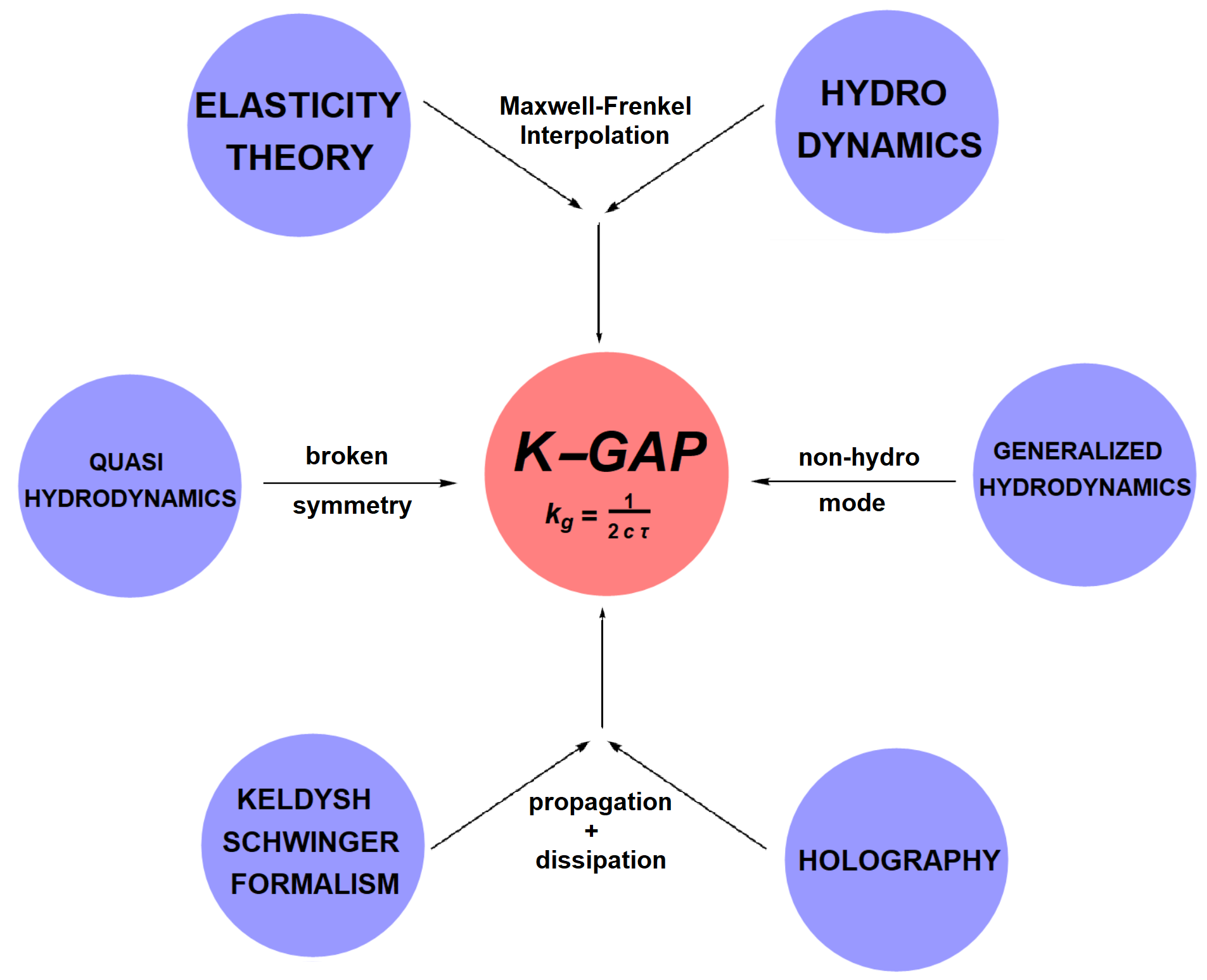}
\caption{Different approaches to gapped momentum states. The arrows indicate essential steps in these approaches.}
\label{final}
\end{figure}

We have reviewed several theoretical approaches to describe GMS. These approaches are summarized in Fig. \ref{final}.

In liquids, an equation leading to the $k$-gap emerged in the early work of Frenkel \cite{frenkel} who developed Maxwell idea \cite{maxwell} of liquid viscoelasticity and generalized the Navier-Stokes equation to endow the system with elastic response. As shown more recently \cite{ropp}, this generalization results in the $k$-gap equation

\begin{equation}
\boxed{\omega^2\,+\,\frac{i\,\omega}{\tau}\,-\,c^2\,k^2\,=\,0} \nonumber
\label{god}
\end{equation}

Interestingly, the same equation can be derived by starting with the elastic wave equation and, using Maxwell-Frenkel idea, generalizing it to endow the system with viscous response \cite{pre-l}. This implies that liquid dynamics and GMS in particular occupy a symmetric place between elastic and viscous description as discussed in section \ref{sss}.

Traditionally, the starting point of liquid description has been hydrodynamics, which is subsequently generalized to include the solid-like response at large $k$ and $\omega$. This approach similarly predicts the $k$-gap \cite{boon}. A similar generalization was discussed for dielectric systems \cite{madden1984consistent}. The main step to connect hydrodynamics with GMS is the introduction of non-hydrodynamic damped modes which result in the formulation of generalized hydrodynamics \cite{boon,hansen,march1,baluca}. The latter procedure is equivalent to the introduction of dissipation, quantified by the relaxation timescale $\tau$ which, combined with a hydrodynamic diffusive mode, produces the GMS as discussed in section \ref{due}.

More recently, a ``quasihydrodynamics'' approach has been developed \cite{Grozdanov:2018fic} (section \ref{due}) which adds an important step: the non-hydrodynamic response and timescale $\tau$ appear as the result of the soft breaking of a global symmetry. This enables one to generalize the GMS beyond the realm of traditional viscoelasticity and into more recent area of physics such as relativistic hydrodynamics, electromagnetism and field theories with anomalies.

In the quantum case, the interplay between propagating and dissipative terms and the emergence of GMS is discussed in the Keldysh-Scwhinger theory (section \ref{tre}). This technique suggests a connection between GMS and systems with propagating Goldstone's bosons in the presence of dissipation (see, e.g., Ref. \cite{Minami:2018oxl}).

Finally, the $k$-gap master equation \eqref{god} emerges in several holographic models (section \ref{quattro}) where the previous ideas are applied in the context of strongly-coupled quantum field theories. The important ingredient here is the presence of a non-hydrodynamic mode which collides with a hydrodynamic diffusive mode, producing the $k$-gap effect.

As Fig. \ref{final} illustrates, the key ingredients leading to GMS often include (a) the combination of elasticity and dissipation, (b) non-hydrodynamic modes and their collision with hydrodynamic modes or (c) softly broken symmetries. These ingredients emerge in different theories reviewed here.

As we clarify the origin of underlying physical ideas and equations related to GMS in different areas, we start to see analogies (and perhaps analogies between analogies) as well as area-specific differences. We believe this is conducive to better and deeper understanding of the physics involved. In this process, several interesting questions have emerged which we briefly list below.

Interestingly, the first experimental evidence for the $k$-gap in the form of a gapped dispersion relation has come not from the liquid state but from strongly-coupled plasma \cite{pg-exp}. On the other hand, there are currently no dispersion relations with the $k$-gap in liquids obtained directly from experiments. The experiments in liquids involve inelastic neutron or X-ray scattering at high temperature and are challenging as a result. As techniques are improving, it would be interesting to observe the gap directly in the transverse dispersion relation and subsequently extend experiments to higher temperatures so that the evolution of the gap can be observed in detail and compared with predictions from molecular dynamics simulations. It will also be interesting to observe gapped dispersion relations in supercritical systems, involving high-pressure experiments.

On the theoretical side, there are several interesting questions that need to be explored and understood. First, it is intriguing to ask whether the evolution of the $k$-gap could result in a phase transition, possibly non-conventional? In liquids, increasing the $k$-gap to the zone boundary $k_z$ results in the disappearance of quasi-harmonic transverse modes. This suggests that ($k_z-k_g$) may play the role of the order parameter which becomes zero when all transverse modes disappear, corresponding to the loss of system's shear rigidity at all available $k$-points. Exploring this idea in condensed matter systems and other areas and subsequently comparing the results might bring about new effects and advance our physical understanding.

Second, it is intriguing to explore the interpretation \cite{Grozdanov:2018fic} relating the emergence of the $k$-gap to the presence of a weakly broken (global) symmetry. Identifying such a symmetry would be important and fascinating. It is currently unclear what is the definition of such a global symmetry, which is usually related to a dual, and non-canonical, description of the system \cite{Grozdanov:2016tdf}. More generally, it would be interesting to construct a more universal theoretical picture of GMS using the quasihydrodynamics framework \cite{Grozdanov:2018fic}. Once this is developed, we can explore its application to challenging problems in condensed matter physics including understanding of the challenging problem of liquid dynamics and thermodynamics. Another area of application includes superconducting states with relaxation. A finite density of vortices provides the relaxation time $\tau_V$ which will appear in the $k$-gapped dispersion relation. A hydrodynamic description at low momenta has been developed \cite{Davison:2016hno}, and it would be interesting to study GMS in this system.

Third, in most cases the $k$-gap appears beyond the controlled hydrodynamic limit $k/T \ll 1$. Nevertheless, substantial effort has been devoted in the last years to extend the definition of non-convergent series like hydrodynamics beyond their perturbative description \cite{Aniceto:2015mto,Basar:2015ava,Heller:2015dha,Tinti:2018nrp,Tinti:2018qfb,Buchel:2016cbj}. In particular, all-order hydrodynamics and viscoelasticity \cite{Baggioli:2018bfa} can be resummed using Borel and Pad\'e methods leading to interesting properties. It is conceivable to expect that those re-summation techniques could be able to predict and confirm the presence of a momentum gap beyond the hydrodynamic approximation.

Fourth, the $k$-gap dispersion relation is present in several holographic models which, because of their bottom-up nature, are defined for large $N$ and at strong coupling. Similarly, the key to dynamics and collective modes in liquids are strong intermolecular interactions. It would be interesting to explore the role of strong coupling in the emergence of the $k$-gap in holographic models. More generally, this extends to strongly-coupled field theories where many interesting problems exist and where new approaches to those problems need to be developed.

Finally, it would be interesting to understand the essential ingredient of GMS from the point of view of general theory and seek a connection between GMS and hydrodynamics as an effective field theory. Recently, hydrodynamics has been revealed to be at work in unexpected situations such as the motion of electrons in very clean metals \cite{Lucas:2017idv} or the dynamics of phonons in 2D materials \cite{Cepellotti2015}. It will be interesting to determine whether the $k$-gap is present in those systems and ascertain potential experimental consequences in, for example, graphene or Weyl semimetals.

\section*{Aknowledgments}
We are grateful to Saso Grozdanov, Napat Poovuttikul, Alex Buchel, Oriol Pujolas, Amadeo Jimenez Alba, Andy Lucas, Niko Jokela, Jay Armas, Akash Jain, Karl Landsteiner, Paul Romatschke, Martin Ammon, Nabil Iqbal, Michel Cot\'e, Normand Mousseau, Jaakko Nissinen, Ganapathy Baskaran, Mike Blake, Tomas Andrade, Alessio Zaccone, Peter Arseev and Stas Yurchenko for discussions.
We thank especially Alex Buchel, Saso Grozdanov, Laurence Noirez and Alessio Zaccone for reading a preliminary version of this paper and providing useful feedback.\\
K. Trachenko is grateful to EPSRC for support. M. Baggioli acknowledges the support of the Spanish Agencia Estatal de Investigacion through the grant “IFT Centro de Excelencia Severo Ochoa SEV-2016-0597 and thanks NORDITA and the University of Helsinki for the warm hospitality during the final stages of this work.

\section*{Appendix A}

We use the abbreviated notation, which means operations with complex variables:
\begin{align*}
&\int \mathfrak{D}z \exp\left[-zAz +az\right]=
\iint \mathfrak{D}(\Re z)\mathfrak{D}(\Im z)\exp\left[-\iint \mathrm{d}{\bf x}\mathrm{d}{\bf y}\,z_{\bf x}^*A_{\bf x-y}z_{\bf y}+ \right. \\ &\left.
\int \mathrm{d}{\bf x} \,a_{\bf x}^*z_{\bf x}+\int\mathrm{d}{\bf x}\,z_{\bf x}^*a_{\bf x}\right]=\dfr{1}{\det A}\exp\left[\iint\mathrm{d}{\bf x}\mathrm{d}{\bf y}\, a_{\bf x}^*A^{-1}_{\bf x-y}a_{\bf
y}\right]=\dfr{\exp\left[a^*A^{-1}a\right]}{\det A}
\end{align*}
where $\int \mathfrak{D}z$ denotes the functional integration over $z$ field, and $z^*$ field is the complex conjugate to $z$.

\section*{Appendix B}

An useful too is the algebra of bosonic coherent states \cite{berezin,K34}. In this formulation, the state of a many body system can be presented with the bosonic annihilation and creation operators, $\hat b$ and $\hat b^{\dag}$, which operate in the space of the boson occupation numbers $n$ in the following way:
\begin{gather*}
\hat b|n\rangle =\sqrt{n}|n-1\rangle, \quad \hat b^{\dag}|n\rangle =\sqrt{n+1}|n+1\rangle
\end{gather*}
where the number states $|n\rangle$ form a complete orthonormal basis: $\langle n|n'\rangle=\delta_{n,n'}$, and $\sum\limits_n |n \rangle \langle n|=\hat 1$. By acting on an arbitrary basis state, one may check the following relations
\begin{gather*}
\hat b^{\dag}\hat b|n\rangle =n|n\rangle, \quad \hat b\hat b^{\dag}|n\rangle =\sqrt{n+1}|n\rangle,\quad [\hat b,\,\hat b^{\dag}]=\hat 1
\end{gather*}
Therefore, the coherent state of the system, parameterized by a complex number $\phi $, is defined as eigenstates of the annihilation operator with the eigenvalue $\phi $
\begin{gather*}
\hat b|\phi\rangle =\phi|\phi\rangle, \quad \langle \phi | \hat b^{\dag} =\phi^*\langle \phi |
\end{gather*}
where the star denotes complex conjugation. Then, the matrix elements in the coherent state basis of any normally ordered operator $\hat H(\hat b^{\dag},\,\hat b)$ are given by
\begin{gather*}
\langle\phi|\hat H(\hat b^{\dag},\,\hat b)|\phi'\rangle =\hat H(\phi^*,\,\phi')\langle\phi|\phi'\rangle
\end{gather*}
One can check that the following linear superposition of the pure number states:
\begin{gather*}
|\phi\rangle=\sum\limits_{n=0}^{\infty}\dfr{\phi^n}{\sqrt{n!}}|n\rangle=\sum\limits_{n=0}^{\infty}\dfr{\phi^n}{\sqrt{n!}}(\hat b^{\dag})^n|0\rangle=e^{\phi \hat b^{\dag}}|0\rangle
\end{gather*}
where $|0\rangle $ is the vacuum state, is the required eigenstate of the operator $\hat b$. Upon Hermitian conjugation, one
finds $\langle \phi|=\sum\limits_{n}\langle n|{\phi^*}^n/\sqrt{n!}$.

Note that the coherent states are not mutually orthogonal: their set forms an overcomplete basis. The overlap of two coherent states is given by
\begin{gather*}
\langle \phi|\phi'\rangle=\sum\limits_{n,n'=0}^{\infty}\dfr{{\phi^*}^n{\phi'}^{n'}}{\sqrt{n!n'!}}\langle n|n'\rangle=\sum\limits_{n=0}^{\infty}\dfr{(\phi^*\phi')^n}{\sqrt{n!}}=e^{\phi^*\phi'}
\end{gather*}
where we employed the orthonormality of the pure number states. One may express resolution of unity in the coherent states basis. It takes the following
form:
\begin{gather*}
\hat 1=\int\mathfrak{D}\phi e^{-|\phi|^2} |\phi\rangle\langle \phi |
\end{gather*}
where $\mathfrak{D}\phi=\mathfrak{D}(\Re \phi)\mathfrak{D}(\Im \phi)$ denotes the functional integration over real and imaginary parts of $\phi$ field.
The action of the time derivation operator on the state function, important for us, is
\begin{align*}
&\partial_t|\phi\rangle=\sum\limits_{n=0}^{\infty}\dfr{\partial_t\phi^n}{\sqrt{n!}}|n\rangle=
(\partial_t\phi)\sum\limits_{n=0}^{\infty}\dfr{n\phi^{n-1}}{\sqrt{n!}}|n\rangle=
(\partial_t\phi)\sum\limits_{n=0}^{\infty}\dfr{\sqrt{n}\phi^{n-1}}{\sqrt{(n-1)!}}|n\rangle=\\&=
(\partial_t\phi)\hat b^{\dag}\sum\limits_{n=0}^{\infty}\dfr{\phi^{n-1}}{\sqrt{(n-1)!}}|n-1\rangle=
(\partial_t\phi)\hat b^{\dag}|\phi\rangle= \phi^*(\partial_t \phi )|\phi\rangle
\end{align*}
Therefore,
\begin{gather*}
\langle \phi|\partial_t|\phi\rangle=\phi^*(\partial_t \phi )\langle \phi|\phi\rangle =\phi^*(\partial_t \phi )e^{|\phi|^2}
\end{gather*}

\section*{Appendix C}

The equilibrium state does not change with time, and the time instant $ t = \infty $ can always be ``shifted'', therefore $\langle \phi_{eq}, \infty|=\langle \phi_{eq}, \infty | \phi_{eq}',
\infty+\tau \rangle \langle \phi_{\infty}', \infty+\tau |=\langle \phi_{eq}, \infty | e^{-\frac i{\hbar}\tau E_{eq}\phi_{eq}\phi_{\infty}}$, where $E_{eq}$ is the equilibrium state energy.
In this case we can express the transition probability as follows:
\begin{align*}
&\langle \phi_{eq}, \infty |\phi_0, 0 \rangle = \langle \phi_{eq}|\mathcal{\hat U}_{\infty} |\phi_0 \rangle = \\&=
\int \mathfrak{D}\phi_{eq}\mathfrak{D}\phi
\exp \left[\int\limits^{\infty}_{0}\mathrm{d}t\,(\hbar^{-1}\mathcal{L}(\phi)-\phi\partial_t\phi) -\right.  \left.
\hbar^{-1}\tau E_{eq}\phi_{eq}\phi_{\infty} -\hbar^{-1}\tau E_{eq}\phi_{eq}\phi_{eq}\right]
\end{align*}
We note that the obtained transition probability does not depend on the choice of the integration boundary with respect to time corresponding to the final equilibrium state. If we shift the
final time, $ t = \infty \to t = \infty + \tau $, the result of the averaging will not change:
\begin{align*}
& \langle \phi_{eq}', \infty+\tau |\phi_0, 0 \rangle =
\int \mathfrak{D}\phi_{eq}'\mathfrak{D}\phi_{eq}\mathfrak{D}\phi
\exp \left[ \int\limits^{\infty}_{0}\mathrm{d}t\,(\hbar^{-1}\mathcal{L}(\phi)- \right. \\ &\left.
\phi\partial_t\phi ) -\hbar^{-1}\tau E_{eq}\phi_{eq}\phi_{\infty}-\hbar^{-1}\tau E_{eq}\phi_{eq}\phi_{eq}'- \hbar^{-1}\tau E_{eq}\phi_{eq}'\phi_{eq}' \right]
  \\ & =
\int \mathfrak{D}\phi_{eq}\mathfrak{D}\phi \exp \left[\int\limits^{\infty}_{0}\mathrm{d}t\,(\hbar^{-1}\mathcal{L}(\phi)-\phi\partial_t\phi) -\hbar^{-1}\tau E_{eq}\phi_{eq}\phi_{\infty} -\right. \\ & \left.
\hbar^{-1}\tau E_{eq}\phi_{eq}\phi_{eq}\right]=\langle \phi_{eq}, \infty |\phi_0, 0 \rangle
\end{align*}
Therefore, this probability does not depend on shifting of time.

\section*{Appendix D}

The boson system energy in the frequency representation is:
\begin{align*}
&\mathcal{E}=\int\limits_{-\infty}^{\infty}\mathrm{d}\omega\,\hbar\omega \rho(\omega)= \dfr 12\int\limits_{-\infty}^{\infty}\hbar\omega \left(1+\dfr{2}{e^{\beta\hbar\omega}-1}\right)\mathrm{d}\omega=
\\&=
\dfr 12\int\limits_{-\infty}^{\infty}\mathrm{d}\omega\,\hbar\omega \left(\dfr{e^{\beta\hbar\omega}+1}{e^{\beta\hbar\omega}-1}\right)=
\dfr 12\int\limits_{-\infty}^{\infty}\mathrm{d}\omega\,\hbar\omega \,\mbox{coth}\left(\dfr{\hbar\omega}{2kT}\right)
\end{align*}
Therefore the equilibrium density of states for the boson system is:
\begin{gather*}
\rho (\omega)=\dfr 12\,\mbox{coth}\left(\dfr{\hbar\omega}{2kT}\right)
\end{gather*}
\section*{Appendix E}
In this Appendix, we provide details related to the calculation of the shear QNM spectrum for the holographic model \cite{Andrade:2013gsa} presented in section \ref{sechol}.

The transverse or shear perturbations are encoded in the fluctuations $a_x,\,h_{tx}\equiv u^2 \delta g_{tx},\,h_{xy}\equiv u^2\delta g_{xy},\,\delta \phi_x,\,\delta g_{xu}$. Assuming the radial gauge, \textit{i.e.} $\delta g_{xu}=0$, and using the ingoing Eddington-Finkelstein coordinates
\begin{equation}
ds^2=\frac{1}{u^2} \left[-f(u)dt^2-2\,dt\,du + dx^2+dy^2\right]
\end{equation}
the remaining equations read
\begin{align}
&-2h_{tx}+u\,h_{tx}'-i\,k\,u\, h_{xy}-\left( k^2\,u+2\,i\,\omega\right)\,\delta \phi_x+u\,f\, \delta \phi_x''
\nonumber\\&+\left(-2\,f+u\,(2 i \omega+f')\right)\delta \phi_x'\,=\,0
\,\,;\nonumber\\[0.1cm]
&2\,i\,m^2\,u^{2}\omega\,\delta \phi_x+u^2\, k\,\omega\, h_{xy}+\left(6+k^2\,u^2 \, -6 f+2uf'\right) \, h_{tx}
\nonumber\\&+\left(2\,u\,f-i\,u^2\omega\right)h_{tx}'-u^2\,f\,h_{tx}''\,=\,0 \,\,;
\nonumber\\[0.1cm]
&-iku^2h_{tx}'-2\,i \, k \,m^2 \,u^{2}\delta \phi_x+2 h_{xy}\left(3+i\,u \, \omega-3f+uf' \right)
\nonumber\\&+2i\,k\,u\,h_{tx}-\left(2i\,u^2 \, \omega-2uf+u^2\,f'\right)h_{xy}'-u^2\,f\, h_{xy}''\,=\,0\,\,;
\nonumber\\[0.1cm]
&2\,h_{tx}'-u\,h_{tx}''-2m^2\,u \, \delta\phi_x'+ik\,u\,h_{xy}'\,=\,0\,\,.
\end{align}
where we set the momentum $k$ along the $y$ direction: this freedom comes from the fact that the model enjoys SO(2) symmetry in the $(x,y)$ subspace.

These equations can be solved numerically. The asymptotics of the various bulk fields close to the UV boundary $u=0$ are:
\begin{align}
&\delta \phi_x\,=\,\phi_{x\,(l)}\,(1\,+\,\dots)+\,\phi_{x\,(s)}\,u^{3}\,(1\,+\,\dots)\,,\quad\nonumber\\ &h_{tx}\,=h_{tx\,(l)}\,(1\,+\,\dots)\,+\,h_{tx\,(s)}\,u^{3}\,(1\,+\,\dots)\,,\nonumber\\ &h_{xy}\,=h_{xy\,(l)}\,(1\,+\,\dots)\,+\,h_{xy\,(s)}\,u^{3}\,(1\,+\,\dots)\,.
\end{align}
In these coordinates, the ingoing boundary conditions at the horizon are automatically satisfied by regularity at the horizon. It follows that the retarded Green's functions can be defined as:
\begin{align}\label{greenF}
&\mathcal{G}^{\textrm{(R)}}_{T_{tx}T_{tx}}\,=\,\frac{2\,\Delta-d}{2}\,\frac{h_{tx\,(s)}}{h_{tx\,(l)}}\,=\,\frac{3}{2}\frac{h_{tx\,(s)}}{h_{tx\,(l)}}\,,\nonumber\\
&\mathcal{G}^{\textrm{(R)}}_{T_{xy}T_{xy}}\,=\,\frac{2\,\Delta-d}{2}\,\frac{h_{xy\,(s)}}{h_{xy\,(l)}}\,=\,\frac{3}{2}\frac{h_{xy\,(s)}}{h_{xy\,(l)}}\,.
\end{align}
where spacetime dependences are omitted for simplicity and the conformal dimension of the stress tensor is simply $\Delta=3$. From the poles of the Green functions defined as the zero of the leading terms in the UV expansions, we can read off the QNMs frequency at finite momentum.


\bibliographystyle{unsrt}

\end{document}